\documentclass[12pt]{article}
\usepackage{a4wide}
\usepackage{graphicx}
\usepackage{amssymb}
\usepackage{amsmath}
\usepackage{slashed}
\usepackage{cite}
\usepackage{multicol}
\usepackage{braket}
\usepackage{bmpsize}
\usepackage{xcolor}

\newcommand{\be}{\begin{equation}}
\newcommand{\ee}{\end{equation}}
\newcommand{\ba}{\begin{eqnarray}}
\newcommand{\ea}{\end{eqnarray}}

\begin{document}

\allowdisplaybreaks

\begin{titlepage}
\begin{flushright}
\end{flushright}
\vfill
\begin{center}
{\Large\bf Loop enhancement of direct detection cross section in a fermionic dark matter model}
\vfill
{\bf Khadije Rahi Maleki and Karim Ghorbani}\\[1cm]
{Physics Department, Faculty of Sciences, Arak University, Arak 38156-8-8349, Iran}
\end{center}
\vfill
\begin{abstract}
We investigate the effect of one loop quantum corrections on the elastic 
scattering of dark matter off the nucleon in a fermionic dark matter model. The model introduces two new singlet 
fermions and a singlet scalar. The fermions communicate with the SM particles through a Higgs portal. 
It is found that some viable regions in the parameter space respecting the bounds from
the observed relic density, the Higgs invisible decay width, and direct detection experiment, 
will be shrunk significantly when one loop effects are taken into account. 
The regions already resided below the neutrino floor, partly may come into regions which are
testable by the current or future direct detection experiments.
In addition, some regions being viable at tree level, may be excluded when quantum corrections are included.

\end{abstract}
\vfill
\vfill
{\footnotesize\noindent }

\end{titlepage}

\section{Introduction}
\label{int}
The thermal production of weakly interacting massive particles (WIMPs) as dark matter (DM) candidates 
is a quite natural and ubiquitous mechanism \cite{Lee:1977ua,Steigman:1984ac,Arcadi:2017kky,Bergstrom:2000pn,Steigman:2012nb,Leane:2018kjk}. 
The particle nature of dark matter will be demystified if its direct interaction with ordinary matter 
shows up in the so called direct detection (DD) experiments. 
On the theory side, the DM interaction with atoms may be so weak that it lies below 
the neutrino floor (NF) or not being detectable in the current DD experiments.
The point is that when we present theoretical prediction of the DM-nucleon interaction, 
it is fare to know how much the computations are accurate in the perturbation theory. 

There are models with dark matter candidates which may evade strong bounds from 
DD experiments. These models can be classified into two types. The first type occurs
when the scattering amplitude at tree level and at zero momentum transfer vanishes by virtue 
of a symmetry breaking pattern. An example is the complex scalar DM model, wherein the DM 
candidate is a pseudo-Goldstone boson after a softly broken symmetry
\cite{Barger:2010yn,Gonderinger:2012rd,Gross:2017dan}. 
In the same category, it is demonstrated that via a scale symmetry breaking in a model with 
scalar dark matter, tree level DD cross section decreases significantly \cite{Ghorbani:2022muk}.
The second type contains models in which the scattering cross section of DM off nucleons 
is velocity or momentum suppressed, thus, escaping the DD bounds. Some instances are 
thermal DM candidates having a pseudo-scalar type interaction with nucleons \cite{Ghorbani:2014qpa,Berlin:2015wwa,Jia:2015uea,Fan:2015sza,Yang:2016wrl,DuttaBanik:2016jzv,Baek:2017vzd,Ghorbani:2017qwf,Ghorbani:2017jls,YaserAyazi:2018pea,Abe:2019wjw,Ghorbani:2016edw,DiazSaez:2021pmg,Abe:2019wku,Kozaczuk:2015bea,Abe:2020obo,Matsumoto:2018acr}. 
When DM-nucleon scattering cross section is suppressed at tree-level, it deems reasonable to 
include loop corrections. This may bring regions below the neutrino floor within reach of 
the present or future DD experiments. Works in this direction have been growing and the present
findings generally show that the quantum corrections modify the viable parameter space 
significantly, see references \cite{Li:2018qip,Herrero-Garcia:2018koq,Hisano:2018bpz,Han:2018gej,Azevedo:2018exj,Ishiwata:2018sdi,Ghorbani:2018pjh,Ertas:2019dew,Li:2019fnn,Chao:2019lhb,Glaus:2019itb,Borschensky:2020olr,Chao:2020fme,Bell:2018zra,Abe:2018emu}. 

There may be other situations that loop effects of DM-nucleon scattering 
become salient. This is when the coupling involving in the DM-nucleon scattering cross section 
has small effect on the DM annihilation cross section, and in fact 
a second coupling mainly controls the size of the relic density.
These types of models may be extensions to the simplest simplified DM models which are 
excluded almost entirely by the current direct detection experiments. The focus in this work is 
on a fermionic DM model with two fermion WIMPs, one of which playing the role of DM. 
As shown in \cite{Ghorbani:2018hjs}, a large viable parameter space is available in this model
satisfying the observed relic abundance and respecting the DD bounds. Working at tree level DM-nucleon scattering, 
it is found that parts of the parameter space are below the neutrino floor, 
and there are regions which respect the latest DD bounds. 
Now, by incorporating one loop effects, the arising question is that how the 
regions placing below the neutrino floor or the regions which are allowed by DD upper 
limits, will change to become regions above the neutrino floor or 
excluded regions, respectively, .   

The paper has the following structure. The dark matter model including 
two fermionic WIMPs which communicate with the SM particles via scalar-Higgs portal
are introduced in sec.\ref{model}. A brief discussion is given in 
sec.\ref{recap} about the annihilation cross section and the tree level
DD cross section of the DM particle.
The invisible decay of the SM Higgs is introduced in sec.\ref{invisible}.
In sec.\ref{DDtreelevel}, by imposing constraints from observed relic 
abundance and upper limits from DD experiments we show the viable parameter 
space by considering only tree level DD cross. 
We introduce in sec.\ref{renormalization} the renormalization program for 
the present model, and find expressions for the counter terms needed 
to cancel the relevant divergences at one loop level.
We collect the formulas representing different loop contributions to the DD cross section at one loop order
in sec.\ref{DDCS}. In sec.\ref{results} our results are presented. The conclusion is given in 
sec.\ref{conclusion}.

\section{Model}
\label{model}
We describe here a renormalizable extension to the SM with two extra Dirac 
fermion fields $\chi_1$ and $\chi_2$, which are singlet under gauge symmetry 
of the SM. The new fermions communicate with the SM particles via a real singlet scalar $\varphi$.  
The scalar potential of the model accommodating the singlet scalar and the SM Higgs reads
\begin{equation}
\label{potential}
V(\varphi, H) =\mu _{H}^2 H^\dagger H +  \lambda_H (H^\dagger H)^2
+  \lambda_1 \varphi H^\dagger H  +  \lambda_2 \varphi^2 H^\dagger H 
+ \frac{1}{2} m^2 \varphi^2 + \lambda_0 \varphi + \lambda_3 \varphi^3 + \lambda \varphi^4,
\end{equation}
The singlet scalar field gets a zero vacuum expectation value (vev), and the Higgs doublet in the unitary gauge 
is parameterized around its vacuum as
\begin{equation}
\label{2}
\qquad      H=\begin{pmatrix} 0  \\  \frac{{{v}_{H}}+{h}^{\prime}}{\sqrt{2}}  \\ \end{pmatrix},
\end{equation}
where ${{v}_{H}}$  is the vacuum expectation value of the Higgs field with ${v}_{H}=246$ GeV.
We choose $\mu_{H}$ and $\lambda_0$ such that at tree level the tadpole terms 
for the fields, s and $h^\prime$, become zero; i.e., $t_{\varphi} = t_{H} = 0$.
The fermion fields $\chi_1$ and $\chi_2$ transform under ${{\mathbb{Z}}_{2}}$ symmetry as 
$\chi_i \to -\chi_i$. The particles of the standard model interact with 
the DM only through the Higgs portal. In this work we set $\lambda =\lambda_3 = 0$. 
The renormalizable Lagrangian containing the interactions of the new fermions with the 
singlet scalar is as follows
\begin{equation}
\label{4}
\mathcal{L}_{\text{Dark}}= \kappa_1 \varphi \bar{\chi}_1 \chi_1
+\kappa_2 \varphi \bar{\chi}_2 \chi_2 + (\kappa_{12} \varphi \bar{\chi}_1 \chi_2 + h.c ) \,.
\end{equation}
The mass matrix of the scalars is not diagonal, so in the following we obtain the physical 
masses and eigenstates. By taking double derivative of the potential with respect to ${h}^{\prime}$ and 
$\varphi$, the elements of the mass matrix are obtained 
\begin{equation}\label{10}
{{M}^{2}}=\left( \begin{matrix}
m_{\varphi}^2 = m^2 + \frac{1}{2} \lambda_2 v_H^2  &  m_{\varphi h^\prime}^2 = \lambda_1 v_H  \\
m_{\varphi h^\prime}^2 = \lambda_1 v_H  & m_{h^\prime}^2= \frac{1}{2} \lambda_H v_H^2 \\
\end{matrix} \right) \,.
\end{equation}
The mass eigenstates that diagonalize the mass matrix are defined 
\begin{equation}
\label{11}
h= s_\omega \varphi  + c_\omega h^\prime, ~~~ 
s= c_\omega \varphi  - s_\omega h^\prime \,,
\end{equation}
where $s_\omega = \sin \omega$ and $c_\omega = \cos \omega$, 
and the mixing angle, $\omega$, is given by
\begin{equation}
	\tan \,2\omega =\frac{2m_{\varphi h^\prime}^2}{m_{h^\prime}^2-m_{\varphi}^2} \,.
\end{equation}
The mass eigenvalues are obtained as 
\begin{equation}
	m_{h,s}^{2}=\frac{m_{\varphi}^2 + m_{h^\prime}^2}{2}\pm \frac{1}{2} (m_{\varphi}^2-m_{h^\prime}^2) \sec 2\omega \,.
\end{equation}
In our numerical computations we take $m_h= 125$ GeV as the SM Higgs mass, and $m_s$ 
is the physical mass of the singlet scalar being a free parameter in our model.
The couplings $\lambda_H$ and $\lambda_1$  are obtained in terms 
of the mixing angle and the physical masses of the scalars,
\begin{equation}
\label{14}
\lambda_H = \frac{m_s^2 \sin^2 \omega + m_h^2 \cos^2 \omega}{2 v_H^2},~~~
\lambda_1 = \frac{m_s^2-m_h^2}{2v_H} \sin 2\omega  \,.
\end{equation}
A set of independent free parameters in our model is: $m_1, m_2, m_s, \lambda_2, \kappa_1, \kappa_2, \kappa_{12}, \omega$.
We may define the mass difference between the two fermions as $\Delta = m_2 - m_1$, 
assuming that the light fermion is $\chi_1$, being our dark matter candidate with mass $m_{\text{DM}}$.
Without lose of generality, we can take $\kappa_2 \sim 0$. The size of its one loop correction 
is then $\delta \kappa_2 \sim \kappa_{12}^2/(16\pi^2)$, which is still quite small for 
$\kappa_{12} \sim {\cal O}(1)$. 
This will simplify our calculations. 
We end up having seven independent free parameters: $m_1, m_2, m_s, \lambda_2, \kappa_1, \kappa_{12}, \omega$. We will use $m_1 = m_{\text{DM}}$ interchangeably.
The quartic couplings of the potential are constrained theoretically by requiring the stability
of the potential. The stability conditions are $\lambda_H > 0$, $\lambda > 0$, 
and in case $\lambda_2 < 0$, then $\lambda \lambda_H > \lambda^2_2/4$.

\section{Annihilation Cross Section vs DD Cross Section}
\label{recap}
A detailed discussion is laid out on the relic density calculations and the tree-level DD cross section in \cite{Ghorbani:2018hjs} for the present model. 
Here we provide a short recap. There are two ways through which fermion DM can annihilate. 
(1) Through s-channel; by mediating $h$ or $s$ scalars, where DM may annihilate to the SM particles, a pair of Higgs, and a pair of singlet scalars. 
(2) Through t/u-channel; by mediating $\chi_1$ or $\chi_2$, where DM may annihilate to a pair of Higgs or a pair of singlet scalars. The annihilation cross section formulas are provided in Appendix A.

The condition where $\kappa_{12} = 0$, reduces the model to the simplest scenario. 
In this case, the annihilation cross section is a combination of terms each of which 
is proportional to $\kappa_1^2, \kappa_1^3$ or $\kappa_1^4$.
On the other hand, the elastic scattering cross section is proportional to $\kappa_1^2$.
Therefore, in order to get the relic density right, large coupling $\kappa_1$ is required, 
and this will give rise to quite a large direct detection cross section. 
As demonstrated in \cite{Ghorbani:2018hjs}, unless the mixing angle is quite small, the entire parameter space of the simplest fermionic model
(other than a resonance region with $m_{\text{DM}} \sim m_s/2$)
is excluded by DD experiments. 

For a given mixing angle, when $\kappa_{12} \ne 0$ then the DM annihilation cross section finds some new 
contributions as a function of $\kappa_{12}^4, \kappa_{12}^2 \kappa_1^2$, and so on. 
However, the DD cross section at tree level in this case remains intact, being proportional to $\kappa_1^2$.
Now it is possible for the two cross sections to move in opposite directions. 
To evade DD bounds (demanding small DD cross section), small $\kappa_1$ 
is required, and at the same time to have 
large enough annihilation cross section, terms proportional to $\kappa_{12}^4$ 
will dominate when sizable $\kappa_{12}$ of ${\cal O}$(1) is picked out.    

\section{Invisible Higgs Decay}
\label{invisible}
In this model, if kinematically allowed, the SM Higgs may decay invisibly as: $h \to ss$, $h\to \chi_1 \chi_1$, and $h\to \chi_1 \chi_2$. 
These new decay channels will alter the theoretical decay width of the Higgs,
\begin{equation}
\begin{split}
\Gamma^{\text{tot}}_{h} &= \cos^2(\omega)~\Gamma^{\text{SM}}_{h} 
+ \Theta(m_h -2m_s) \Gamma(h\to ss) + \Theta(m_h -2m_{\chi_1}) \Gamma(h\to \chi_1 \chi_1) 
\\&
+ \Theta(m_h -m_{\chi_1}-m_{\chi_2}) \Gamma(h\to \chi_1 \chi_2) \,,
\end{split}
\end{equation}
where $\Theta$ is the step function, and $\Gamma^{\text{SM}}_{h}$ is the Higgs decay width
computed within the SM. The Higgs decay to a pair of singlet scalars has the width
\begin{equation}
\Gamma(h\to s s) = \frac{{\cal A}^2}{128\pi m_h}\sqrt{1-4m^2_s/m^2_h} \,,
\end{equation}
where, the coupling ${\cal A}$ is given by 
\begin{equation}
 {\cal A} = (2\sin \omega -3 \sin^2 \omega) \lambda_1 + (6 \sin^2 \omega - 2 \cos \omega) v_H \lambda_2
    - 6 \sin^2 \omega \cos \omega~v_H \lambda_H \,.
\end{equation}
The decay width for the Higgs decay to two identical fermions is 
\begin{equation}
 \Gamma(h \to \chi_1 \chi_1)  = \frac{\kappa^2_1 m_h \sin^2 \omega}{8 \pi} (1-4 m^2_{\chi_1}/m^2_h)^{3/2} \,.
\end{equation}
When the Higgs particle decays to $\chi_1 \chi_2$, its decay width reads
\begin{equation}
 \Gamma(h\to \chi_1 \chi_2) = \frac{\kappa_{12}^2 \sin^2 \omega}{8\pi m^3_h}
 [m^2_h-(m_{\chi_1}+m_{\chi_2})^2]^{3/2} [m^2_h-(m_{\chi_1}-m_{\chi_2})^2]^{1/2} \,.
\end{equation}
There is an experimental upper limit on the branching ratio of the invisible Higgs decay at 
the 95\% CL, as $Br(h \to \text{invisible}) \lesssim 0.18$ \cite{CMS:2018yfx}. 
The mass of the Higgs is measured to be $\sim 125$ GeV, and its total decay width is 
$\Gamma_{\text{Higgs}} = 3.2^{+2.8}_{-2.2}$ MeV \cite{ParticleDataGroup:2020ssz}. 
The constraint from the invisible Higgs decay becomes more effective for 
small mass of the singlet scalar and fermions, as well as, for large mixing angle.

\section{DD Cross Section at Tree Level}
\label{DDtreelevel}
\begin{figure}
\hspace{-.5cm}
\begin{minipage}{.52\textwidth}
\includegraphics[width=.71\textwidth,angle =-90]{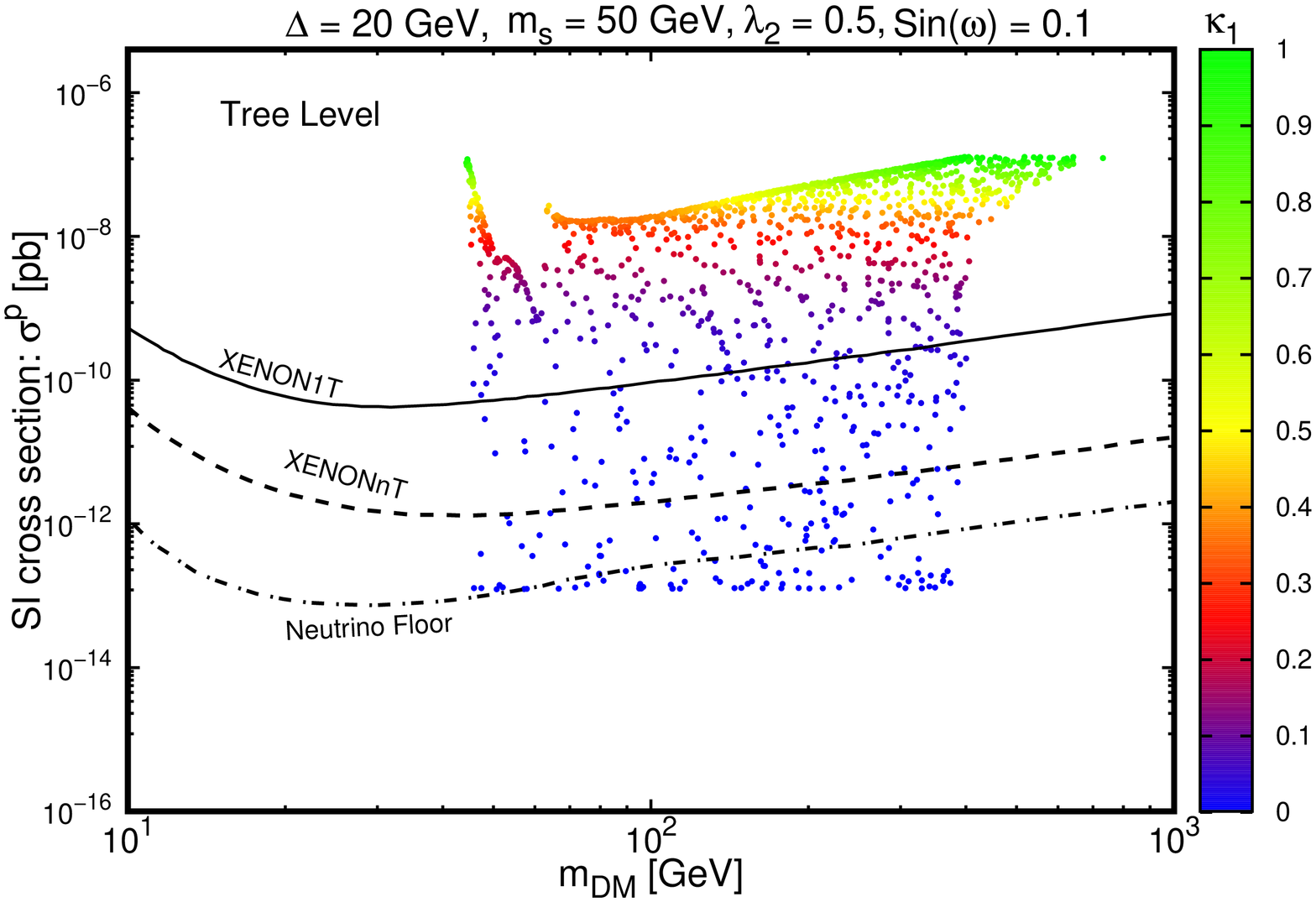}
\end{minipage}
\begin{minipage}{.52\textwidth}
\includegraphics[width=.71\textwidth,angle =-90]{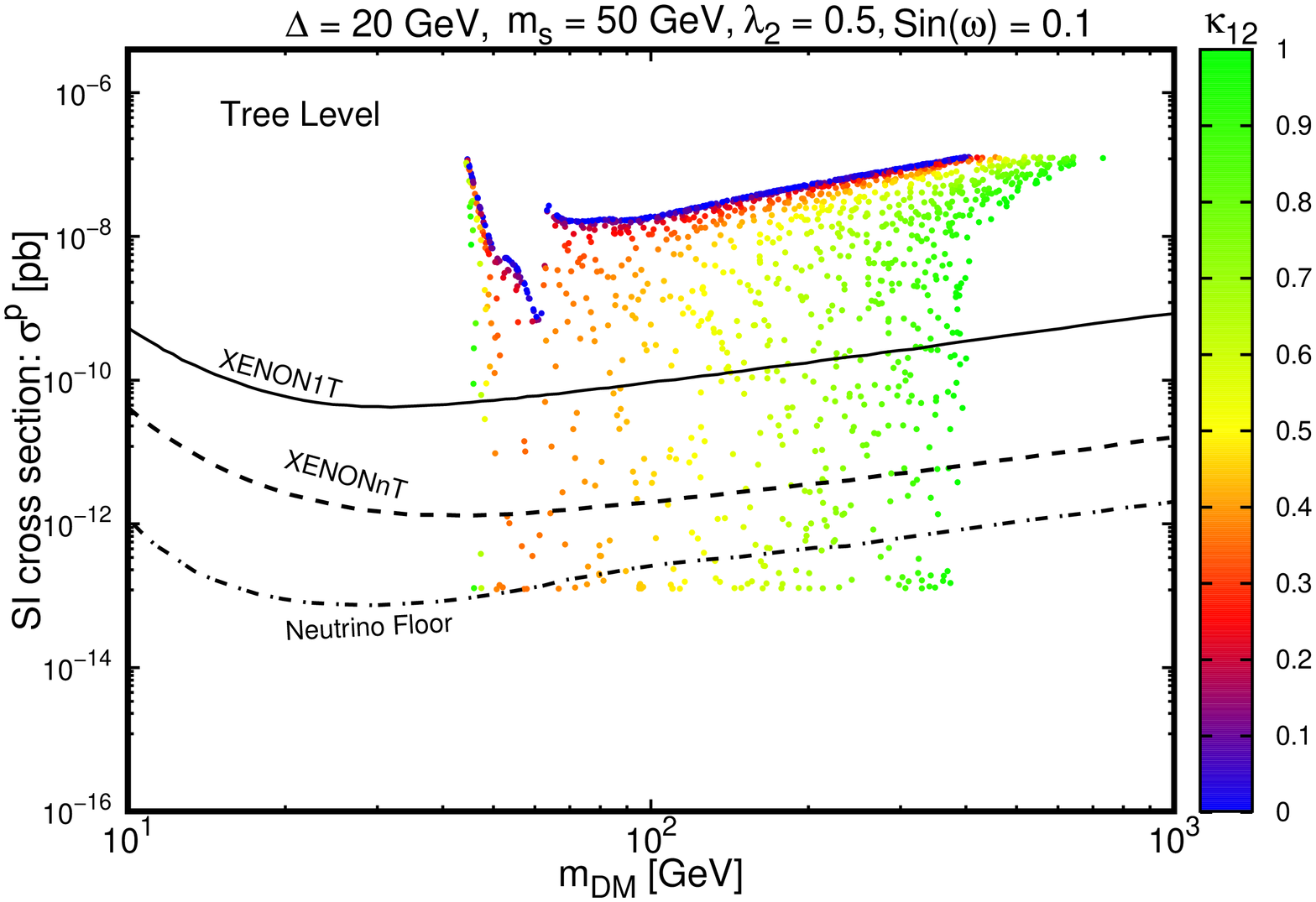}
\end{minipage}
\caption{Tree-level DD cross section is shown as a function of DM mass 
for $m_s = 50$ GeV and $\Delta = 20$ GeV. In the left (right) panel, 
$\kappa_1$ ($\kappa_{12}$) is shown in the vertical color spectrum.
All the points respect the observed relic density and invisible Higgs decay bound. 
Upper limits from XENON1t and projected XENONnT are placed. 
The neutrino floor is also shown.} 
\label{CrossTree-ms50}
\end{figure}

\begin{figure}
\hspace{-.5cm}
\begin{minipage}{.52\textwidth}
\includegraphics[width=.71\textwidth,angle =-90]{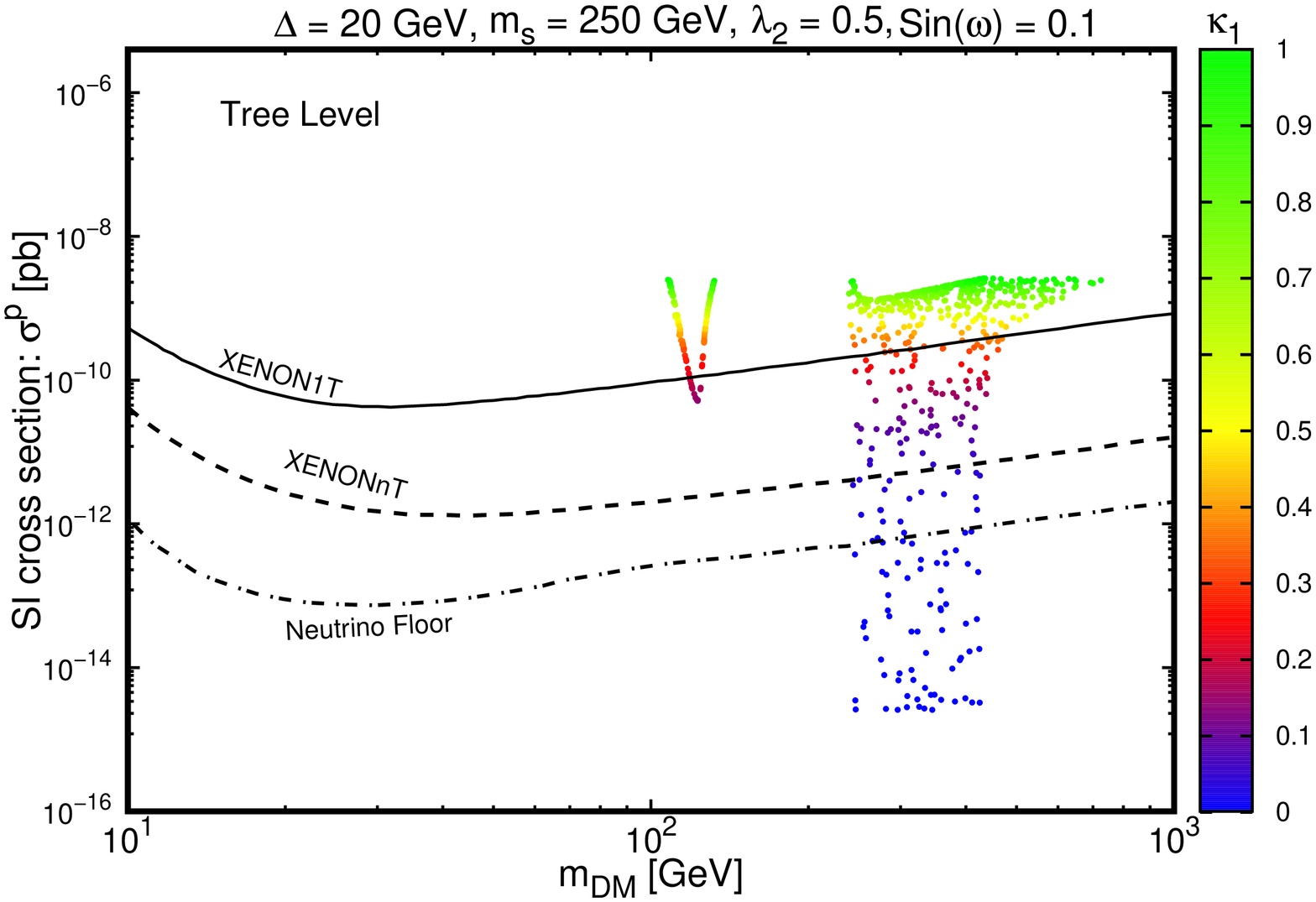}
\end{minipage}
\begin{minipage}{.52\textwidth}
\includegraphics[width=.71\textwidth,angle =-90]{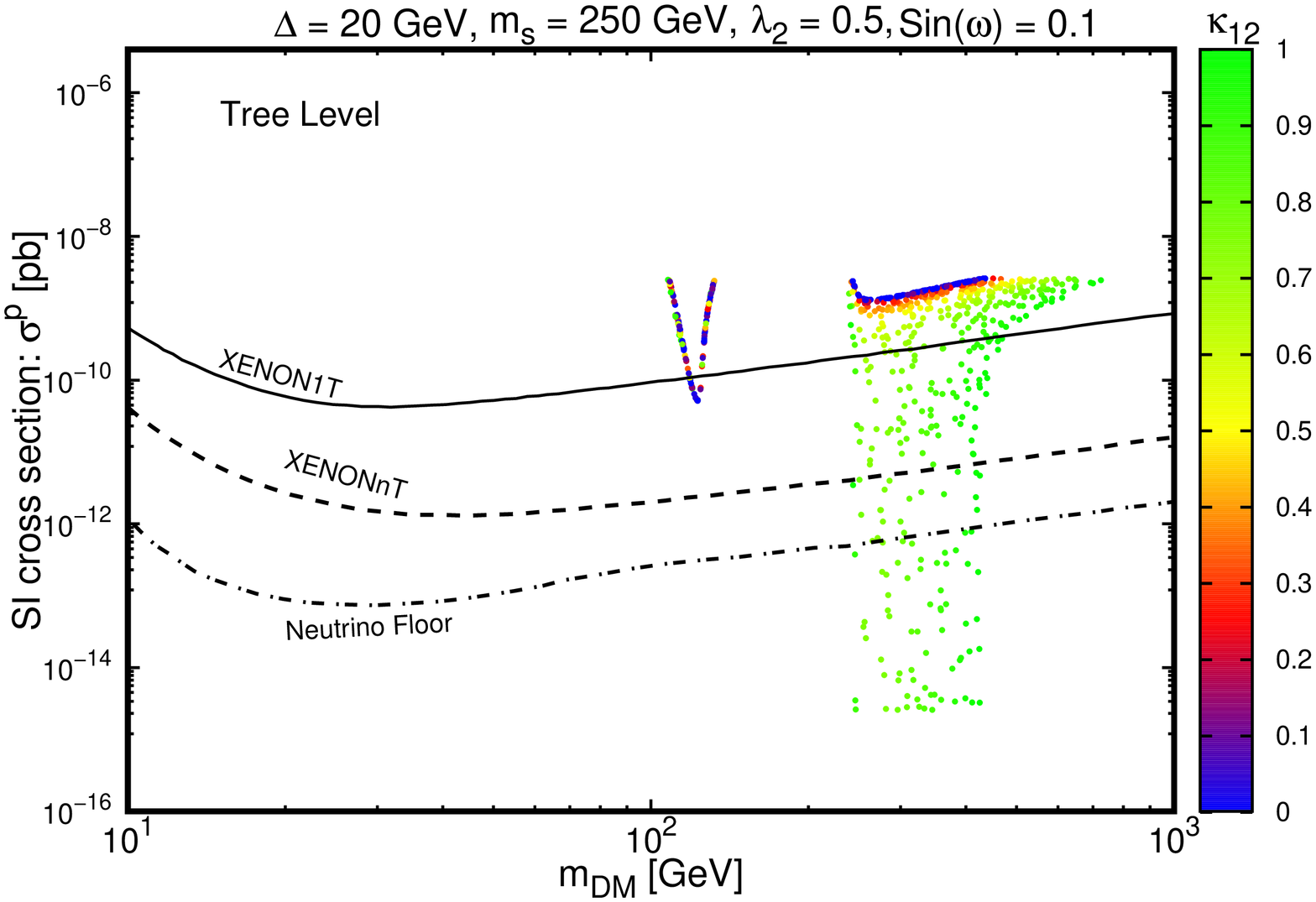}
\end{minipage}
\caption{The same as in Fig.~\ref{CrossTree-ms50}, with $m_s = 250$ GeV.} 
\label{CrossTree-ms250}
\end{figure}

In this section we present our results concerning the DD cross section at 
tree level in perturbation theory. 
The following constraints are considered in the present and next sections:
The upper limit from XENON1T experiment \cite{XENON:2018voc}, and projected limit 
from XENONnT (20 ty) at 90\% CL are imposed \cite{XENON:2020kmp}.
The lower bounds on the DM-nucleon scattering cross section is set by the neutrino 
floor \cite{Billard:2021uyg}. The neutrino floor is a bound below which the detection of 
DM is very difficult. The predicted relic abundance by the model for each 
point in the parameter space is subject to the observed value $\Omega h^2 \sim 0.12$ \cite{Planck:2018vyg}.   
\begin{figure}
\centering
\includegraphics[width=.15\textwidth,angle =0]{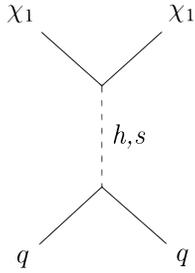}
\caption{Feynman diagram for DM-quark elastic scattering at tree level.}
\label{DD-Tree}
\end{figure}
The DD cross section at tree level is spin-independent (SI) in this model. 
The relevant Feynman diagram for this process is shown in Fig.~\ref{DD-Tree}. 
As mentioned in the previous section, the DD cross section at tree level depends 
only on one coupling, $\kappa_1$. 
At the limit of zero momentum transfer we arrive at the following formula 
for the the scattering amplitude as
\begin{equation}
 {\cal M}^{LO}_{\chi_1 \chi_1} = {\cal C} \bar \chi_1 \chi_1 \bar q q \,,
\label{Tree-Amplitude}
 \end{equation}
where the effective coupling ${\cal C}$ is 
\begin{equation}
 {\cal C} = \kappa_1 \sin(2\omega) \frac{m_q}{2 v_H} (\frac{1}{m_h^2}-\frac{1}{m_s^2}) \,.
\end{equation}
The elastic scattering cross section of DM-proton 
is described by this formula 
\begin{equation}
 \sigma^p = \frac{m_p^4 m_{\text{DM}}^2 \kappa_1^2 \sin^2(2\omega)}{\pi(m_p + m_{\text{DM}})^2 v_H^2} 
 (\frac{1}{m_h^2}-\frac{1}{m_s^2})^2 f^2\,,
\end{equation}
where $m_p$ stands for the proton mass, and $f \sim 0.28$ is the hadronic form factor.
Let us take a numerical look at the tree-level DM-nucleon scattering cross section. 
The range of the parameters used in our scan over the parameter 
space are, 10 GeV $< m_{\text{DM}} <$ 2 TeV, 
$ 0.001 < \kappa_1 < 1$, and $0.001 < \kappa_{12} < 1$. The rest of the free parameters
in the scan are fixed as, $\lambda_2 = 0.5$ and $\sin \omega = 0.1$.
The results are shown in Fig.~\ref{CrossTree-ms50} and Fig.~\ref{CrossTree-ms250} for 
the DM-proton cross section as a function of DM mass 
for $m_s = 50$ GeV and $m_s = 250$ GeV, respectively. 
In both cases the mass difference between the two fermions is, $\Delta = 20$ GeV.

In both figures, it is evident that there are points with small DD cross section which 
reside below the XENONnT bound and the neutrino floor. 
When $m_s = 50$ GeV, a broader range of viable DM candidates are found with respect to the case when 
$m_s = 250$ GeV. The reason is that when $m_s$ is smaller, then annihilation of DM to a pair of singlet scalars is possible with smaller DM candidates and this will affect the range of the viable parameter space.
As expected, the lower DD cross section, the smaller coupling $\kappa_1$ is picked out. 
The other involved coupling, $\kappa_{12}$, is quite large in the low DD cross section regions.
This latter coupling has to be large in order to control the size of the theoretical relic density in 
such a way to satisfy the observed density.

Next, we redo our scan with 50 GeV $< m_s < 250$ GeV, and let the 
couplings $\kappa_1$ and $\kappa_{12}$ take smaller values in the range, $0.0001 < \kappa_1 < 1$ 
and $0.0001 < \kappa_1 < 1$. The other parameters are kept the same as before. 
\begin{figure}
\hspace{-.5cm}
\begin{minipage}{.52\textwidth}
\includegraphics[width=.71\textwidth,angle =-90]{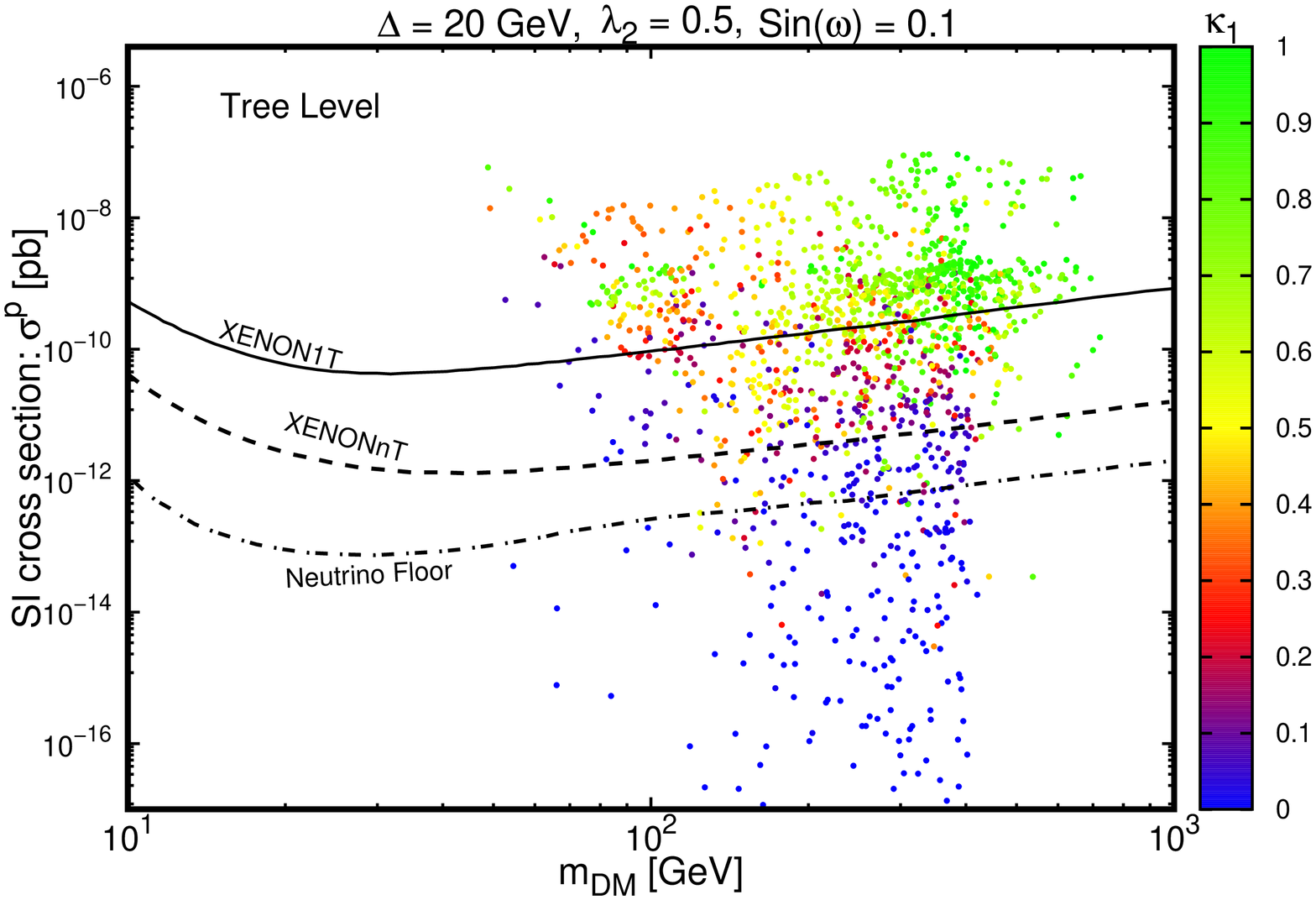}
\end{minipage}
\begin{minipage}{.52\textwidth}
\includegraphics[width=.71\textwidth,angle =-90]{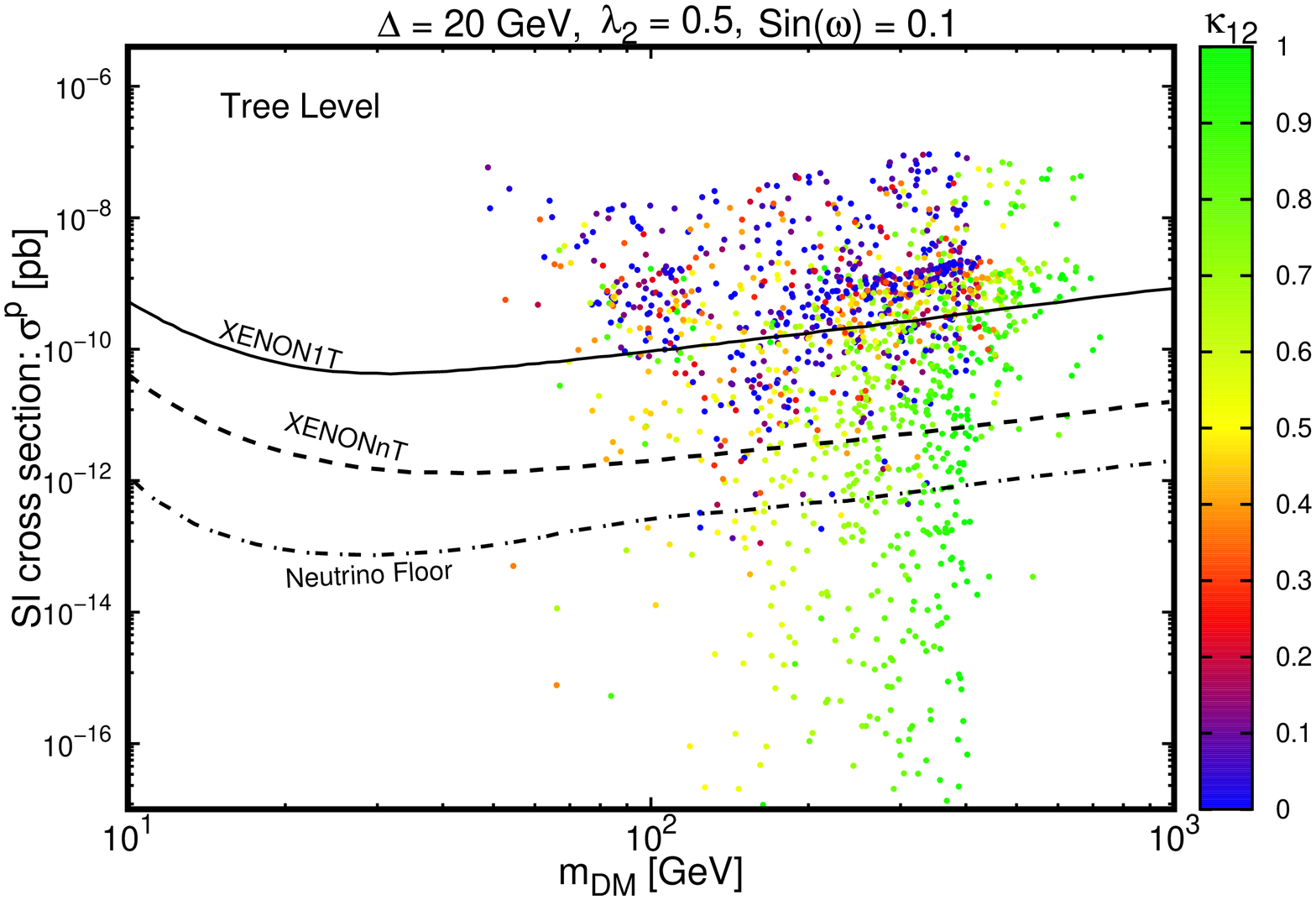}
\end{minipage}
\begin{minipage}{.52\textwidth}
\hspace{4cm}
\includegraphics[width=.71\textwidth,angle =-90]{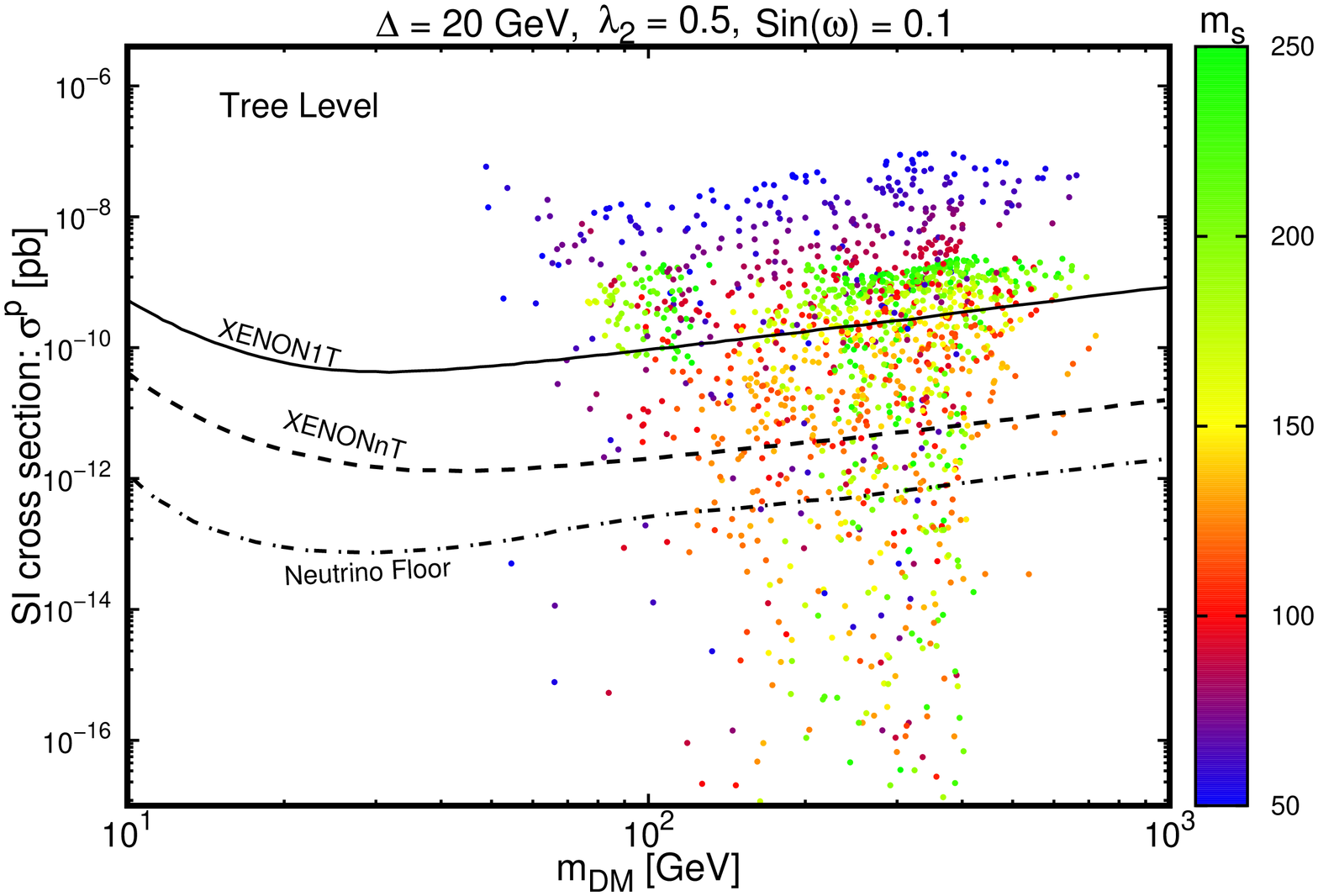}
\end{minipage}
\caption{The same as in Fig.~\ref{CrossTree-ms50}, with 50 GeV $< m_s < 250$ GeV.} 
\label{CrossTree-ms}
\end{figure}
It is expected that by taking smaller $\kappa_1$, smaller DD cross section is achieved which 
goes well below the neutrino floor. This is demonstrated in Fig.~\ref{CrossTree-ms}, wherein the 
cross section of DM-proton scattering is shown as a function of the DM mass. There, we notice 
that a large portion of the parameter space is below the neutrino floor.

So, the main point that motivates the computations of the DD cross section 
at one loop order, is the existence of one loop Feynman diagrams for DM-nucleon 
scattering with purely $\kappa_{12}$ coupling which
will enhance the DD cross section significantly in the regions with small DD cross section.
It is also numerically checked that by varying the couplings $\lambda_2$ and 
the mass difference $\Delta$, the overall picture at tree level 
remains almost the same.

\section{Renormalization of the Model}
\label{renormalization}
We describe the renormalization program at one-loop order in this section. 
There are seven independent parameters in our model which need renormalization, namely, 
$m_1, m_2, m_s, \lambda_2, \kappa_1$,$ \kappa_{12}$, $\omega$. 
In the following section we describe how to pursue the program.

\subsection{One-point and Two-point Functions}

Since we have mixing in the scalar sector at tree level, then the renormalization 
procedure needs a careful treatment. Note that here we will apply the on-shell scheme 
in the renormalization. 
Let us begin with the two scalar tadpoles; $t_\varphi$ and $t_H$.   
The relations between the tadpole counter terms in the two bases are
\begin{equation}
\begin{split}
 \delta t_\varphi = c_\omega ~\delta t_s - s_\omega~\delta t_h, \\
 \delta t_H = s_\omega~\delta t_s + c_\omega ~\delta t_h \,,
\end{split}
\end{equation}
where $c_\omega = \cos(\omega)$ and $s_\omega = \sin(\omega)$.
The wave function renormalization of the fields, $h$ and $s$, in the present of mixing can be formulated in the following way, as introduced in \cite{Kanemura:2016lkz,Glaus:2019itb},
\begin{equation}
 \begin{pmatrix}
   h \\ 
   s     
 \end{pmatrix}
 \to
 \begin{pmatrix}
   1+ \frac{1}{2} \delta Z_{hh}    &   \delta c_{hs} + \delta \alpha \\ 
   \delta c_{sh}-\delta \alpha   &   1+ \frac{1}{2} \delta Z_{ss}
 \end{pmatrix}
 \begin{pmatrix}
   h \\ 
   s     
 \end{pmatrix}
 \,,
\end{equation}
where $\delta c_{hs}$ and $\delta c_{sh}$ are the counter terms of the off-diagonal mass terms.  
Let us begin by the renormalized one-point functions for the physical scalar fields, $h$ and $s$, that we write them in terms of the 1PI diagrams $\Gamma^{\text{1PI}}_h$ and $\Gamma^{\text{1PI}}_s$: $t^r_h = \delta t_h + \Gamma^{\text{1PI}}_h$, $t^r_s = \delta t_s + \Gamma^{\text{1PI}}_s$.
At one loop level, we impose the renormalization condition for the two scalar tadpoles: 
$t^r_s = t^r_h = 0$. This results in $\delta t_s = - \Gamma^{\text{1PI}}_s$ and 
$\delta t_h = - \Gamma^{\text{1PI}}_h$. It is worth mentioning that 
when the renormalized tadepoles vanish at one loop, it implies no shift of the 
vacuum state of the potential.
Next, we express relations for the renormalized two-point functions of the scalar fields,
\begin{equation}
\begin{split}
 \Pi^{r}_{ss}(p^2) &= \Pi^{\text{1PI}}_{ss}(p^2) + \frac{s_\omega^2 \delta t_H}{v_H}
 + \Big[(p^2- m_s^2) \delta Z_{ss} -\delta m_s^2 \Big] \,, \\
\Pi^{r}_{hh}(p^2) &= \Pi^{\text{1PI}}_{hh}(p^2) + \frac{c_\omega^2 \delta t_H}{v_H}
 + \Big[(p^2- m_h^2) \delta Z_{hh} -\delta m_h^2 \Big] \,, \\
\Pi^{r}_{sh}(p^2) &= \Pi^{\text{1PI}}_{sh}(p^2) + \frac{s_\omega c_\omega \delta t_H}{v_H}
 + p^2(\delta c_{hs} +  \delta c_{sh}) + m_h^2 (\delta \omega - \delta c_{hs}) - m_s^2 (\delta \omega + \delta c_{sh})  \,.
\end{split}
\label{2point}
\end{equation}
We choose the one-shell renormalization conditions for the two-point functions as follow,
\begin{equation}
 \Pi^{r}_{ss}(m_s^2) = \Pi^{r}_{hh}(m_h^2) = 0, ~~ \frac{d}{dp^2}\Pi^{r}_{ss}(p^2)|_{p^2=m_s^2} 
 =  \frac{d}{dp^2}\Pi^{r}_{hh}(p^2)|_{p^2=m_h^2} = 0 \,,
\end{equation}
and from these relations four counter terms are determined,
\begin{equation}
\begin{split}
\delta m_s^2 &=  \Pi^{\text{1PI}}_{ss}(m_s^2) + \frac{s_\omega^2 \delta t_H}{v_H},
 ~~ \delta m_h^2 = \Pi^{\text{1PI}}_{hh}(m_h^2) + \frac{c_\omega^2 \delta t_H}{v_H} \\
 ~~\delta Z_{ss} &= -\frac{d}{dp^2} \Pi^{\text{1PI}}_{ss}(p^2)|_{p^2=m_s^2},
 ~~ \delta Z_{hh} = -\frac{d}{dp^2} \Pi^{\text{1PI}}_{hh}(p^2)|_{p^2=m_h^2} \,.
\end{split}
\end{equation}
We can fix three counter terms,  $\delta c_{hs}$, $\delta c_{sh}$ and $\delta \omega$ by 
imposing the conditions
\begin{equation}
 \Pi^r_{sh}(m_h^2) = \Pi^r_{sh}(m_s^2) = 0, ~~ \delta c_{sh} = \delta c_{hs}  \,,
\end{equation}
to find 
\begin{equation}
\begin{split}
\delta \omega &= \frac{1}{2(m_s^2- m_h^2)} \Big[ \Pi^{\text{1PI}}_{sh}(m_h^2) + 
\Pi^{\text{1PI}}_{sh}(m_s^2)  + \frac{s_{2\omega} \delta t_H}{v_H} \Big] \\
\delta c_{hs} &= \frac{1}{2(m_s^2- m_h^2)} \Big[ \Pi^{\text{1PI}}_{sh}(m_h^2) - 
\Pi^{\text{1PI}}_{sh}(m_s^2)  \Big]  \,.
\end{split}
\end{equation}
Now we consider the renormalized two point function of the two fermions, $\chi_1$ and $\chi_2$
\begin{equation}
\begin{split}
\Pi^{r}_{\chi_i \chi_i}(\slashed p) &= \Pi^{\text{1PI}}_{ii}(\slashed p) + \slashed{p}  \delta Z_{ii} -\delta m_i \\
\Pi^{r}_{\chi_i \chi_j}(\slashed p) &= \Pi^{\text{1PI}}_{ij}(\slashed p) \,.
\end{split}
\end{equation}
With the renormalization conditions,
\begin{equation}
 \Pi^{r}_{\chi_i \chi_i}(\slashed p = m_i) = 0 , ~~ 
 \frac{d}{d\slashed p} \Pi^{r}_{\chi_i \chi_i}(\slashed p )|_{\slashed p = m_i} = 0 \,.
 \end{equation}
We can then obtain the two parameters,
\begin{equation}
 \delta Z_{ii} = \frac{d\Pi^{\text{1PI}}_{ii}(\slashed p)}{d\slashed p}|_{\slashed p=m_i}
 ~~~, ~~~ \delta m_i = \Pi^{\text{1PI}}_{ii}(m_i) + m_i \delta Z_{ii} \,.
\label{fermion-2point}
\end{equation}
A comment is appropriate to mention here. It is turned out in \cite{Bojarski:2015kra} that
the mixing angle, $\omega$, remains gauge dependence in the on-shell renormalization scheme.
A suggested in \cite{Pilaftsis:1997dr} to get gauge-independent definition for $\delta \omega$,
one may define it in a physical process, like the decay $h \to \tau \tau$ . 
We have checked this in our numerical results and it is found out that 
the gauge dependence of the one-shell renormalization at one loop order 
is up to about $1 \%$.

\subsection{Renormalization of the Couplings $\kappa_1$ and $\kappa_{12}$}
\begin{figure}
\centering
\includegraphics[width=.45\textwidth,angle =0]{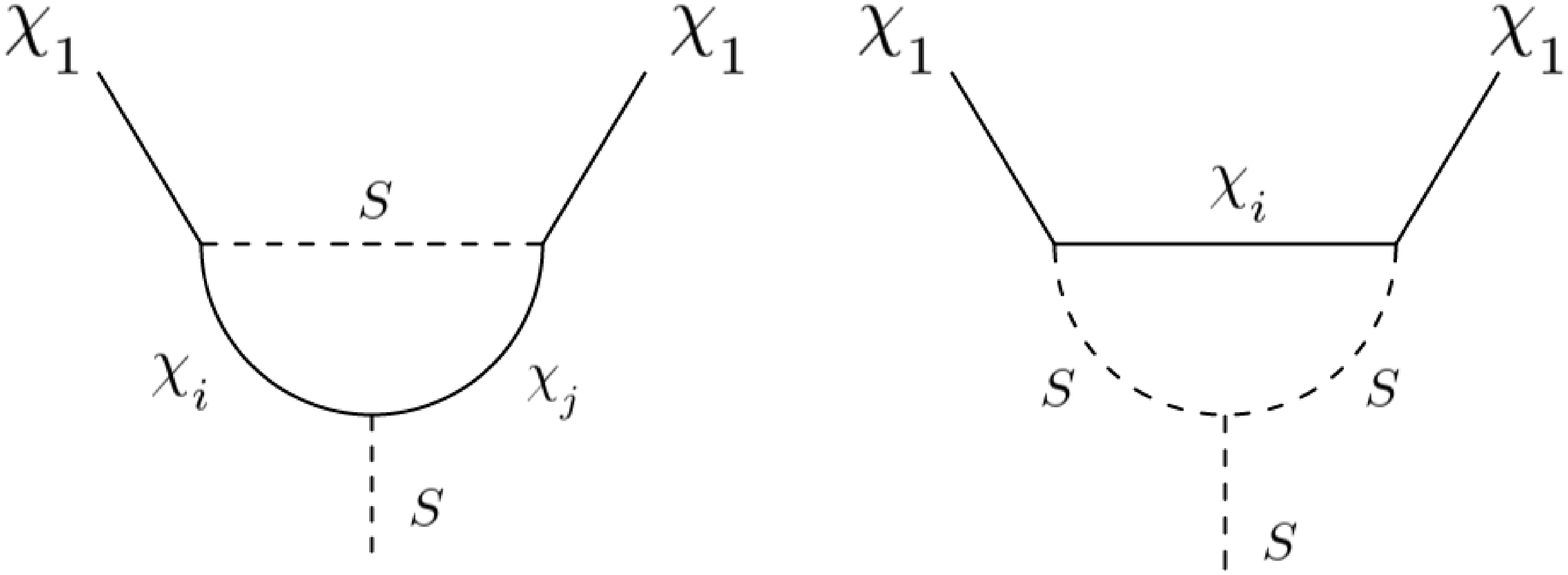}
\caption{Triangle diagrams contributing to $\Gamma^{\text{1PI}}_{h\chi_1 \chi_1}$ and 
$\Gamma^{\text{1PI}}_{s\chi_1 \chi_1}$. $S$ stands for the two scalars $h$ and $s$, and $\chi_i = \chi_1 , \chi_2$.}
\label{Triangle-Feynman}
\end{figure}
Since the ultra-violet divergences are universal we opt for the
minimal subtraction scheme. We will look at two vertices $h\chi_1 \chi_1$ and 
$h\chi_1 \chi_{12}$ to find the relevant counter terms.
We begin by the vertex $h\chi_1 \chi_1$ and write,
\begin{equation}
 \Gamma^{\text{NLO}}_{h \chi_1 \chi_1} = \Gamma^{\text{LO}}_{h \chi_1 \chi_1}
  + \Gamma^{\text{1PI}}_{h\chi_1 \chi_1} + \Gamma^{\text{CT}}_{h\chi_1 \chi_1} \,,
\end{equation}
where $\Gamma^{\text{1PI}}_{h\chi_1 \chi_1}$ indicates the loop correction to the 
triple vertex, and the vertex counter terms are collected in $\Gamma^{\text{CT}}_{h\chi_1 \chi_1}$.
Combining the mixing effects and wave function renormalization the full expression for 
$\Gamma^{\text{CT}}_{h\chi_1 \chi_1}$ reads
\begin{equation}
 \Gamma^{\text{CT}}_{h\chi_1 \chi_1} = \frac{1}{2} \lambda_{h\chi_1 \chi_1} \delta Z_{hh} 
  + \lambda_{h\chi_1 \chi_1} \delta Z_{\chi_1 \chi_1}
  + \lambda_{s\chi_1 \chi_1} (\delta c_{hs} -\delta \omega) 
  +\lambda_{h\chi_1 \chi_2} \frac{\Pi^{\text{1PI}}_{\chi_1 \chi_2}(m_1)}{m_2- m_1}
  + \frac{\partial \lambda_{h\chi_1 \chi_1}}{\partial \kappa_1} \delta \kappa_1 \,,
\end{equation}
where for the couplings we have
\begin{equation}
\lambda_{h\chi_1 \chi_1} = -\sin(\omega) \kappa_1, ~~ \lambda_{s\chi_1 \chi_1} = -\cos(\omega) \kappa_1, ~~\lambda_{h\chi_1 \chi_2} = -\sin(\omega) \kappa_{12} \,.
 \end{equation}
The divergent part of $\delta \kappa_1$ is found as 
\begin{equation}
 \delta \kappa_1|_{\text{div}} = \frac{1}{\sin(\omega)} 
 \Big(\Gamma^{\text{1PI}}_{h\chi_1 \chi_1} + \Gamma^{\text{CT}}_{h\chi_1 \chi_1}|_{\delta \kappa_1}   \Big)|_{\text{div}} \,.
\end{equation}
The relevant Feynman diagrams as triple vertex corrections are shown in Fig.~\ref{Triangle-Feynman}.    
Following the same approach the divergent part of $\delta \kappa_{12}$ cab be found.
The counter term of this triple vertex is
\begin{equation}
\begin{split}
  \Gamma^{\text{CT}}_{h\chi_1 \chi_2} &= \frac{1}{2} \lambda_{h\chi_1 \chi_2} \delta Z_{hh} 
  + \lambda_{h\chi_1 \chi_2} \delta Z_{\chi_2 \chi_2}
  + \lambda_{s\chi_1 \chi_2} (\delta c_{hs} -\delta \omega) 
  +\frac{1}{2} \lambda_{h\chi_2 \chi_2} \frac{\Pi^{\text{1PI}}_{\chi_1 \chi_2}(m_1)}{m_2- m_1}
 \\&
 +\frac{1}{2} \lambda_{h\chi_1 \chi_1} \frac{\Pi^{\text{1PI}}_{\chi_1 \chi_2}(m_2)}{m_1- m_2}
  + \frac{\partial \lambda_{h\chi_1 \chi_2}}{\partial \kappa_{12}} \delta \kappa_{12} \,,
\end{split}
\end{equation}
where, $\lambda_{s\chi_1 \chi_2} = -\cos(\omega) \kappa_{12}$, and $\lambda_{h\chi_2 \chi_2} = 0$, 
since $\kappa_{2} = 0$ as discussed in section \ref{model}. 
Finally, we can get the divergent part of $\delta \kappa_{12}$ by 
\begin{equation}
 \delta \kappa_{12}|_{\text{div}} = \frac{1}{\sin(\omega)} 
 \Big(\Gamma^{\text{1PI}}_{h\chi_1 \chi_2} + \Gamma^{\text{CT}}_{h\chi_1 \chi_2}|_{\delta \kappa_{12}}   \Big)|_{\text{div}} \,.
\end{equation}

\section{DD Cross Section at One Loop}
\label{DDCS}
\begin{figure}
\centering
\includegraphics[width=.5\textwidth,angle =0]{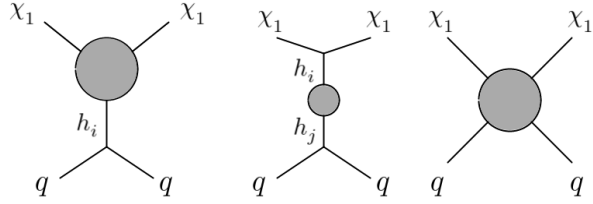}
\caption{One-loop Feynman diagrams including all types of corrections to the DM scattering off the quarks. The left diagram indicates vertex corrections, the diagram in the middle stands for propagator corrections, and the last diagrams shows box corrections. The scalar, $h_i$, indicates scalar $s$ or $h$.}
\label{DDloop}
\end{figure}
In this section we present the amplitude for DM-quark scattering at one loop level 
or next to leading order (NLO). 
The structure of the one loop corrections to the DM scattering off the nucleons are shown as Feynman diagrams in Fig.~\ref{DDloop}. 
The diagrams entail one-loop contributions as triple vertex corrections, internal propagator corrections, and Box diagrams, respectively. 
The scattering amplitude can then be written as
\begin{equation}
 {\cal M}^{\text{NLO}} =  {\cal M}^{\text{LO}} + {\cal M}^{\text{VC}} + {\cal M}^{\text{PC}}
 + {\cal M}^{\text{Box}} \,,
\end{equation}
where ${\cal M}^{\text{VC}}$, ${\cal M}^{\text{PC}}$ and ${\cal M}^{\text{Box}}$ stand
for vertex correction of $\chi_1 \chi_1 h_i$, propagator correction and Box correction, respectively.
We begin with the vertex correction. There are two types of corrections for the vertex 
$\chi_1 \chi_1 h_i$, as already shown in Fig.~\ref{Triangle-Feynman}. 
Including these vertex corrections, it is possible to find an effective scattering amplitude. 
Let's begin with the vertex correction where only one scalar runs in the loop. 
The corresponding scattering amplitude is 
\begin{equation}
\begin{split}
{\cal M}^{\text{VC}}_{(a)} &= 
  \Big[\frac{\alpha_i}{(p_1-p_2)^2-m_{h_i}^2} \Big]  
\Big( \kappa_1^3 \int \frac{d^4l}{(2\pi)^4}   \frac{ (\bar q q)~~\bar \chi(p_{2}) 
(\slashed{p_2}-\slashed{l}+m_1) (\slashed{p_1}-\slashed{l}+m_1) \chi(p_1) }
{[(p_2-l)^2-m_1^2][(p_1-l)^2-m_1^2][l^2-m_{h_j}^2]} 
\\
 &+ 2 \kappa_1 \kappa_{12}^2 \int \frac{d^4l}{(2\pi)^4}   \frac{ (\bar q q)~~\bar \chi(p_{2}) 
(\slashed{p_2}-\slashed{l}+m_1) (\slashed{p_1}-\slashed{l}+m_2) \chi(p_1) }
{[(p_2-l)^2-m_1^2][(p_1-l)^2-m_2^2][l^2-m_{h_j}^2]}  
\Big)  \,,
\end{split}
\end{equation}
where $h_i = s, h$, and  $\alpha_h = (-m_q/v_H) c_\omega$, and $\alpha_s = (m_q/v_H) s_\omega$.
At zero momentum transfer we get the following relation for the effective scattering 
amplitude. Note that the amplitude is obtained in the minimal subtraction scheme,
\begin{equation}
\begin{split}
 {\cal M}^{\text{VC}}_{(a)} &= \frac{m_q \kappa_{1}^3}{16\pi^2 v_H}
\Big[ \frac{4}{\bar \epsilon} + \frac{c_\omega s_\omega^3}{m_h^2} F(m_1,m_h) 
+\frac{s_\omega c_\omega^3}{m_h^2} F(m_1,m_s)
-\frac{s_\omega c_\omega^3}{m_s^2} F(m_1,m_s) 
-\frac{c_\omega s_\omega^3}{m_s^2} F(m_1,m_h) \Big] \\
& \times \bar q q~\bar \chi_1 \chi_1
+ \frac{m_q \kappa_1 \kappa_{12}^2}{16\pi^2 v_H}
\Big[ \frac{4}{\bar \epsilon} + \frac{c_\omega s_\omega^3}{m_h^2} F(m_1,m_2,m_h)
+\frac{s_\omega c_\omega^3}{m_h^2} F(m_1,m_2,m_s)  \\
&- \frac{s_\omega c_\omega^3}{m_s^2} F(m_1,m_2,m_s) 
- \frac{c_\omega s_\omega^3}{m_s^2} F(m_1,m_2,m_h) \Big] 
\bar q q~\bar \chi_1 \chi_1  \,,
\end{split}
\end{equation}
where in the above expression, $F(m_1,m_i) = F(m_1,m_1,m_i)$, and
$1/\bar \epsilon =1/\epsilon -\gamma_E + \log(4\pi)$, $\gamma_E$ being the Euler-Mascheroni constant.
These functions are defined in Appendix B.

The second diagram with vertex corrections in the scattering process involves 
two scalars ruining in the loop. The resulting scattering amplitude is found as
\begin{equation}
\begin{split}
{\cal M}^{\text{VC}}_{(b)} &= 
  \Big[\frac{\alpha_k}{(p_1-p_2)^2-m_{k}^2} \Big]  
\Big( \kappa_1^2 \int \frac{d^4l}{(2\pi)^4}   \frac{ (\bar q q)~~\bar \chi(p_{2}) (\slashed{l}+m_1) \chi(p_1) }
{[(p_2-l)^2-m_{i}^2][(p_1-l)^2-m_{j}^2][l^2-m_1^2]} 
\\
&+ \kappa_{12}^2 \int \frac{d^4l}{(2\pi)^4}   \frac{ (\bar q q)~~\bar \chi(p_{2}) (\slashed{l}+m_2) \chi(p_1) }
{[(p_2-l)^2-m_{i}^2][(p_1-l)^2-m_{j}^2][l^2-m_2^2]} 
\Big)  \,, 
\end{split}
\end{equation}
where the indices $i, j, k$ stand for the scalars $h$ and $s$. 
In the formula above, $\alpha_k$ takes the same definition as before.
The integration will be performed at the limit of zero momentum transfer. 
The effective amplitude is obtained by taking into account all the possible 
triple scalar couplings in the triangle loop. The final result is finite and follows as 
\begin{equation}
\begin{split}
{\cal M}^{\text{VC}}_{(b)} &= \frac{m_q m_1 \kappa_1^2}{16\pi^2 v_H}
\Big[ \frac{c_\omega s_\omega^2}{m_h^2} c_{hhh} G(m_h,m_1) 
+ \frac{s_\omega c_\omega^2}{m_h^2} c_{shh} G(m_h,m_s,m_1) 
+ \frac{c_\omega^3}{m_h^2} c_{hss} G(m_s,m_1) \\
&-\frac{s_\omega^3}{m_s^2} c_{shh} G(m_h,m_1)
-\frac{c_\omega s_\omega^2}{m_s^2} c_{hss} G(m_h,m_s,m_1) 
- \frac{s_\omega c_\omega^2}{m_s^2} c_{sss} G(m_s,m_1) \Big] 
\bar q q~\bar \chi_1 \chi_1  
\\&
+ \kappa_{12}^2 (m_1 \to m_2) \,,
\end{split}
\end{equation}
where $G(m_i,m_1) = G(m_i,m_i,m_1)$, and $G$ functions are given in Appendix B.
As well, the scalar couplings, $c_{ijk}$, are given in Appendix B.

Another type of corrections to the scattering amplitude are due to propagator corrections as shown in Fig.~\ref{DDloop} (the diagram in the middle). These corrections come in as self-energy correction to the scalar propagators $s$ and $h$. At zero momentum transfer the amplitude reads    
\begin{equation}
 {\cal M}^{\text{PC}} = \sum_{i,j=h,s} f_i g_j \frac{\Pi^{r}_{ij}(p^2=0)}{m^2_{h_i} m^2_{h_j}} \frac{m_q \kappa_1}{v_H} \bar q q~\bar \chi_1 \chi_1  \,,
\end{equation}
where $f_s = -\cos \omega$, $f_h = -\sin \omega$, and $g_s = \sin \omega$, 
$g_h = \cos \omega$. The renormalized two-point functions are defined in Eq.~\ref{2point}, 
$\Pi^{r}_{ij}  = \Pi^{r}_{hh},\Pi^{r}_{ss},\Pi^{r}_{sh}$. 

The last contributions to the amplitude are those from box diagrams, see the 
right diagram in Fig.~\ref{DDloop}. Including both t-channel and u-channel, 
the amplitude at zero momentum transfer is written 
\begin{equation}
\begin{split}
{\cal M}^{\text{Box}} &=  (\frac{m_q}{v_H})^2 c_{ijk} d_{jk}
\int \frac{d^4l}{(2\pi)^4} \Big[ \frac{\bar q(p_q) (\slashed p_q +\slashed l + m_q) q(p_q) \bar \chi_1(p_\chi) (\slashed l + \slashed p_\chi + m_{\chi_i}) \chi_1(p_\chi)}
{ (l^2-m_{h_j}^2)(l^2-m_{h_k}^2)((l+p_\chi)^2-
m^2_{\chi_i}) (p_q+l)^2- m_q^2} \\&
-\frac{\bar q(p_q) (\slashed p_q -\slashed l + m_q) q(p_q) \bar \chi_1(p_\chi) (\slashed l + \slashed p_\chi + m_{\chi_i}) \chi_1(p_\chi)}
{ (l^2-m_{h_j}^2)(l^2-m_{h_k}^2)((l+p_\chi)^2-
m^2_{\chi_i}) (p_q-l)^2-m_q^2}
  \Big]  \,,
\end{split}
\end{equation}
where $m_{h_j} = m_h, m_s$, and $m_{\chi_i} = m_{\chi_1}, m_{\chi_2}$. 
The parameter, $c_{ijk}$, is the multiplication of couplings coming from vertices involving 
$\chi_i$ and $h_i$, and the coupling $d_{jk}$ presents couplings which come from vertices with 
quarks. $c_{ijk}$ parameters are, $c_{\chi_1 hh} = \kappa^2_1 \sin^2 \omega$, 
$c_{\chi_1 ss} = \kappa^2_1 \cos^2 \omega$, $c_{\chi_1 hs} = \kappa^2_1 \sin \omega \cos \omega$, $c_{\chi_2 hh} = \kappa^2_{12} \sin^2 \omega$, 
$c_{\chi_2 ss} = \kappa^2_{12} \cos^2 \omega$, $c_{\chi_2 hs} = \kappa^2_{12} \sin \omega \cos \omega$, and $d_{jk}$ parameters are, $d_{hh} = \cos^2 \omega$, $d_{ss} = \sin^2 \omega$,$d_{sh} = \cos \omega \sin \omega$. 
To compute the effective scattering amplitude, we set the quark mass to zero 
in the denominator and when contracting the fermion lines in the numerator, 
will omit terms which generate momentum suppressed operators. 
Thus, what remains in the end is the spin-independent operator $\bar q q \chi_1 \chi_1$. 
Taking into account these approximations, the effective scattering amplitude reads
\begin{equation}
 {\cal M}^{\text{Box}} =  (\frac{m_q}{v_H})^2 c_{ijk} d_{jk} 
 \Big[H_1(m_{\chi_1},m_{\chi_i},m_{h_j},m_{h_k}) - H_2(m_{\chi_1},m_{\chi_i},m_{h_j},m_{h_k}) \Big]
 m_q \bar q q~\bar \chi_1 \chi_1  \,,
\end{equation}
where the functions, $H_{1,2}(m_{\chi_1},m_{\chi_i},m_{h_j},m_{h_k})$, is defined in Appendix B.
We expect that the Box corrections be a subleading contribution to the cross section, because of 
the small extra factor, $m_q/v_H$, arising from the second scalar-quark vertex in the loop diagram. 
A consistency check of our calculation is the cancellation of the divergences in the scattering 
amplitude at one loop order. This is done by using the Mathematica tool 
{\tt Package-X} \cite{Patel:2016fam}, also cross checking our results obtained independently. 

\section{Results}
\label{results}

\begin{figure}
\begin{minipage}{.52\textwidth}
\includegraphics[width=.75\textwidth,angle =-90]{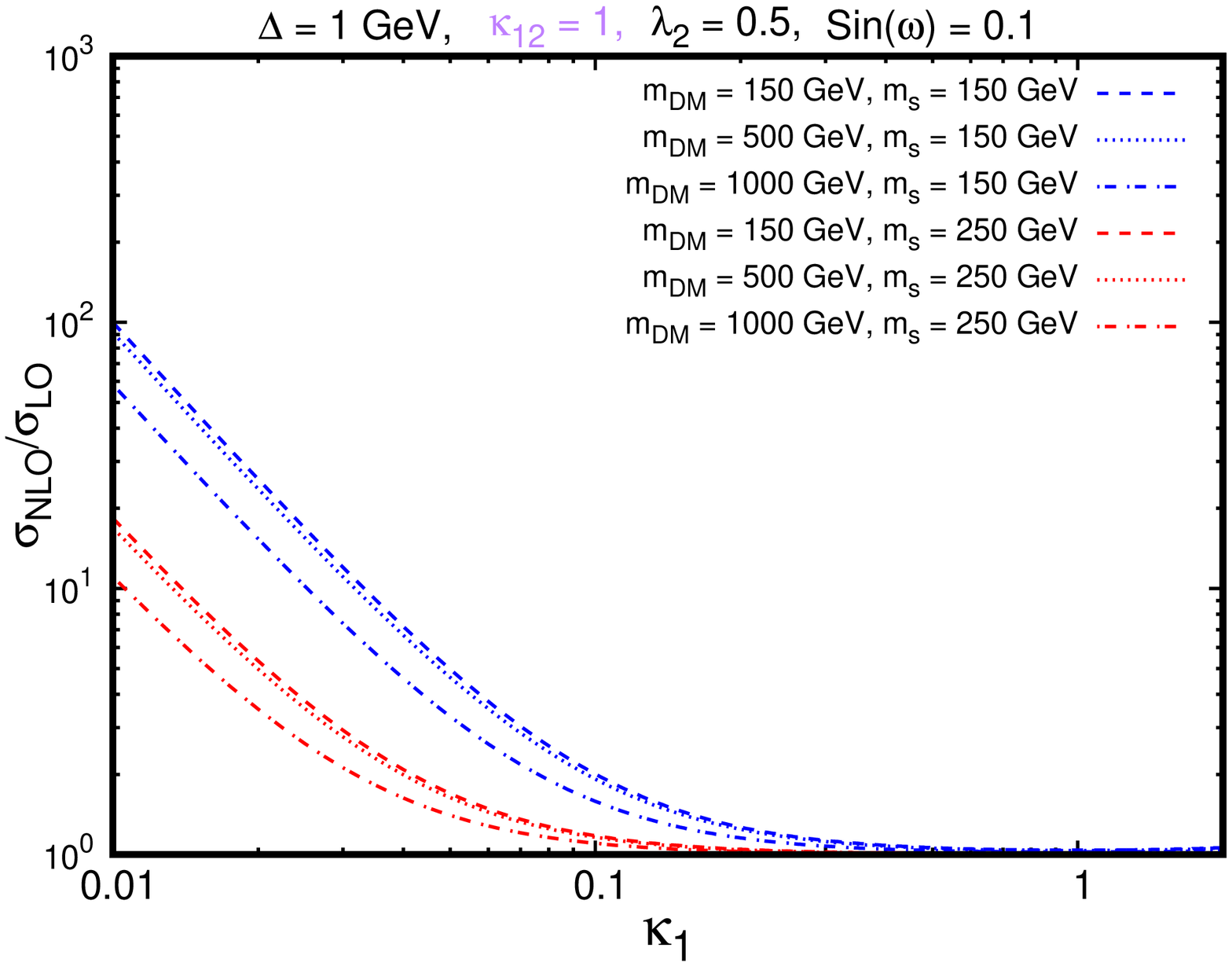}
\end{minipage}
\begin{minipage}{.52\textwidth}
\includegraphics[width=.75\textwidth,angle =-90]{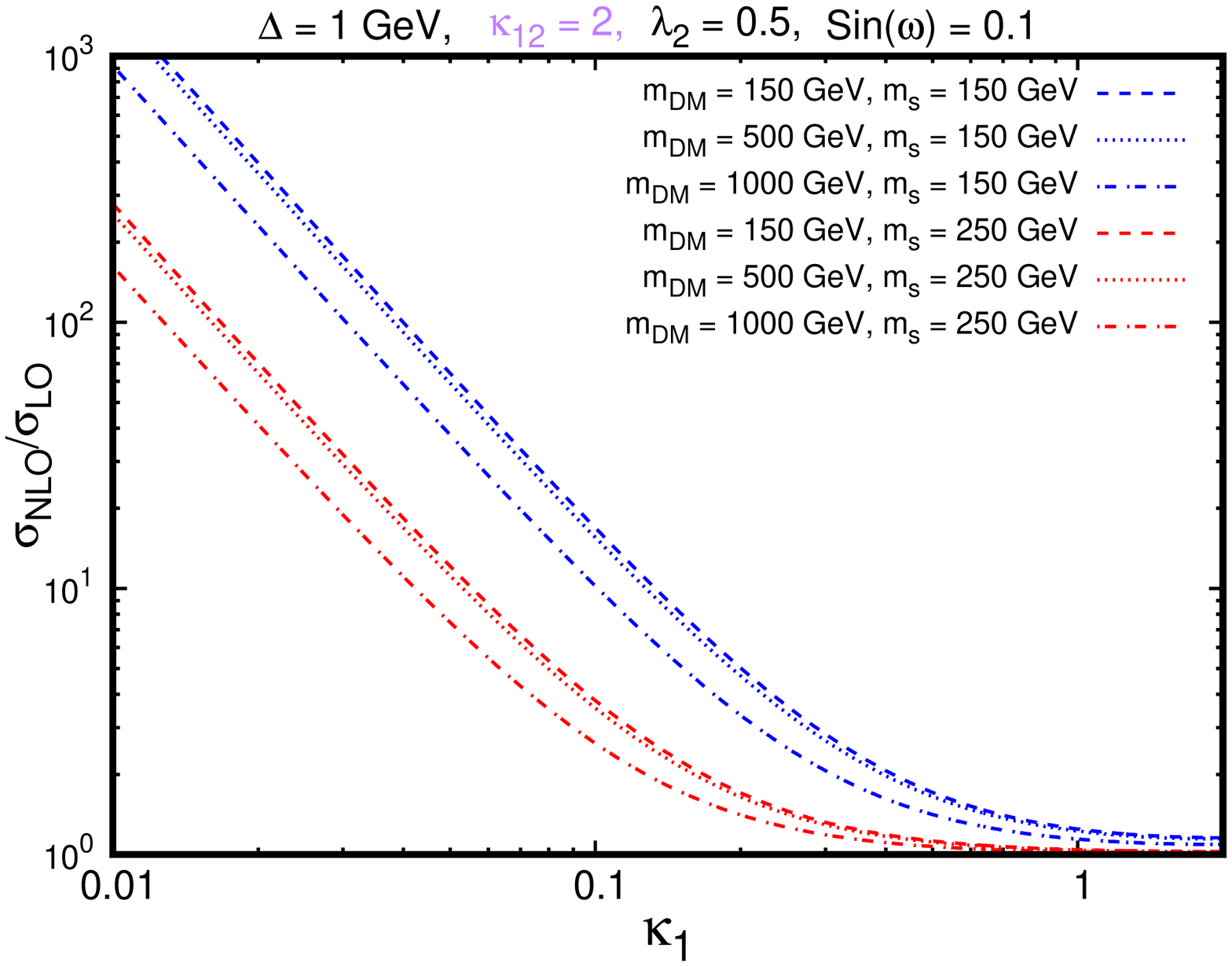}
\end{minipage}
\begin{minipage}{.52\textwidth}
\includegraphics[width=.75\textwidth,angle =-90]{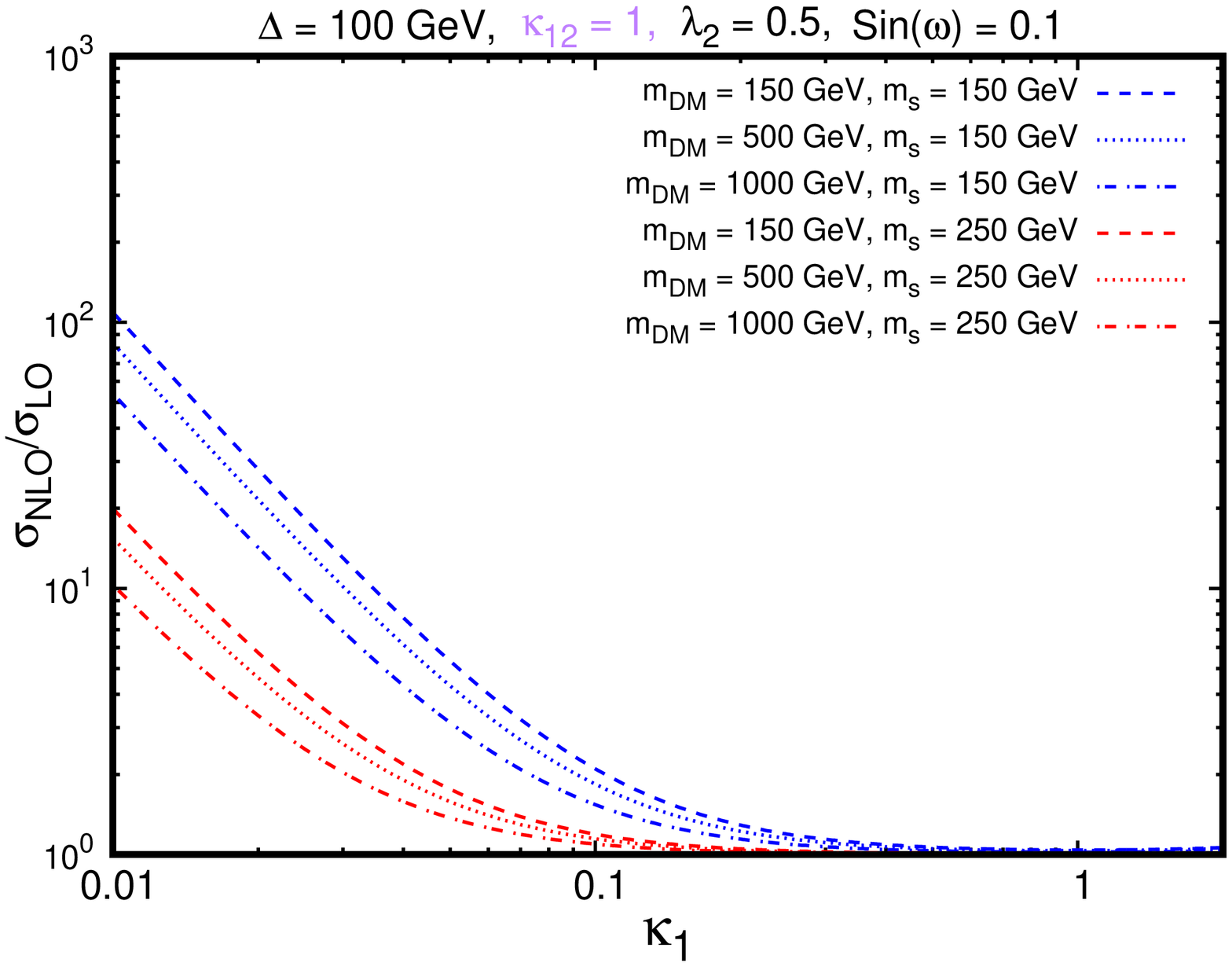}
\end{minipage}
\begin{minipage}{.52\textwidth}
\includegraphics[width=.75\textwidth,angle =-90]{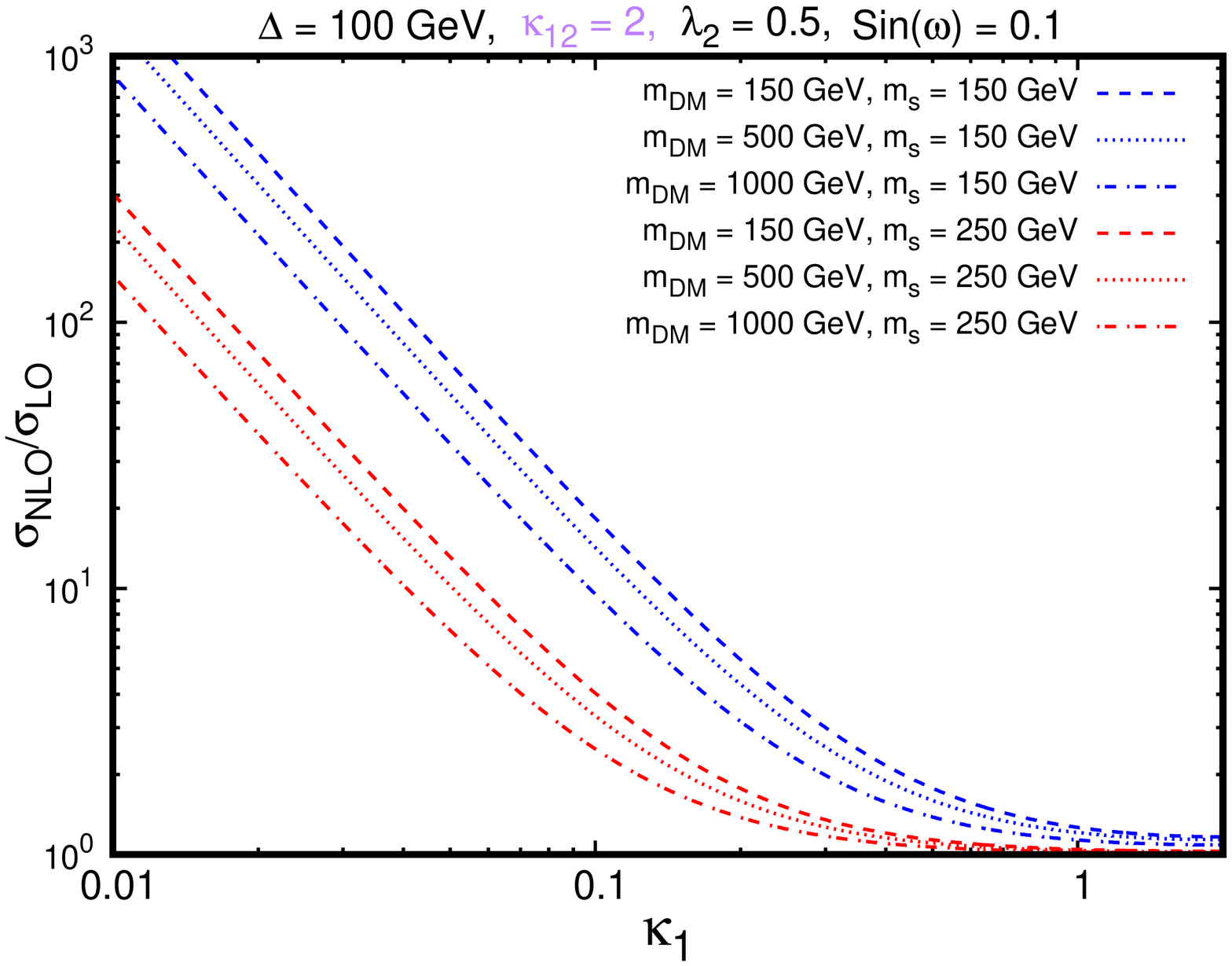}
\end{minipage}
\caption{Shown are the ratio of DD cross section at one-loop to DD cross section at 
tree level as a function of the coupling $\kappa_1$. In all plots, the mixing angle is such that $\sin \omega = 0.1$. Two values are chosen for the mass difference between the fermions, $\Delta = 1, 100$ GeV. The coupling $\kappa_{12} = 1,2$. The constraint from 
the observed relic density is not imposed.} 
\label{Ratio-Sin-0.1}
\end{figure}

\begin{figure}
\begin{minipage}{.52\textwidth}
\includegraphics[width=.75\textwidth,angle =-90]{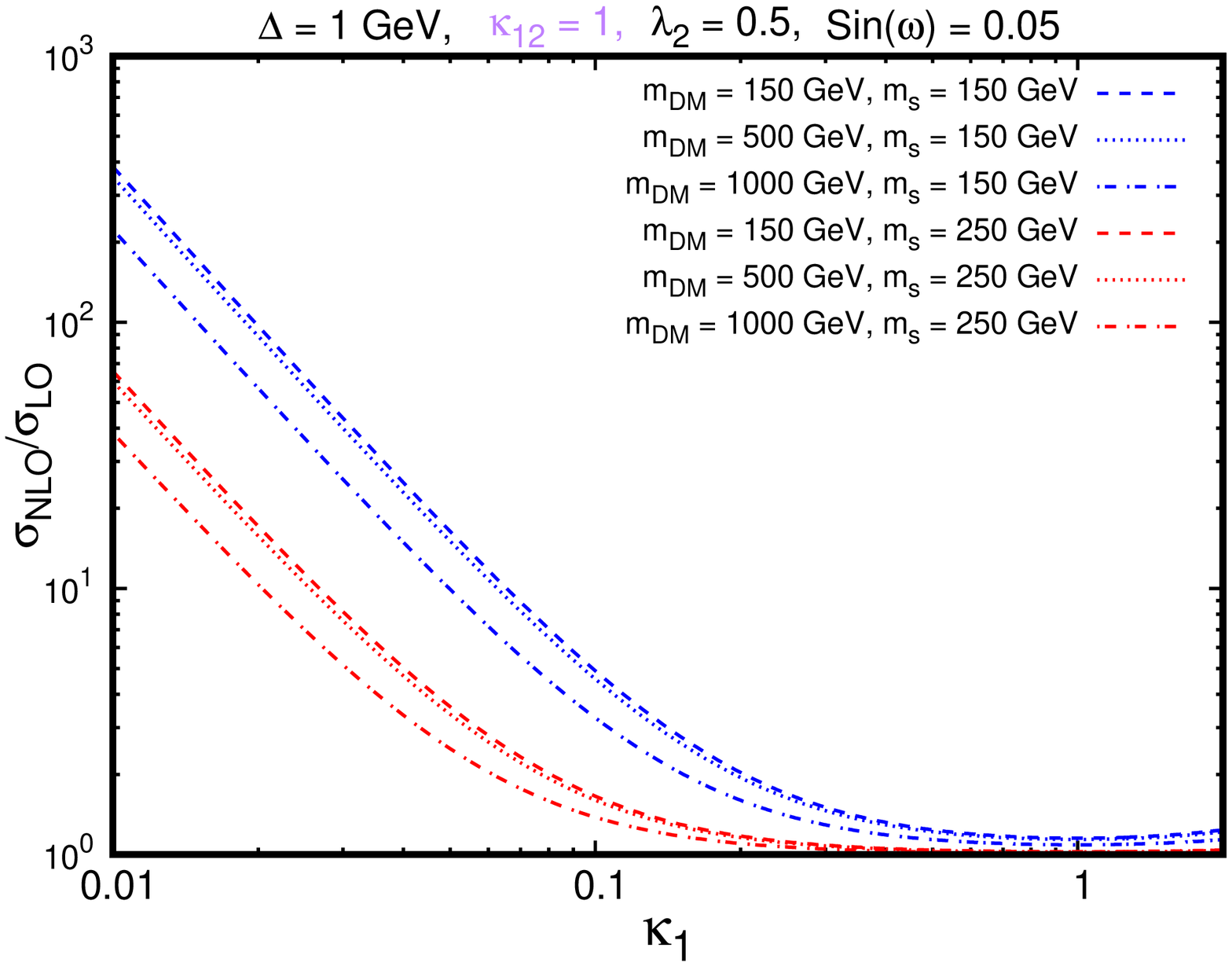}
\end{minipage}
\begin{minipage}{.52\textwidth}
\includegraphics[width=.75\textwidth,angle =-90]{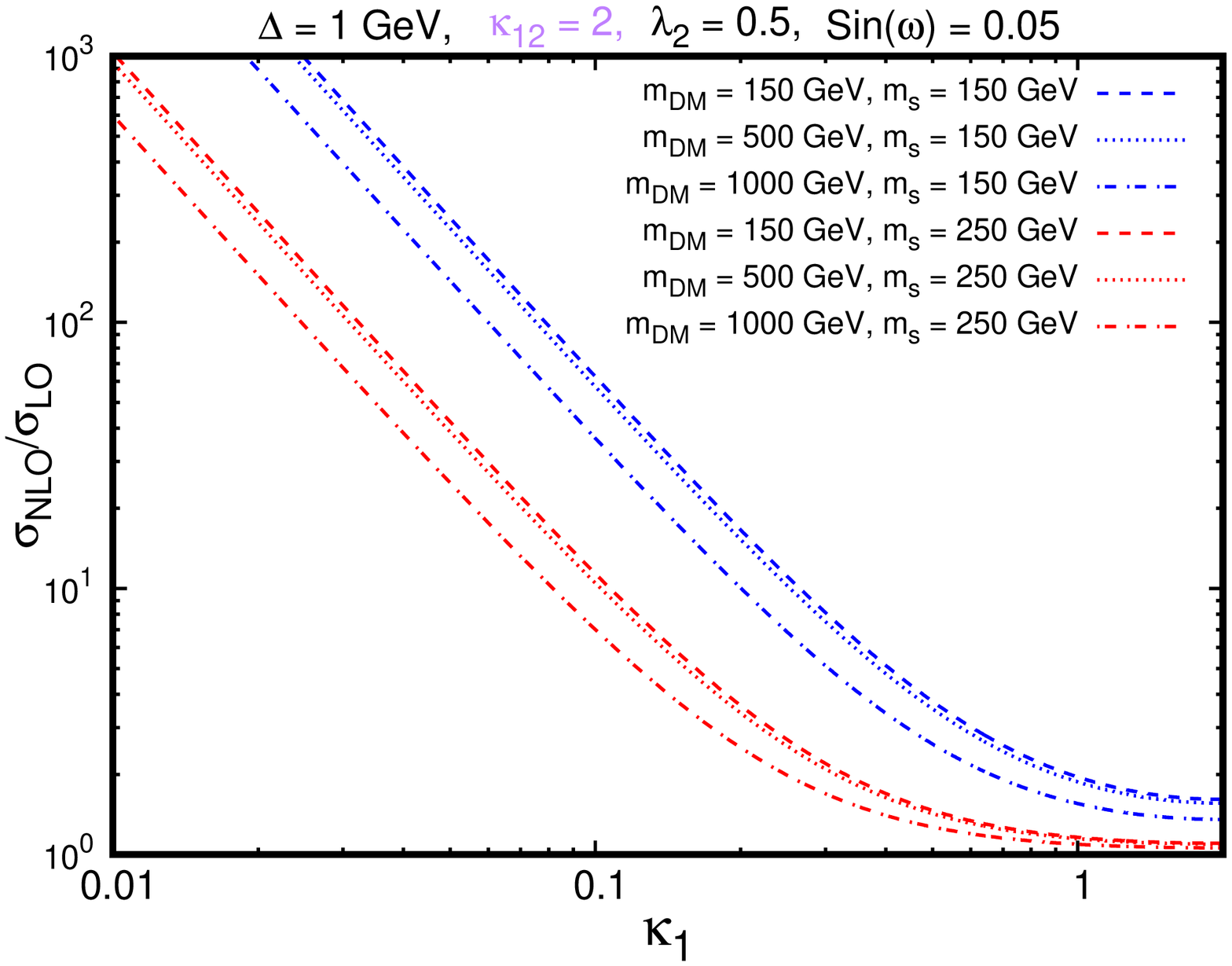}
\end{minipage}
\begin{minipage}{.52\textwidth}
\includegraphics[width=.75\textwidth,angle =-90]{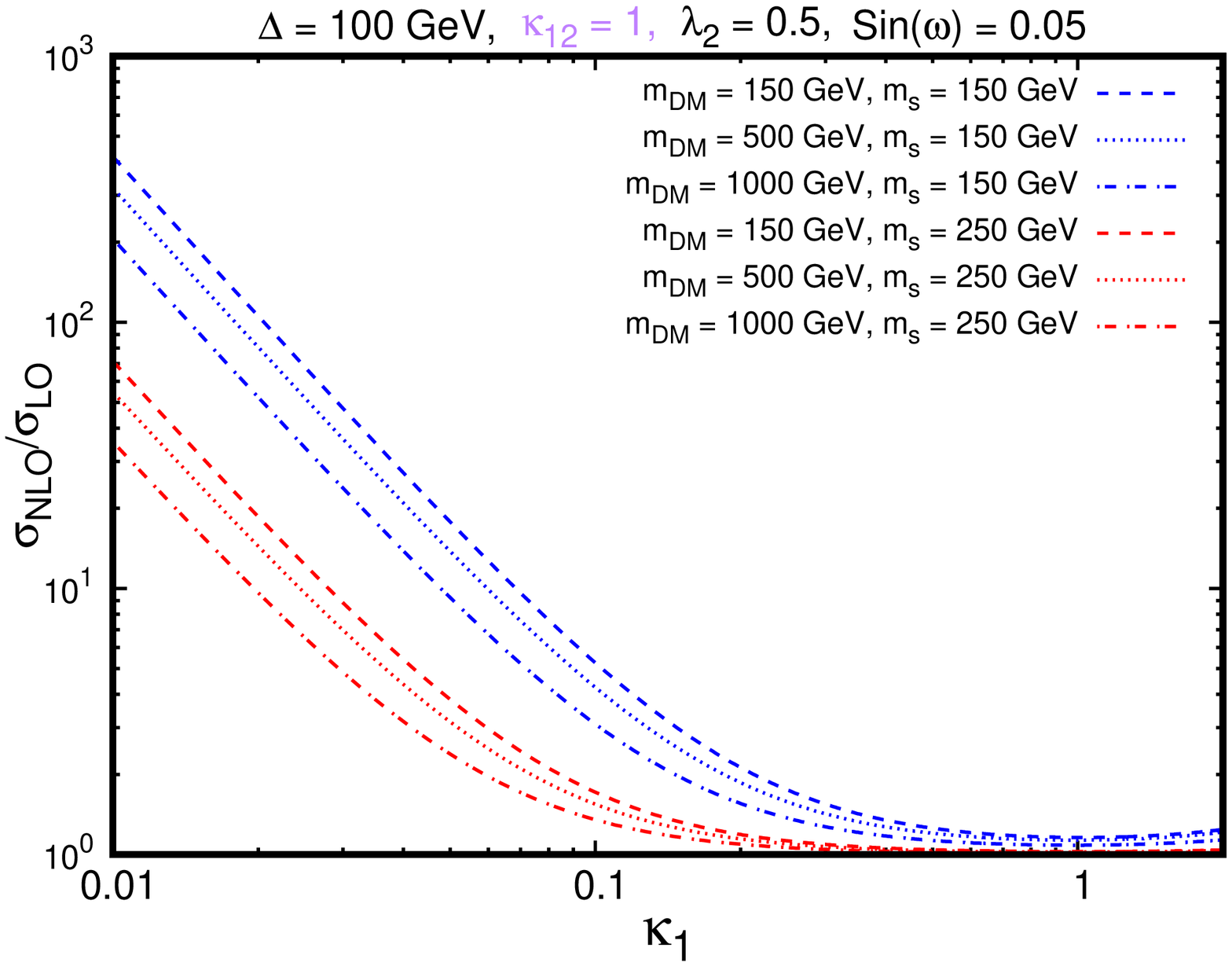}
\end{minipage}
\begin{minipage}{.52\textwidth}
\includegraphics[width=.75\textwidth,angle =-90]{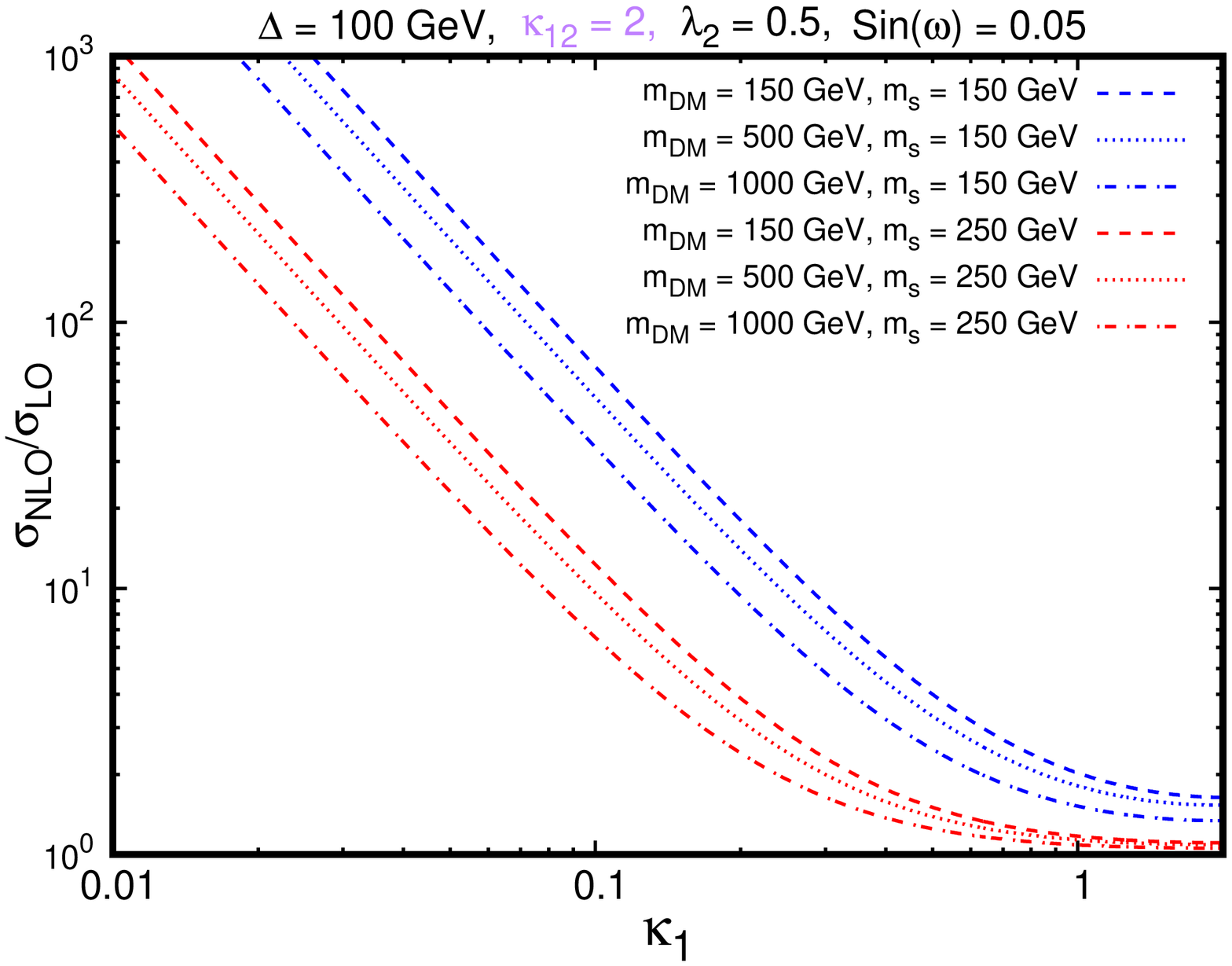}
\end{minipage}
\caption{The same as in Fig.\ref{Ratio-Sin-0.1}, with the mixing angle such that $\sin \omega = 0.05$.} 
\label{Ratio-Sin-0.05}
\end{figure}

Here, we present our numerical results for the DD cross section at one loop level. 
First, we look at the ratio $\sigma^{\text{NLO}}/\sigma^{\text{LO}}$ as a function of the coupling $\kappa_1$, 
while relaxing the bound from the observed relic density. 
The result for the mass splitting $\Delta =1, 100$ GeV, and the mixing angle $\sin \omega = 0.1$, are 
shown in Fig.~\ref{Ratio-Sin-0.1}. We also decrease the mixing angle as $\sin \omega = 0.05$ and present 
the results in Fig.~\ref{Ratio-Sin-0.05}. A general trend in all the figures is that the ratio is approaching unity
with increasing the coupling $\kappa_1$ from 0.01 to 1, for the given values of the coupling $\kappa_{12}= 1,2$.
This result is expected. The DD cross section at one loop depends on both $\kappa_1$ and $\kappa_{12}$, 
while it only depends on the coupling $\kappa_1$ at tree level. For instance, when $\kappa_1 \sim {\cal O}(10^{-2})$
and $\kappa_{12} \sim {\cal O}(1)$ then  $\sigma^{\text{NLO}}/\sigma^{\text{LO}} \propto \kappa_{12}^2/\kappa_{1}^2$. 
Now by increasing $\kappa_1$, the cross section at tree level starts increasing, 
such that when $\kappa_{1} \sim {\cal O}(1)$, then we approach the limit $\sigma^{\text{NLO}} \sim \sigma^{\text{LO}}$.

Another observation is that by decreasing the mixing angle, the ratio increases slightly. The reason is that the cross section at tree level only depends on $\sin \omega$, while the cross section at one loop includes terms depending 
on $\cos \omega$ as well.
 
\begin{figure}
\begin{minipage}{.55\textwidth}
\includegraphics[width=.65\textwidth,angle =-90]{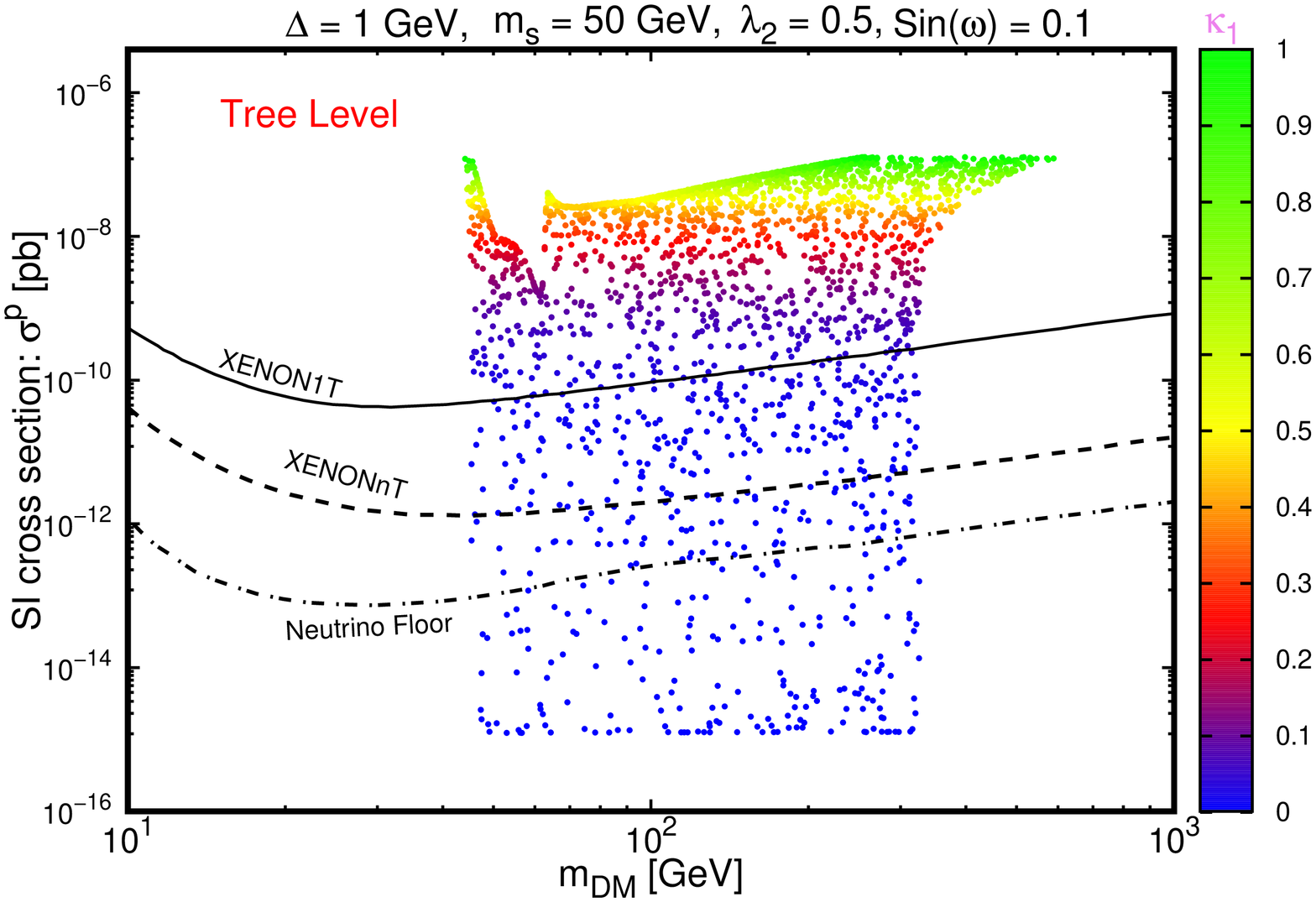}
\end{minipage}
\begin{minipage}{.55\textwidth}
\includegraphics[width=.65\textwidth,angle =-90]{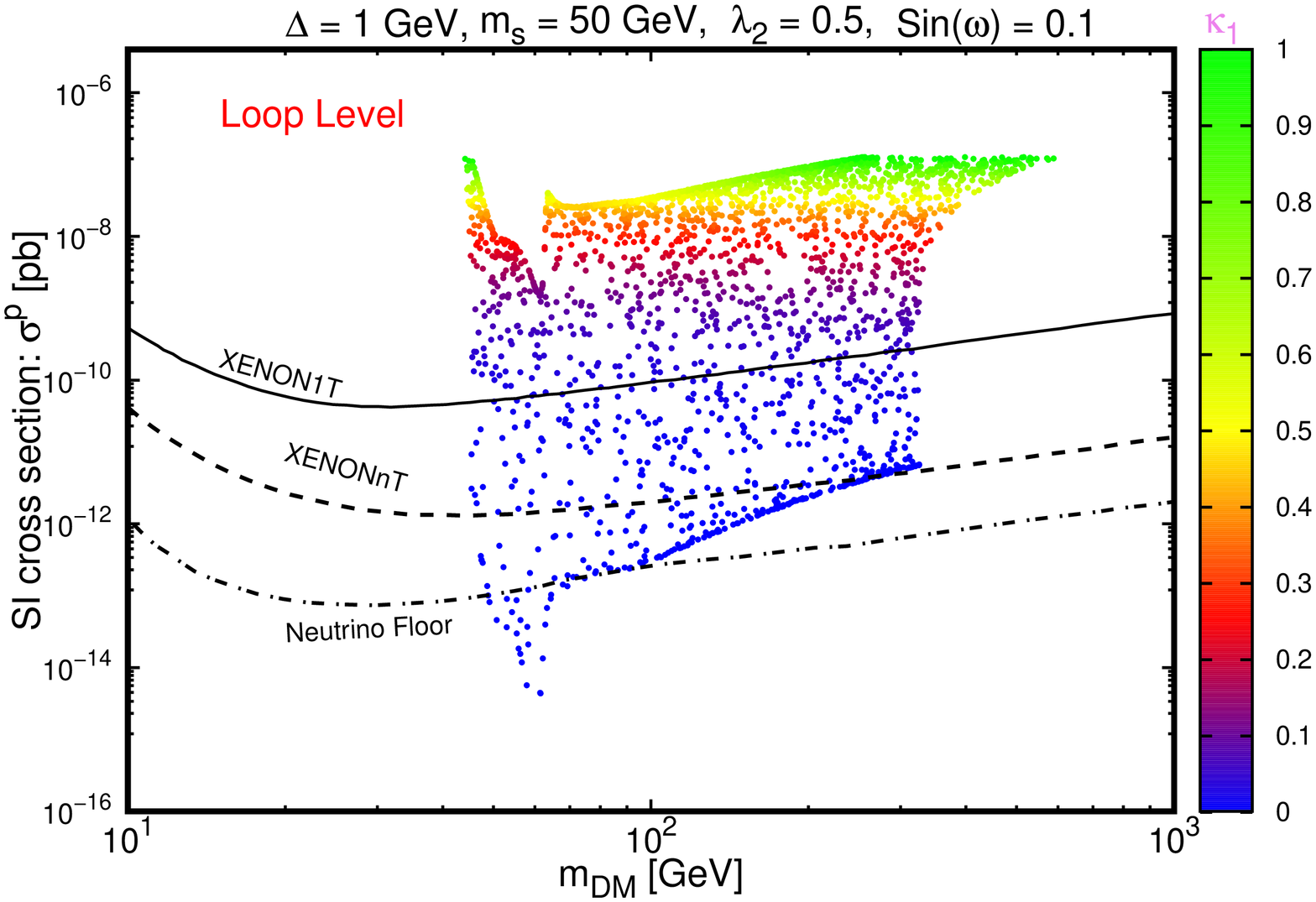}
\end{minipage}
\begin{minipage}{.55\textwidth}
\includegraphics[width=.65\textwidth,angle =-90]{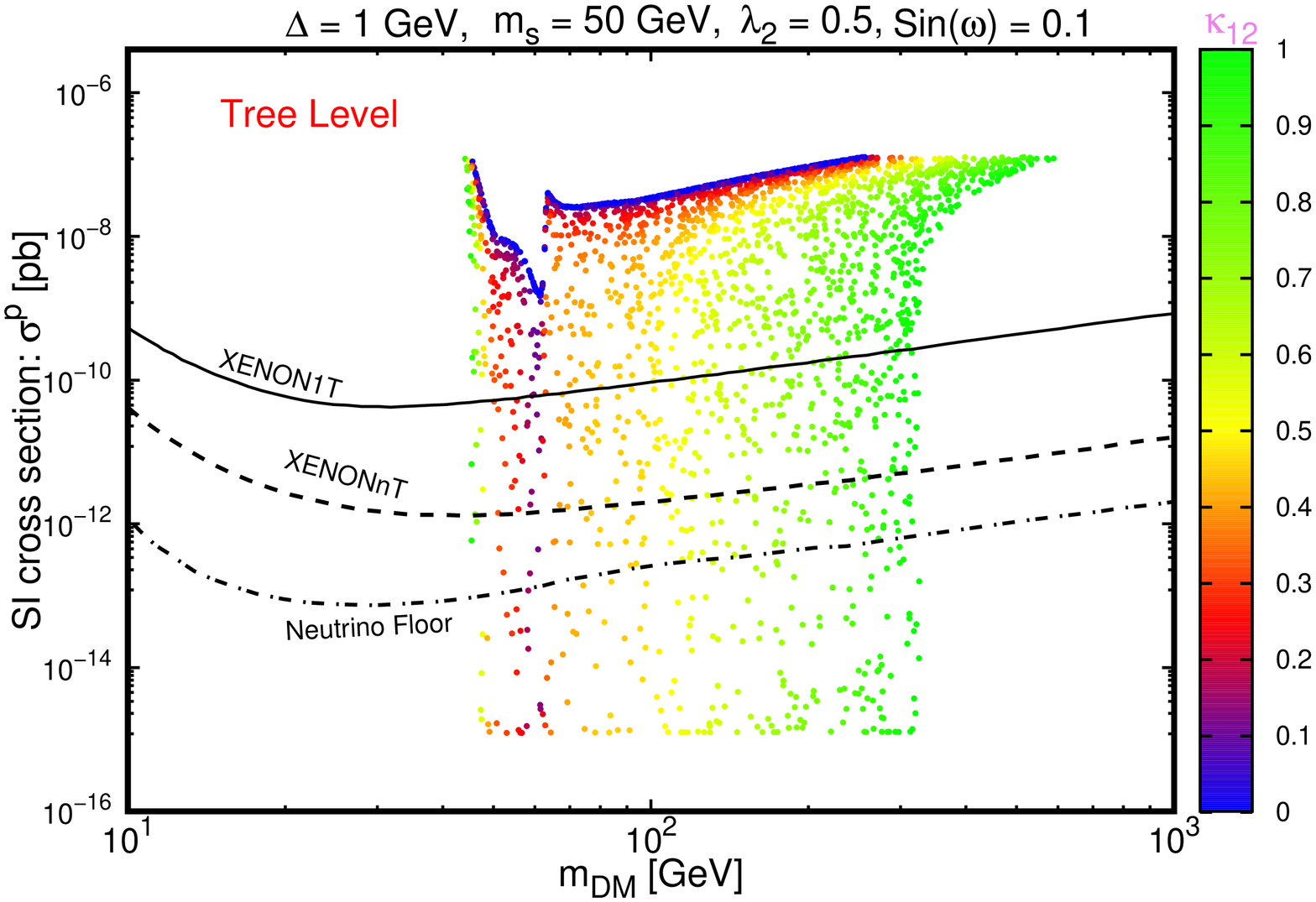}
\end{minipage}
\begin{minipage}{.55\textwidth}
\includegraphics[width=.65\textwidth,angle =-90]{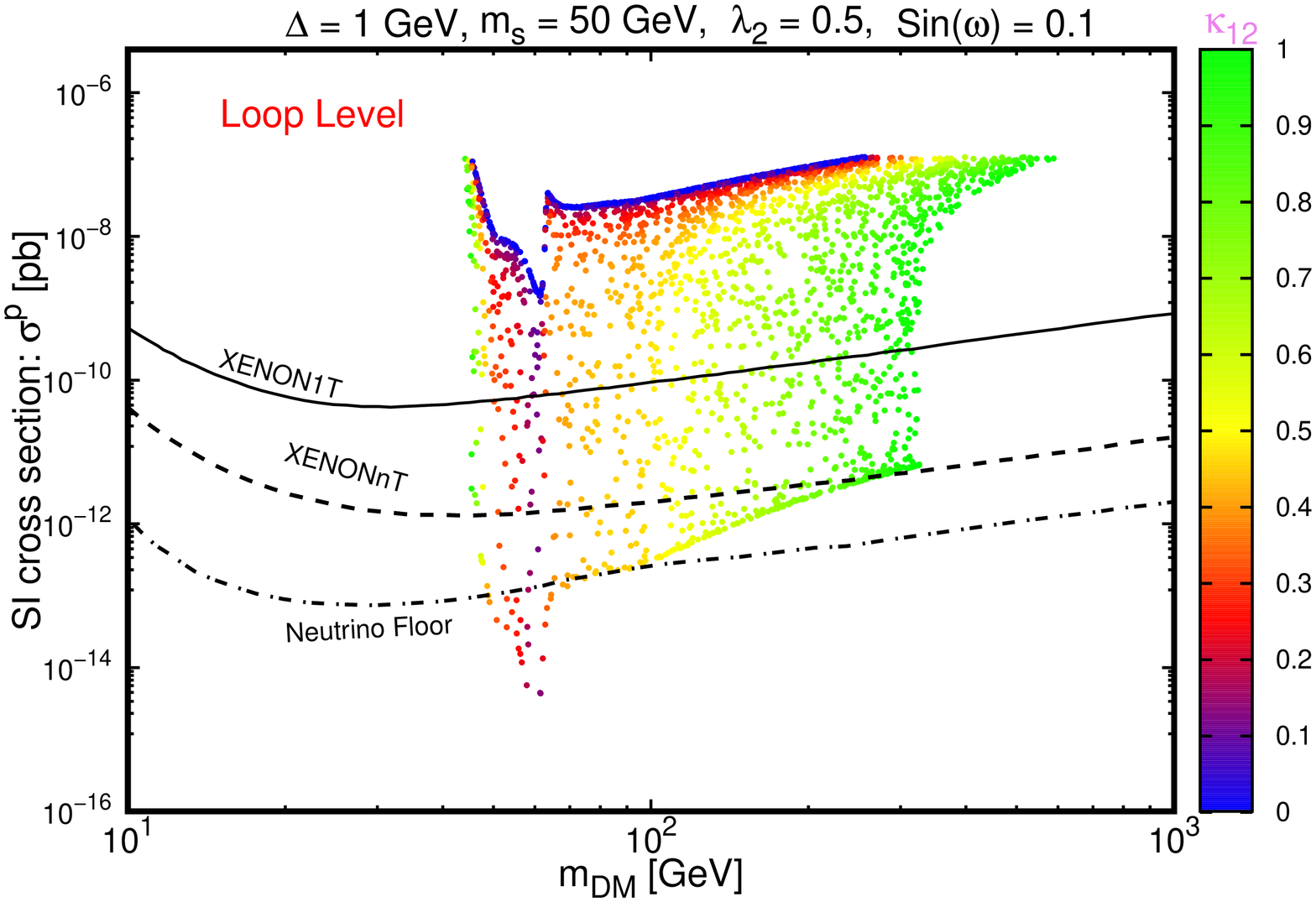}
\end{minipage}
\caption{DD cross section at one loop is shown as a function of DM mass for $m_s = 50$ GeV and $\Delta =1 $ GeV. 
All the points respect the observed relic density. 
Upper limits from XENON1t and projected XENONnT are placed. As such, the neutrino floor is shown.} 
\label{CrossLoop-del1-ms50}
\end{figure}

\begin{figure}
\begin{minipage}{.55\textwidth}
\includegraphics[width=.65\textwidth,angle =-90]{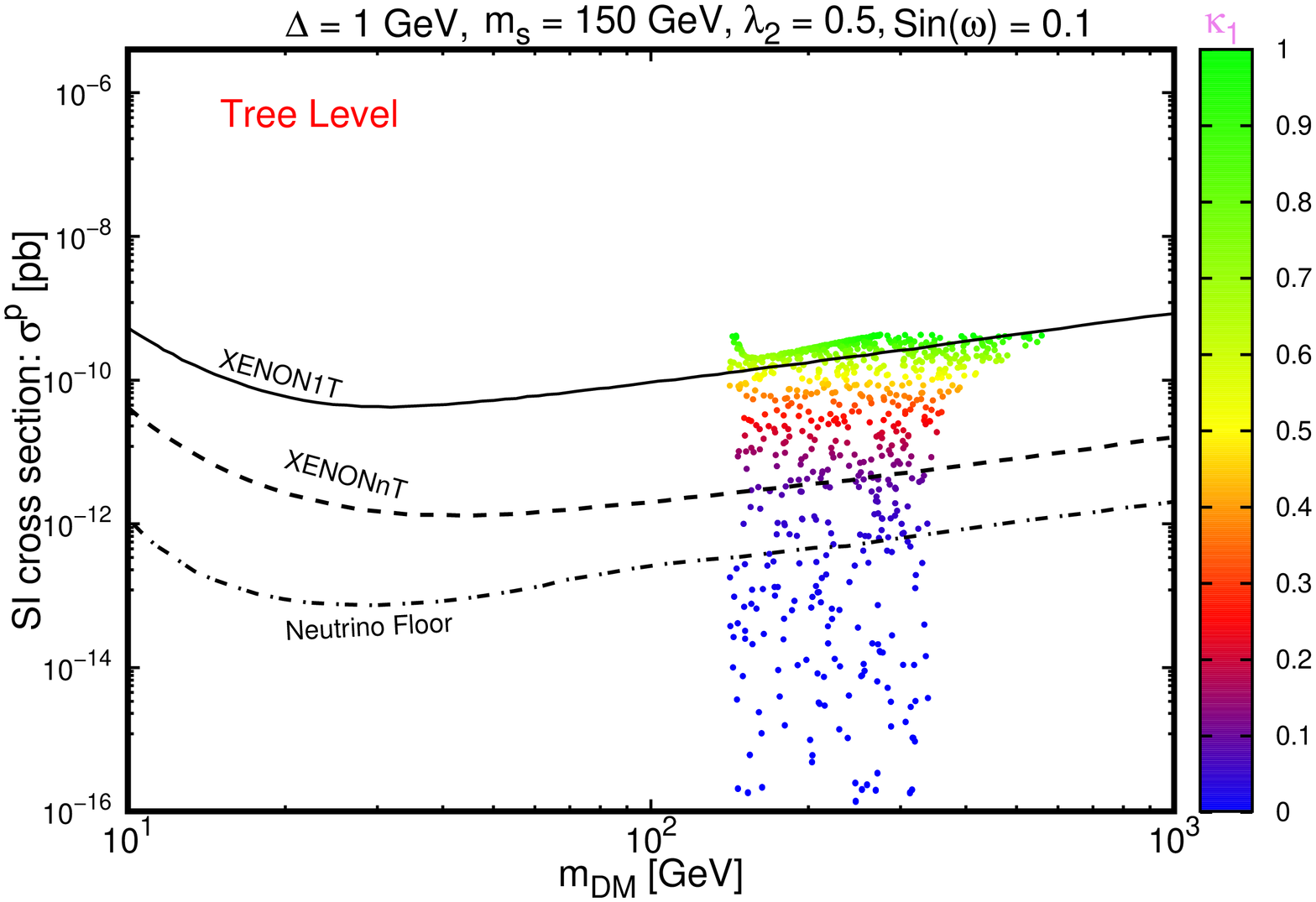}
\end{minipage}
\begin{minipage}{.55\textwidth}
\includegraphics[width=.65\textwidth,angle =-90]{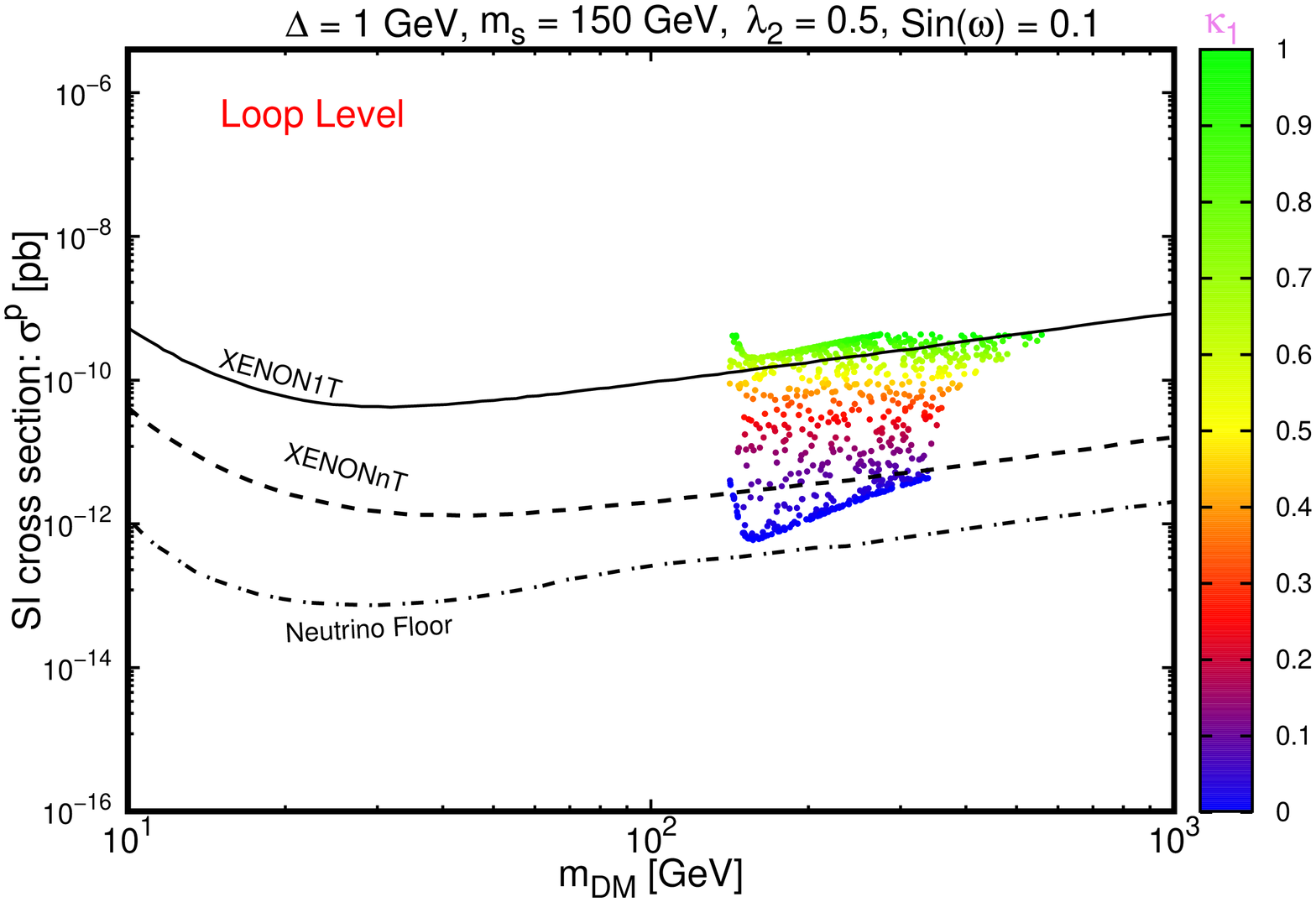}
\end{minipage}
\begin{minipage}{.55\textwidth}
\includegraphics[width=.65\textwidth,angle =-90]{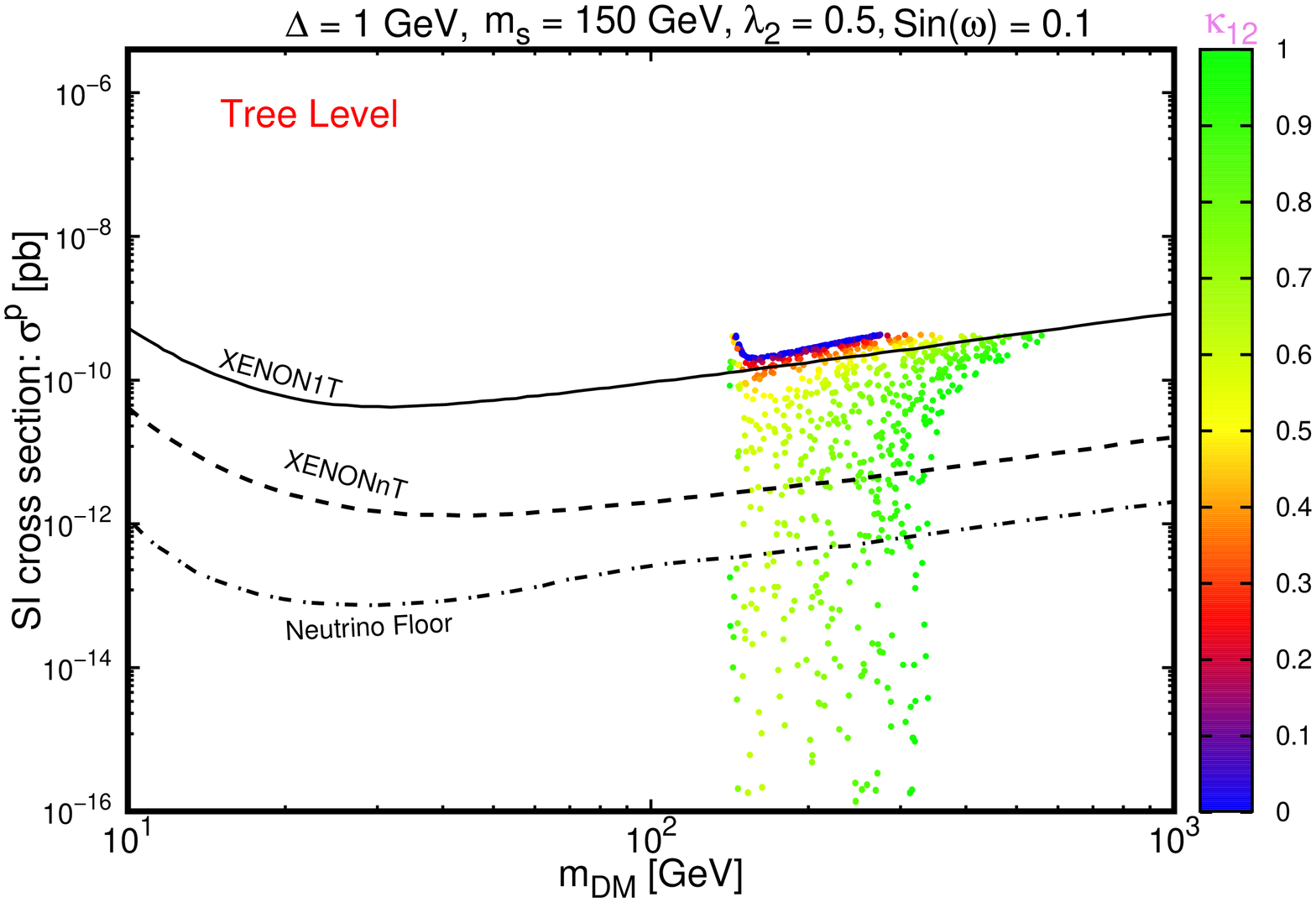}
\end{minipage}
\begin{minipage}{.55\textwidth}
\includegraphics[width=.65\textwidth,angle =-90]{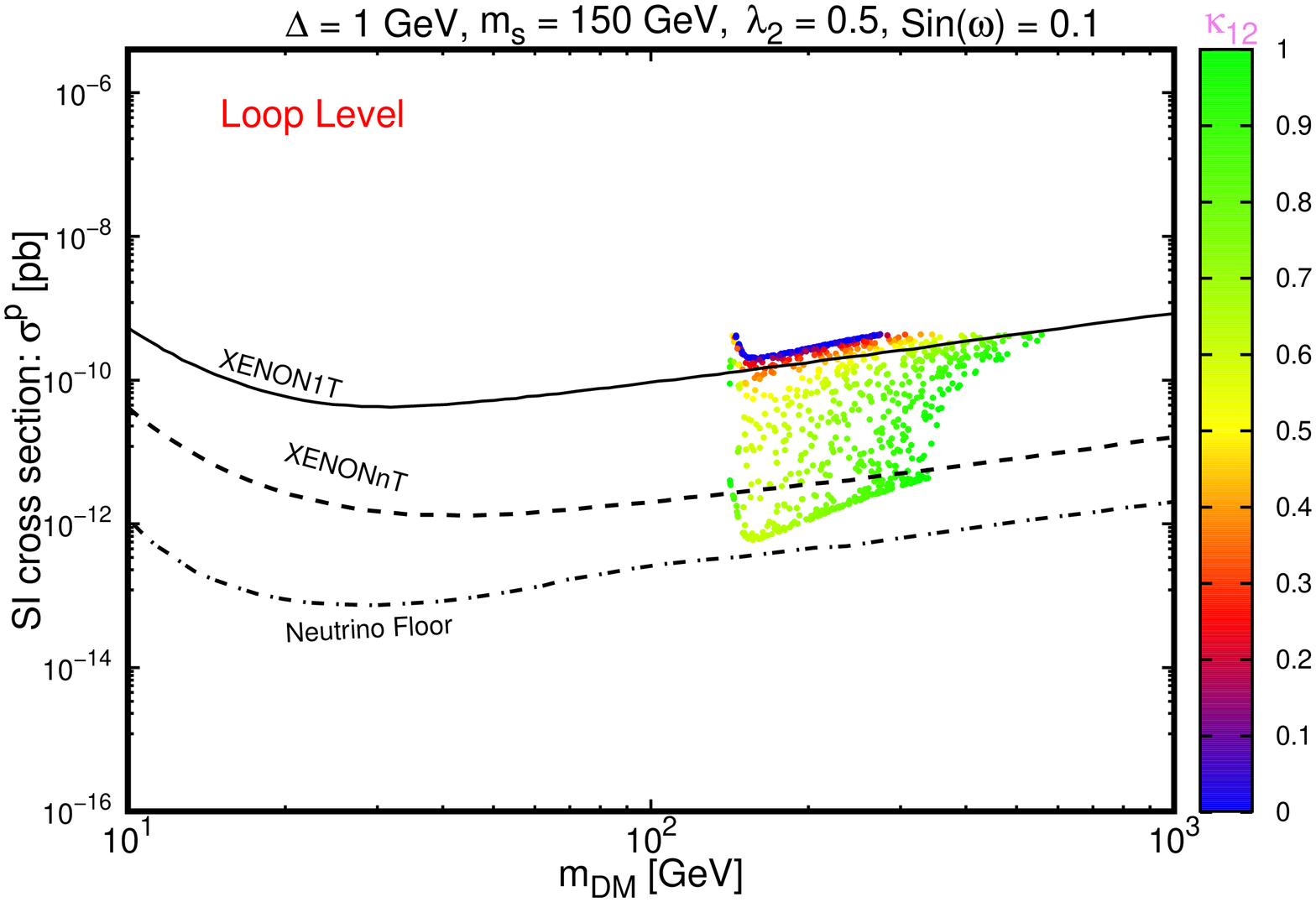}
\end{minipage}
\caption{The same as in Fig.~\ref{CrossLoop-del1-ms50}, with $m_s = 150$ GeV.} 
\label{CrossLoop-del1-ms150}
\end{figure}

\begin{figure}
\begin{minipage}{.55\textwidth}
\includegraphics[width=.65\textwidth,angle =-90]{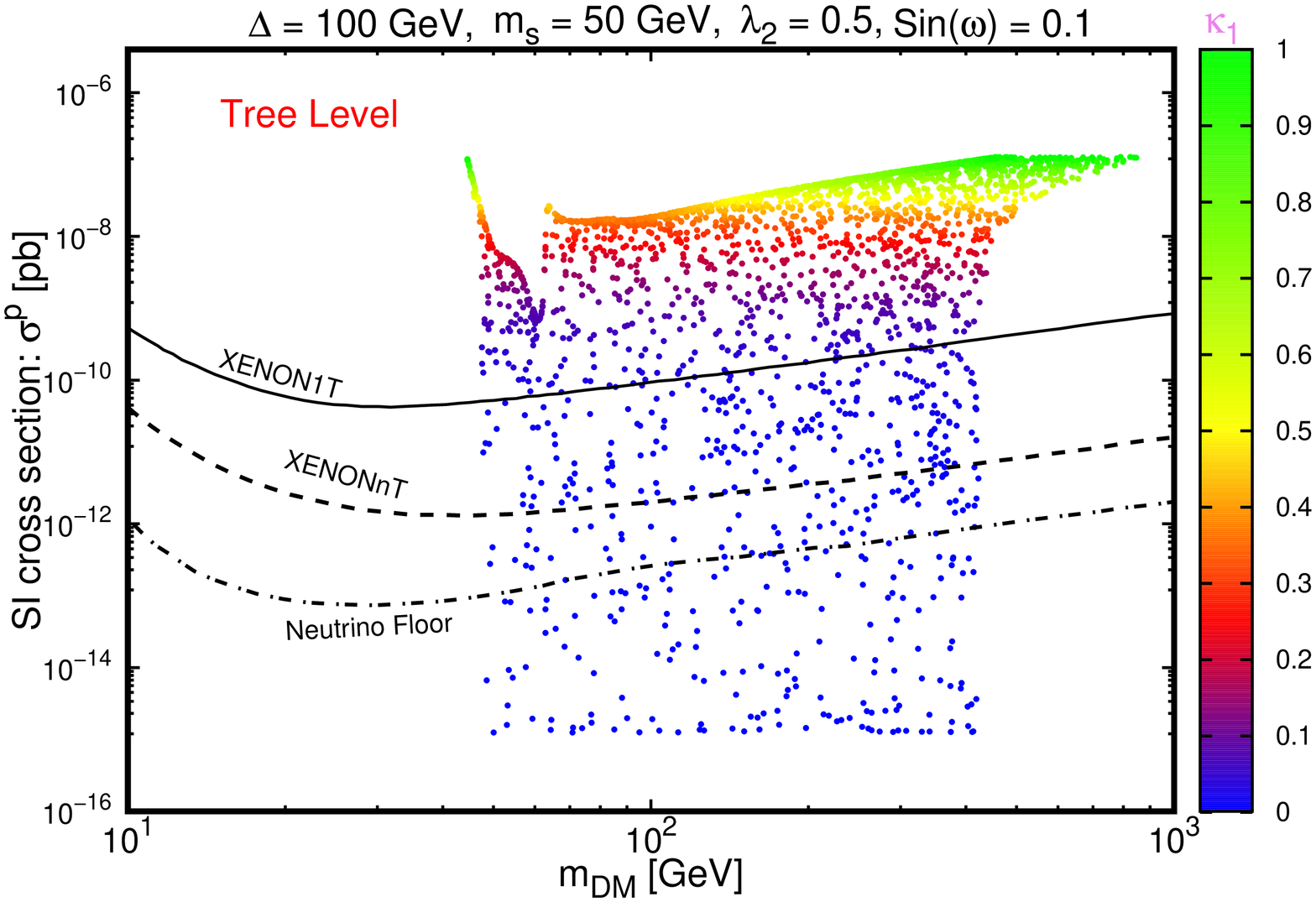}
\end{minipage}
\begin{minipage}{.55\textwidth}
\includegraphics[width=.65\textwidth,angle =-90]{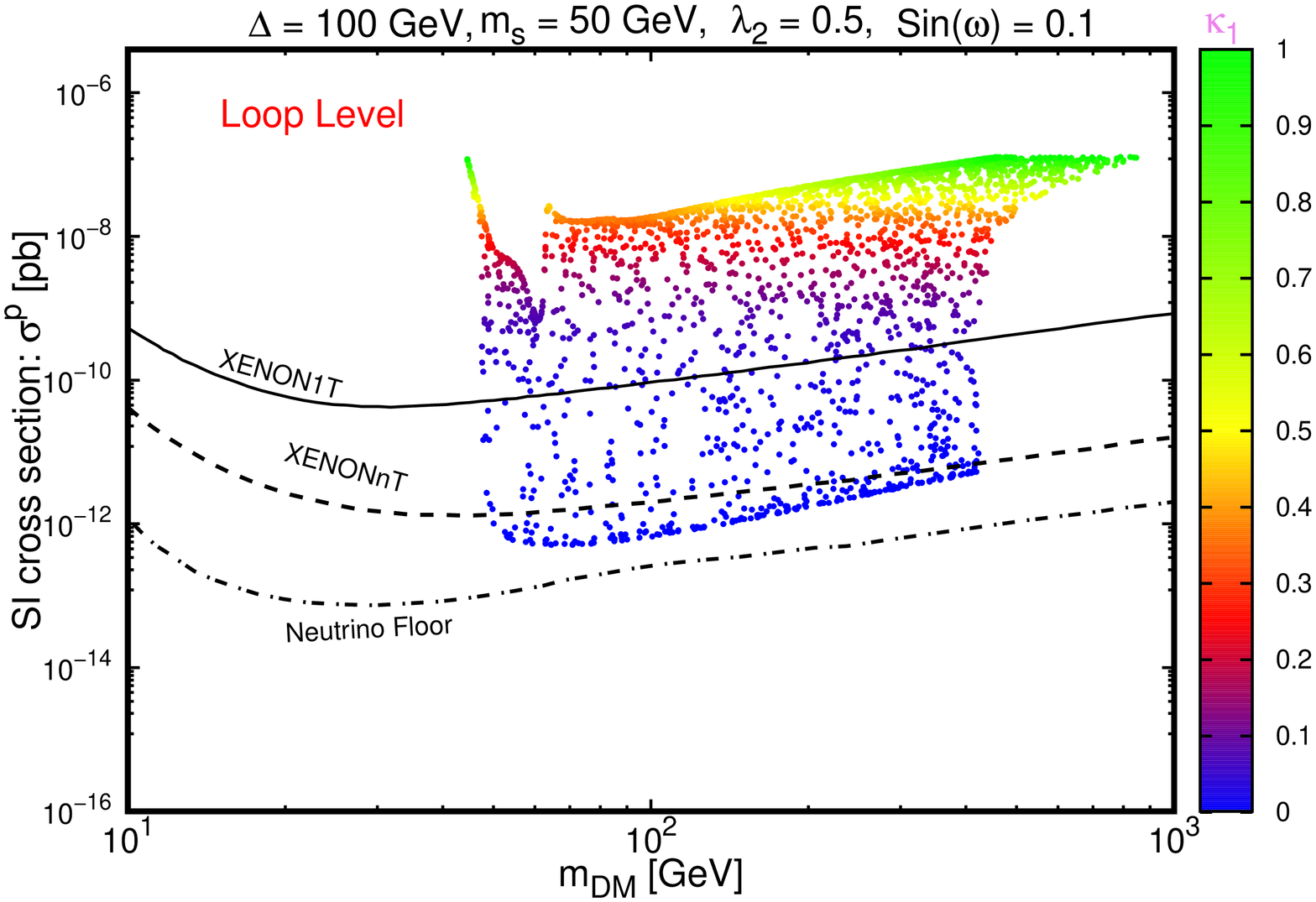}
\end{minipage}
\begin{minipage}{.55\textwidth}
\includegraphics[width=.65\textwidth,angle =-90]{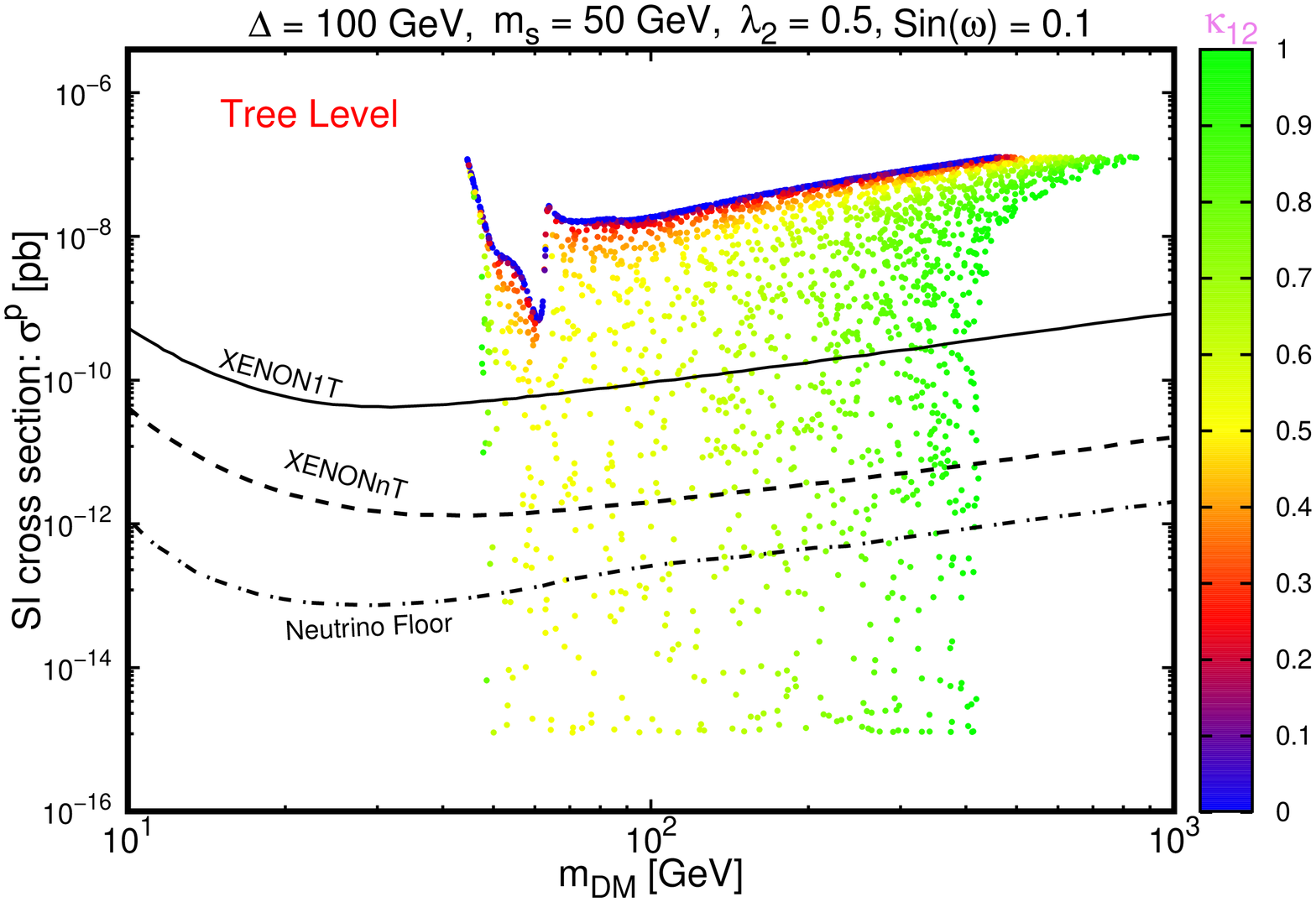}
\end{minipage}
\begin{minipage}{.55\textwidth}
\includegraphics[width=.65\textwidth,angle =-90]{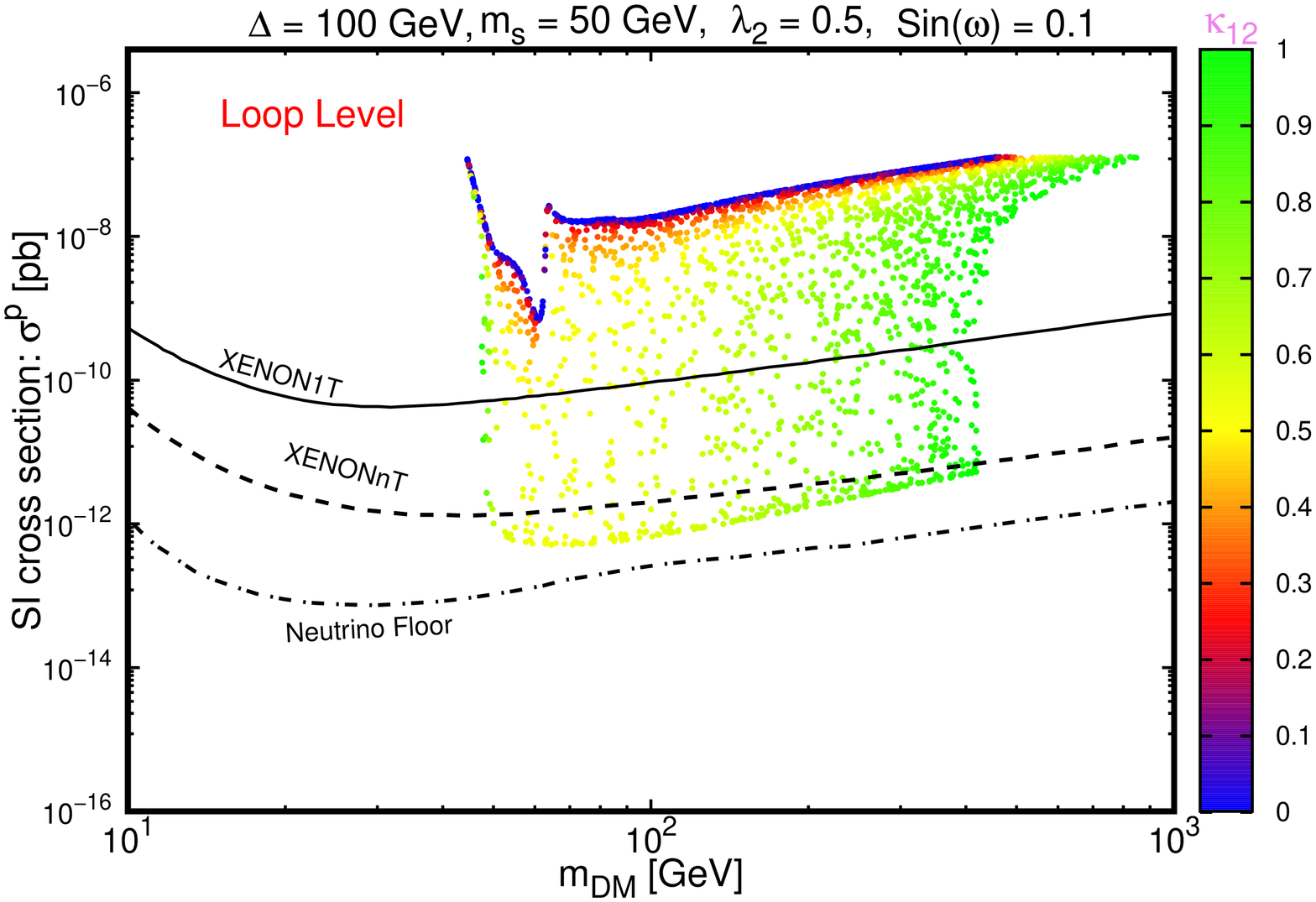}
\end{minipage}
\caption{DD cross section at one loop is shown as a function of DM mass for $m_s = 50$ GeV and $\Delta =100 $ GeV. 
All the points respect the observed relic density. 
Upper limits from XENON1t and projected XENONnT are placed. As such, the neutrino floor is shown.} 
\label{CrossLoop-del100-ms50}
\end{figure}

\begin{figure}
\begin{minipage}{.55\textwidth}
\includegraphics[width=.65\textwidth,angle =-90]{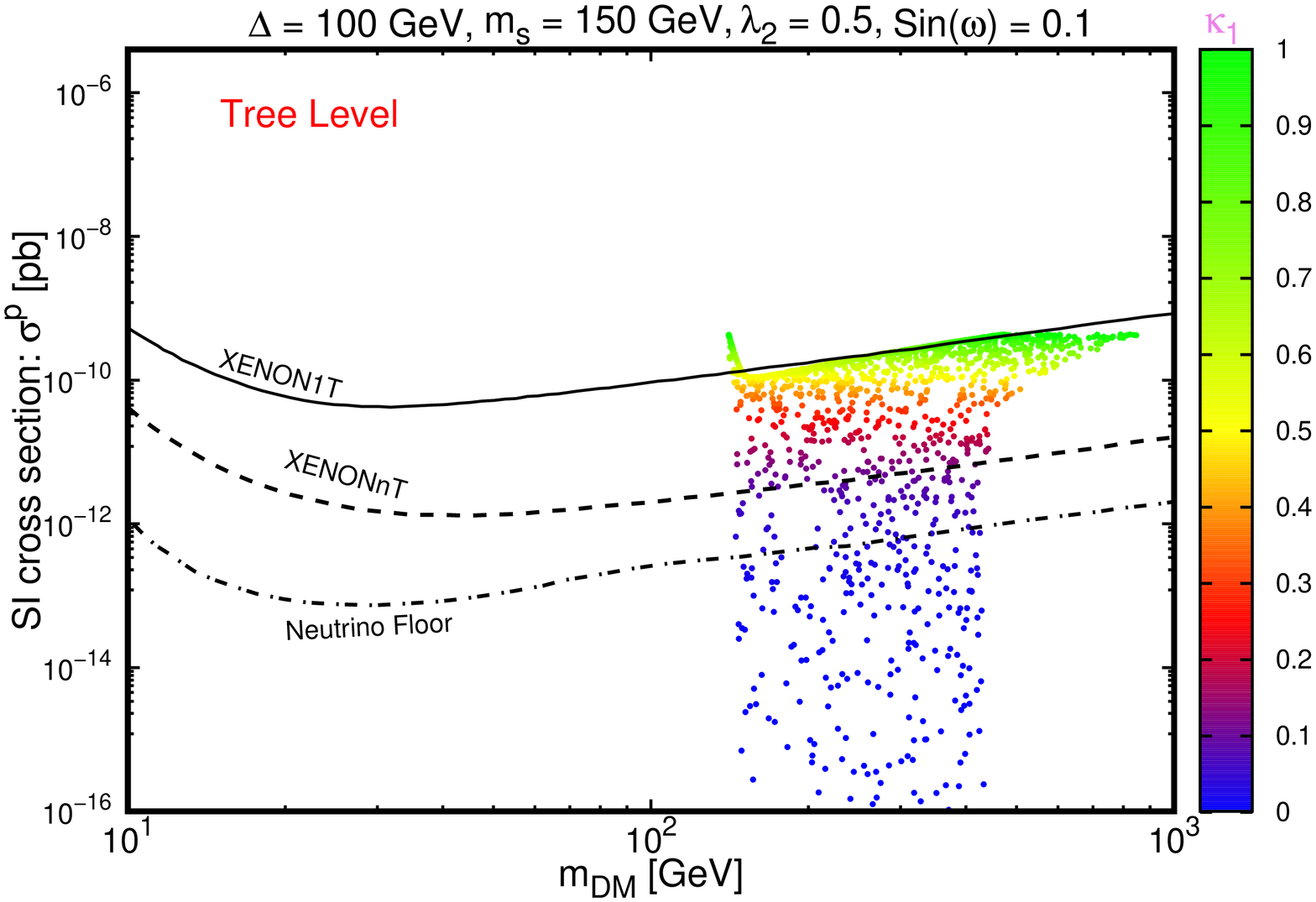}
\end{minipage}
\begin{minipage}{.55\textwidth}
\includegraphics[width=.65\textwidth,angle =-90]{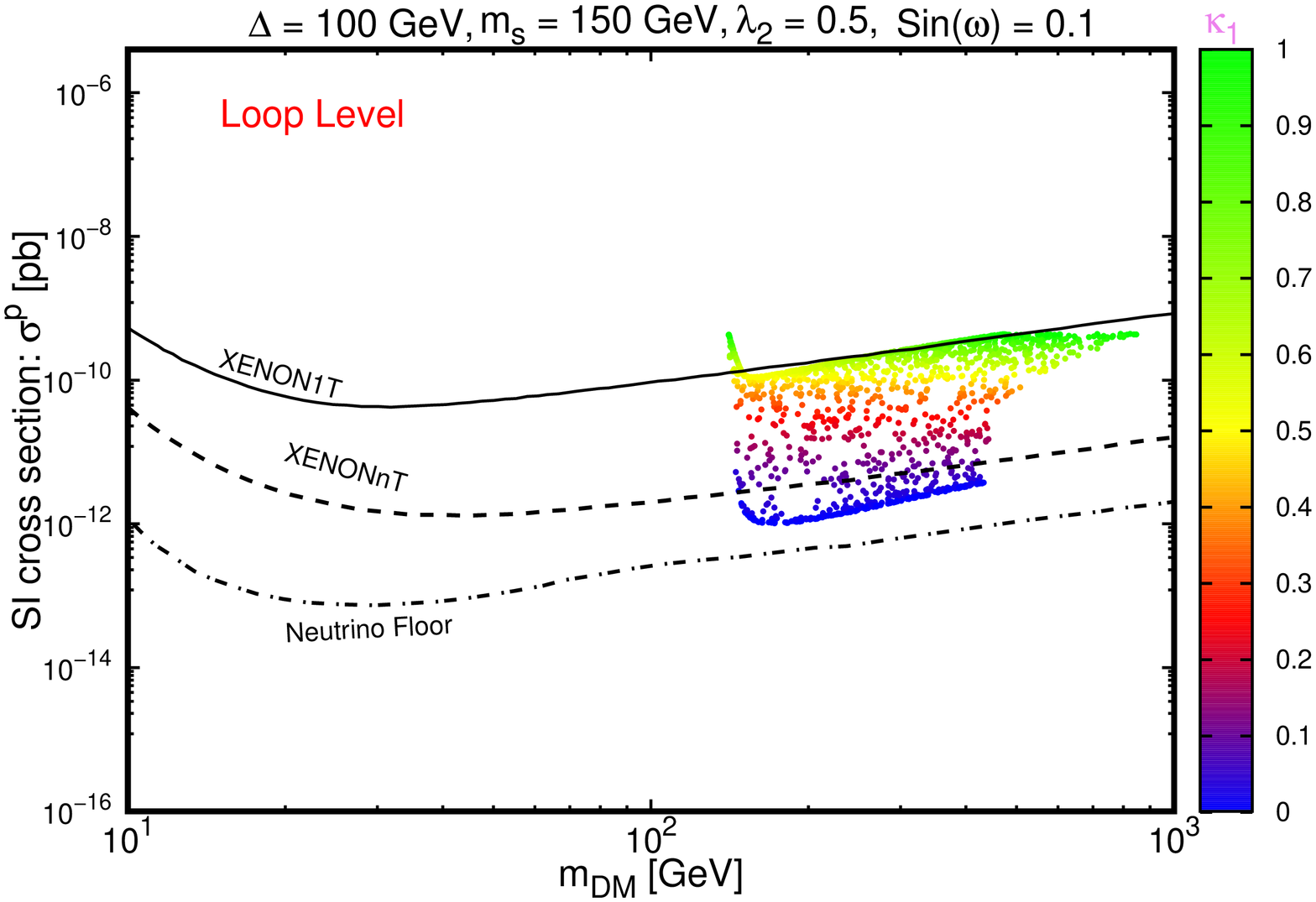}
\end{minipage}
\begin{minipage}{.55\textwidth}
\includegraphics[width=.65\textwidth,angle =-90]{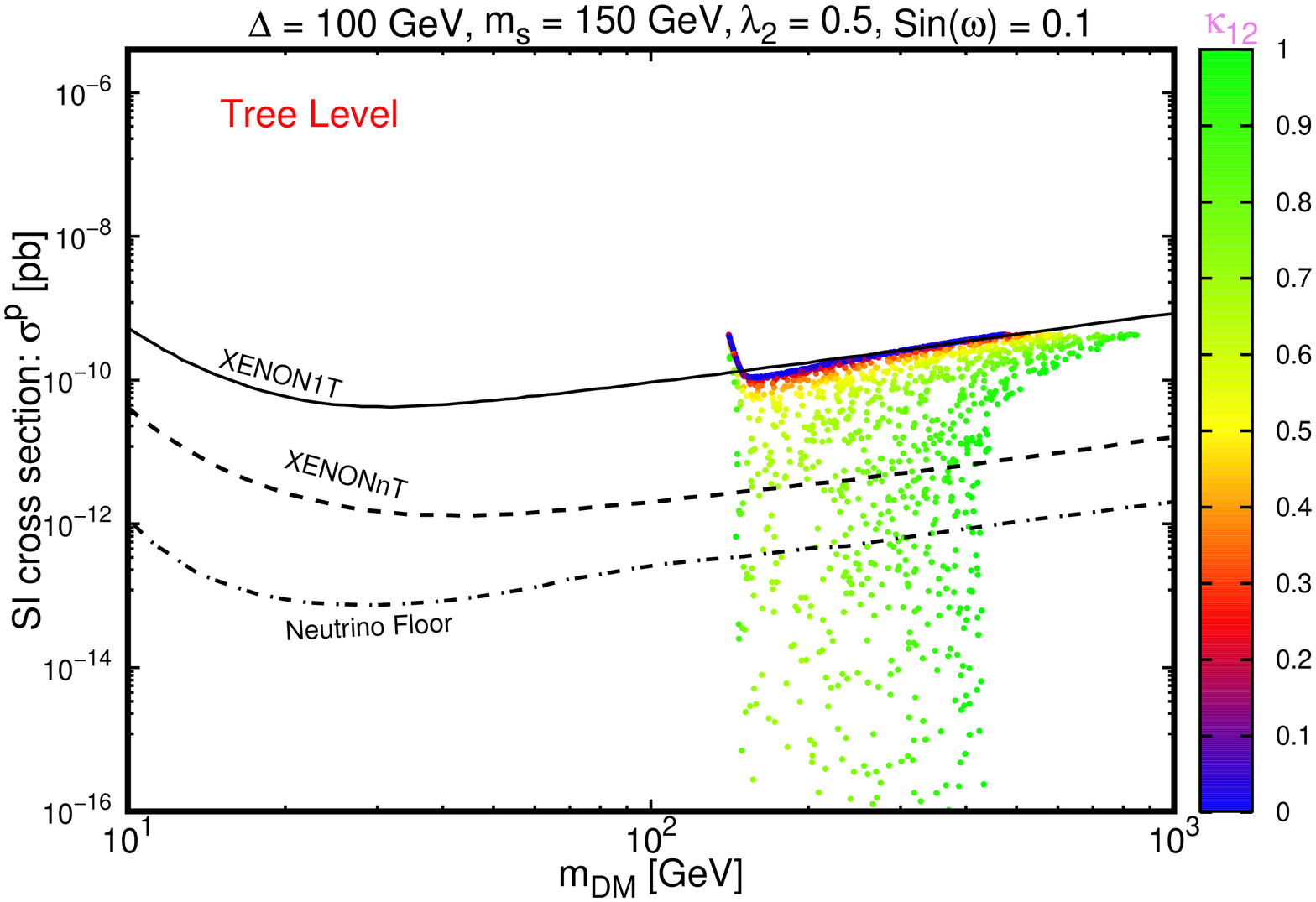}
\end{minipage}
\begin{minipage}{.55\textwidth}
\includegraphics[width=.65\textwidth,angle =-90]{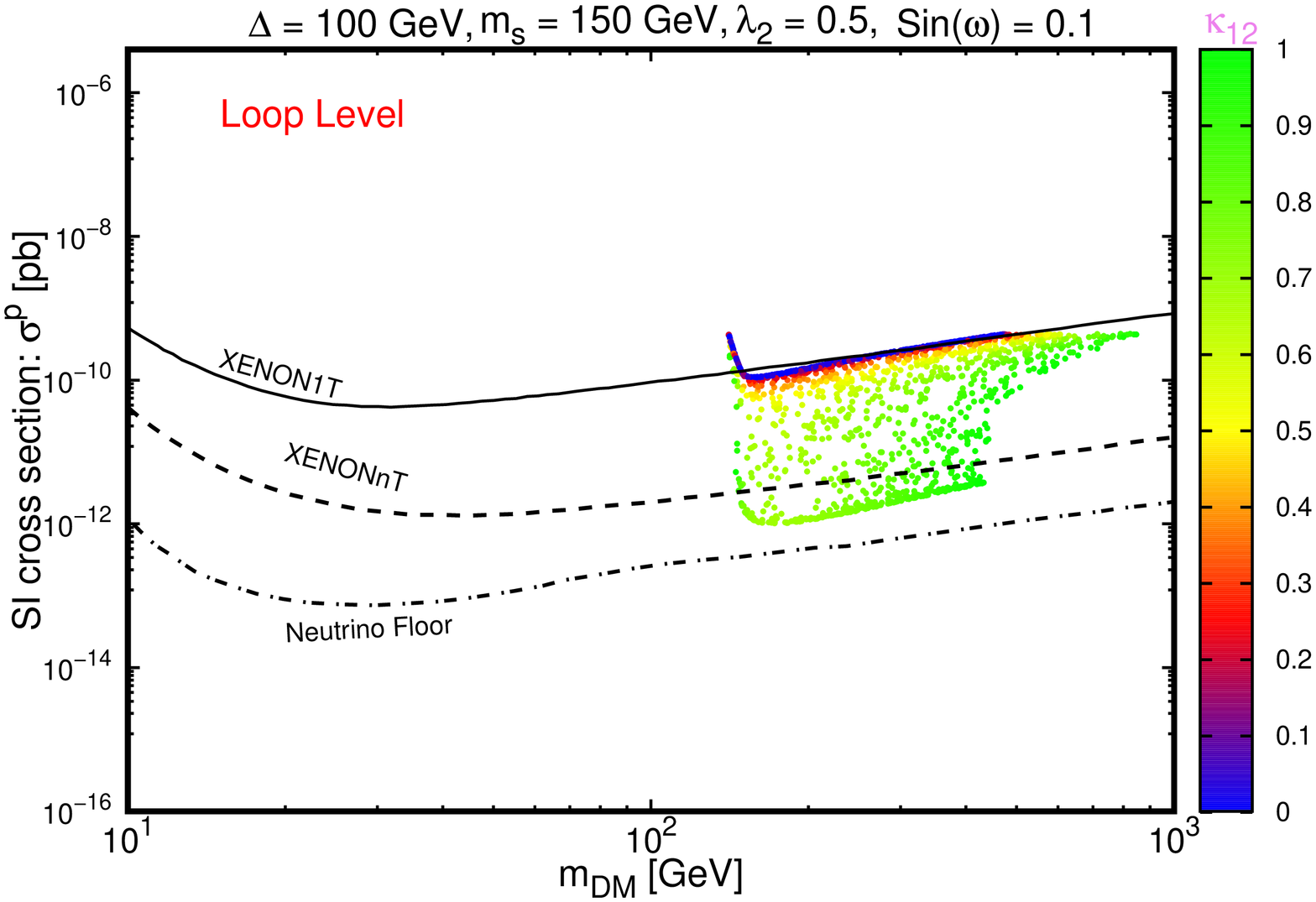}
\end{minipage}
\caption{The same as in Fig.~\ref{CrossLoop-del100-ms50}, with $m_s = 150$ GeV.} 
\label{CrossLoop-del100-ms150}
\end{figure}

Next, we continue our computations to find the regions respecting the observed relic abundance, and bounds from direct detection experiments. First, we keep the same values for the free parameters as those in our computations at tree level; $\lambda_2 = 0.5$ and $\sin(\omega) = 0.1$. 
The singlet scalar masses are fixed at two distinct values, $m_s =50, 150$ GeV. 
The range of the other free parameters in our scan are, $0.001< \kappa_1,\kappa_{12} < 1$ and 
$10~\text{GeV} < m_{\text{DM}} < 2$ TeV.
We show the DD cross section at one loop in terms of the DM mass. 
The results with $\Delta = 1$ GeV and $m_s = 50$ GeV are presented in Fig.~\ref{CrossLoop-del1-ms50},
with $\Delta = 1$ GeV and $m_s = 150$ in Fig.~\ref{CrossLoop-del1-ms150}, 
with $\Delta = 100$ GeV and $m_s = 50$ in Fig.~\ref{CrossLoop-del100-ms50},
and with $\Delta = 100$ GeV and $m_s = 150$ in Fig.~\ref{CrossLoop-del100-ms150}. 
In all the figures, as it was anticipated, we notice an enhancement on the DD cross section 
in the regions with both small $\kappa_1$ and large $\kappa_{12}$. 
It is such that the regions below the neutrino floor shift to the regions which are respected by XENONnT bounds.
In the case with $\Delta = 1$ GeV and $m_s = 50$, a small region around $m_{\text{DM}} 
\sim 60$ GeV still remains below the neutrino floor. 
Another observation is that the regions with DD cross section above the 
neutrino floor get smaller loop corrections. The reason is that in these regions 
the ratio, $\sigma^{\text{NLO}}/\sigma^{\text{LO}}$, 
decreases since $\kappa_1$ grows eventually and one loop and tree level cross sections are almost comparable in size.  

\begin{figure}
\begin{minipage}{.55\textwidth}
\includegraphics[width=.65\textwidth,angle =-90]{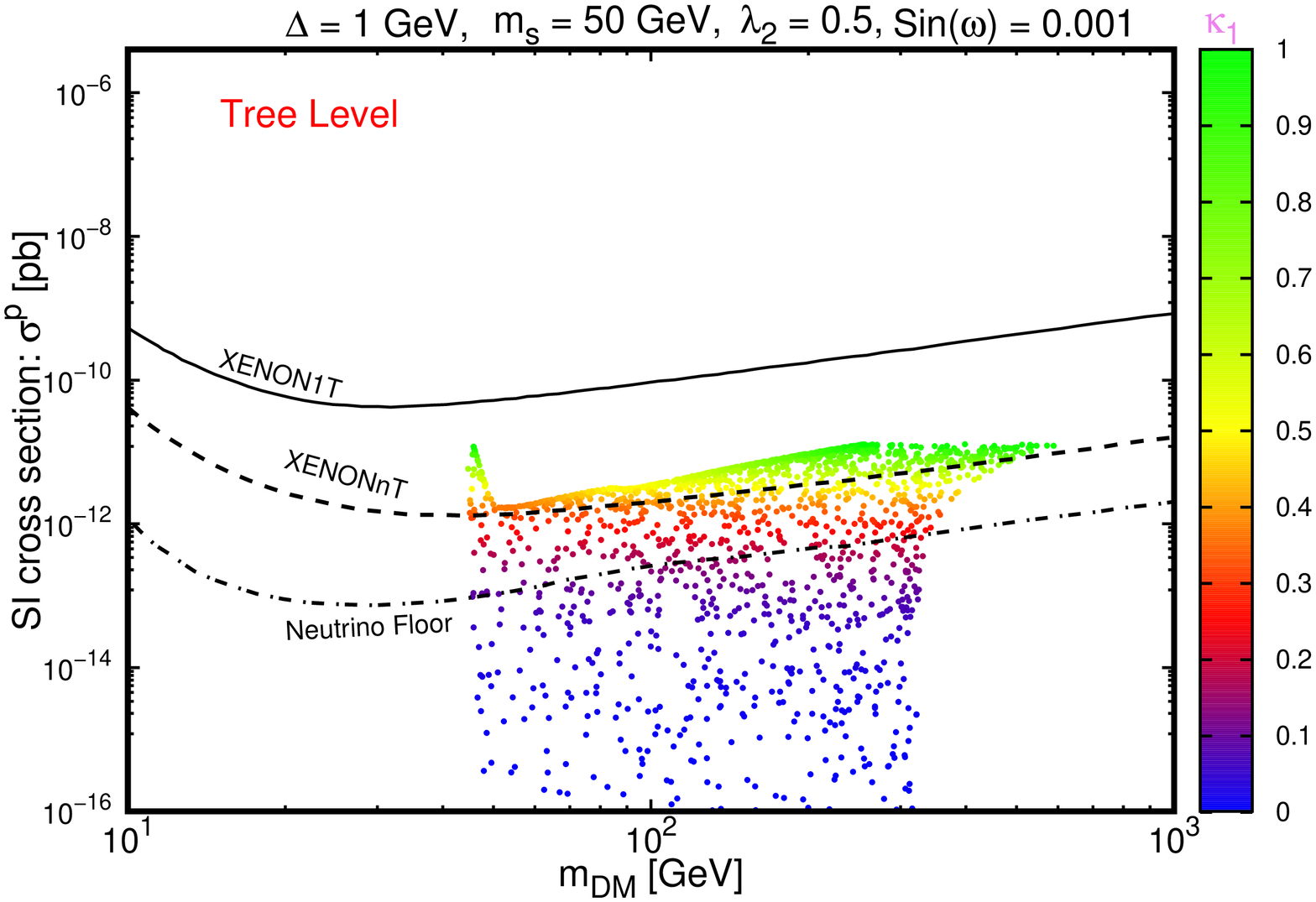}
\end{minipage}
\begin{minipage}{.55\textwidth}
\includegraphics[width=.65\textwidth,angle =-90]{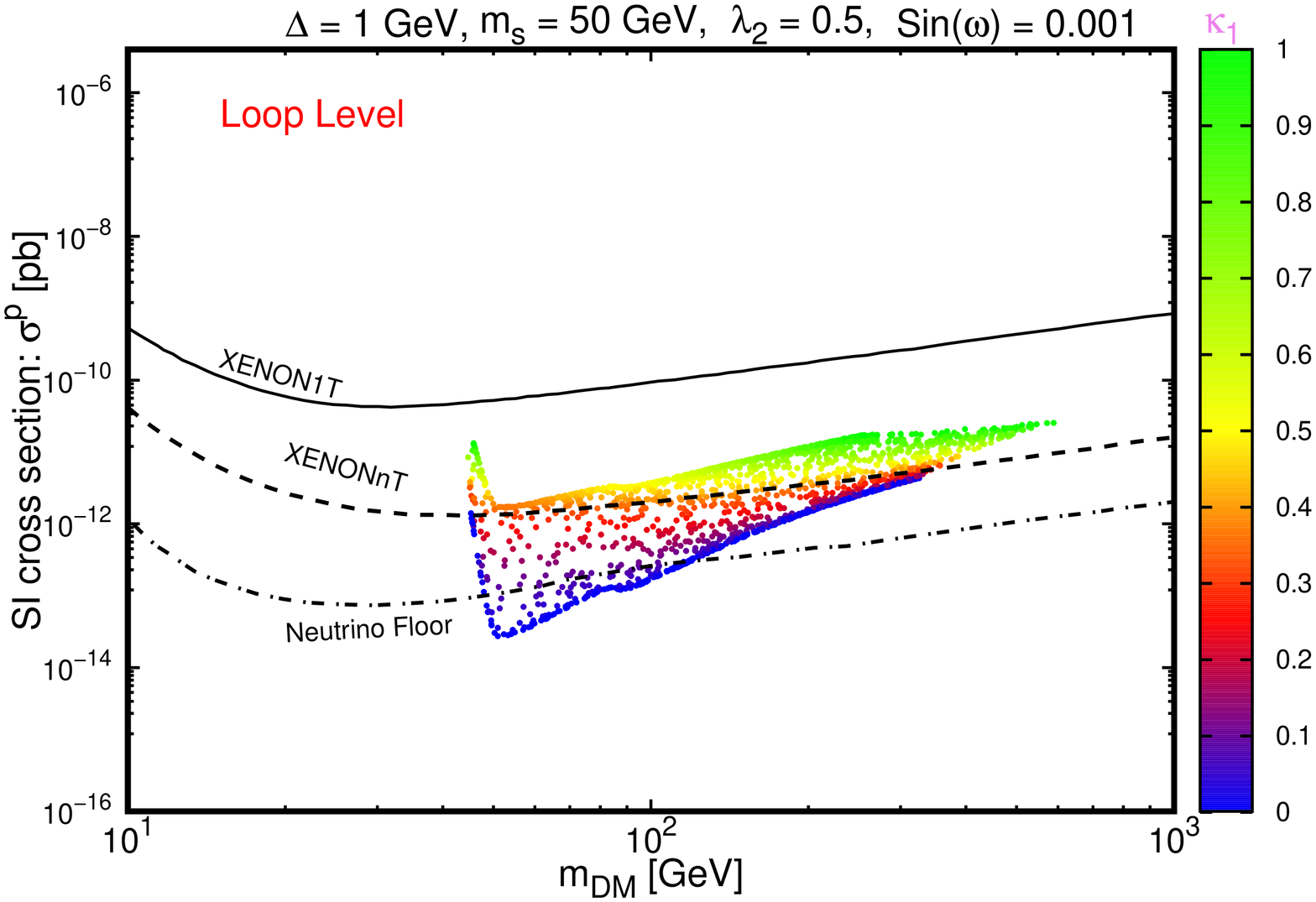}
\end{minipage}
\begin{minipage}{.55\textwidth}
\includegraphics[width=.65\textwidth,angle =-90]{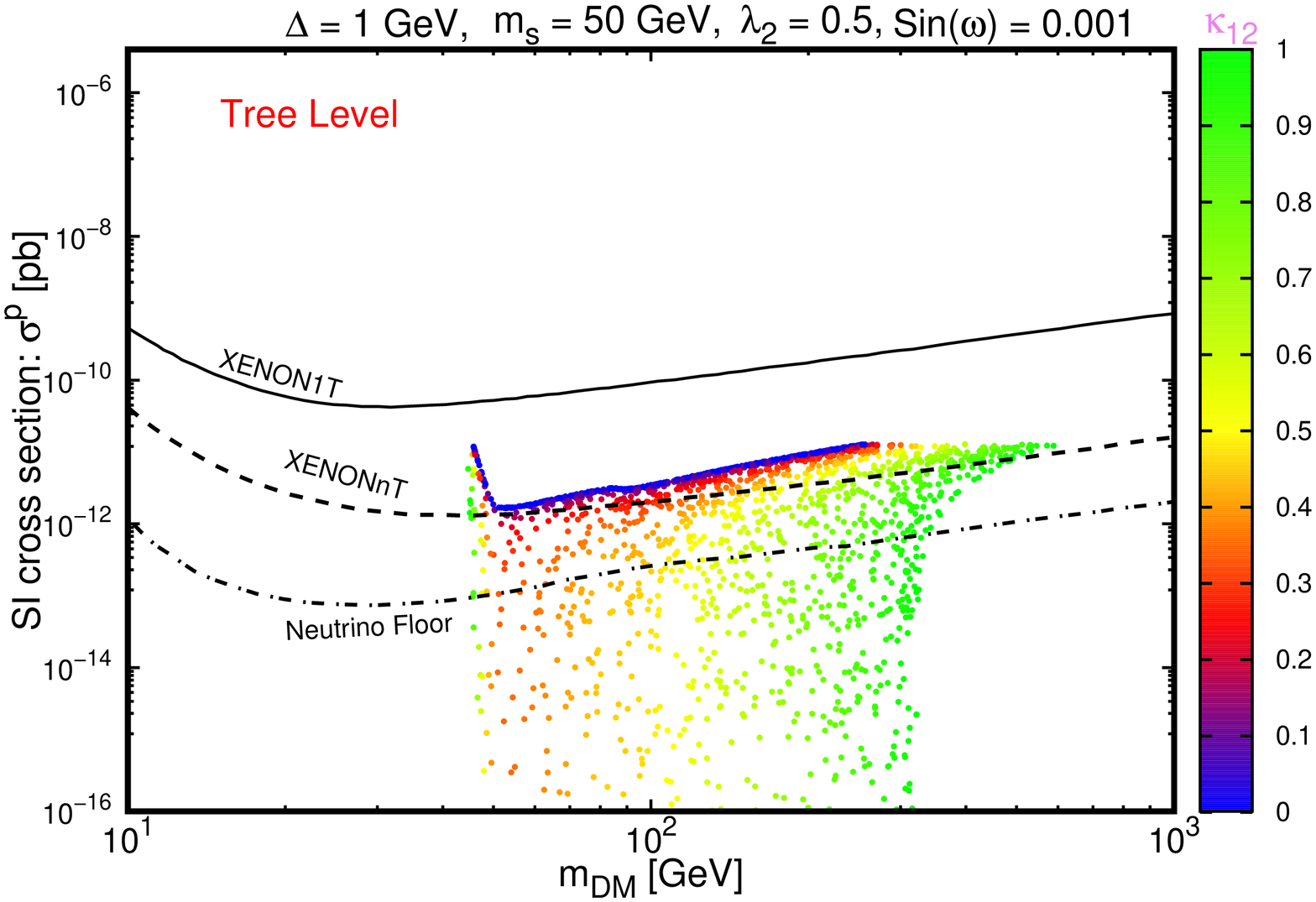}
\end{minipage}
\begin{minipage}{.55\textwidth}
\includegraphics[width=.65\textwidth,angle =-90]{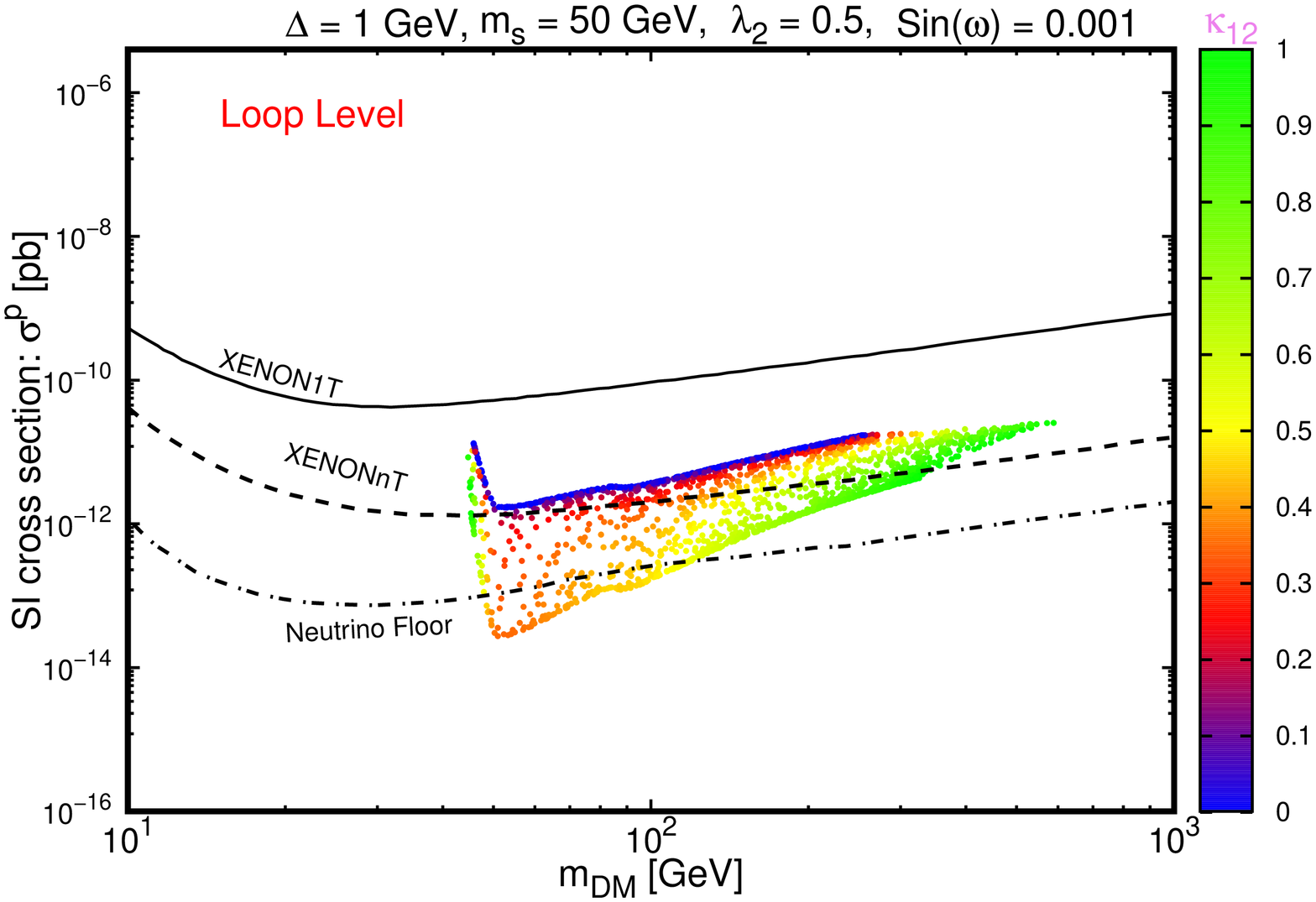}
\end{minipage}
\caption{The same as in Fig.~\ref{CrossLoop-del1-ms50} with the mixing angle as $\sin \omega = 0.001$.} 
\label{CrossLoop-del1-ms50-sin0.001}
\end{figure}

Moreover, we scan regions in the parameter space where the mixing angle is quite small, for instance, $\sin \omega = 0.001$. In this case, we pick out the single scalar mass as $m_s= 50$ GeV, and the mass splitting parameter as $\Delta = 1$ GeV. 
The results presented in Fig.~\ref{CrossLoop-del1-ms50-sin0.001} for this quite smaller 
mixing angle, show that the entire viable parameter space shifts downward 
and resides below the XENON1T bound. 
When $\kappa_{12} \sim 0$, it is evident that the viable points 
lie slightly above XENONnT bounds, while there are regions below the 
neutrino floor at tree level if one chooses non zero value for $\kappa_{12}$.
Again, including the one loop corrections to the DD cross section push a large portion 
of the viable region above the neutrino floor.

\begin{figure}
\begin{minipage}{.55\textwidth}
\includegraphics[width=.65\textwidth,angle =-90]{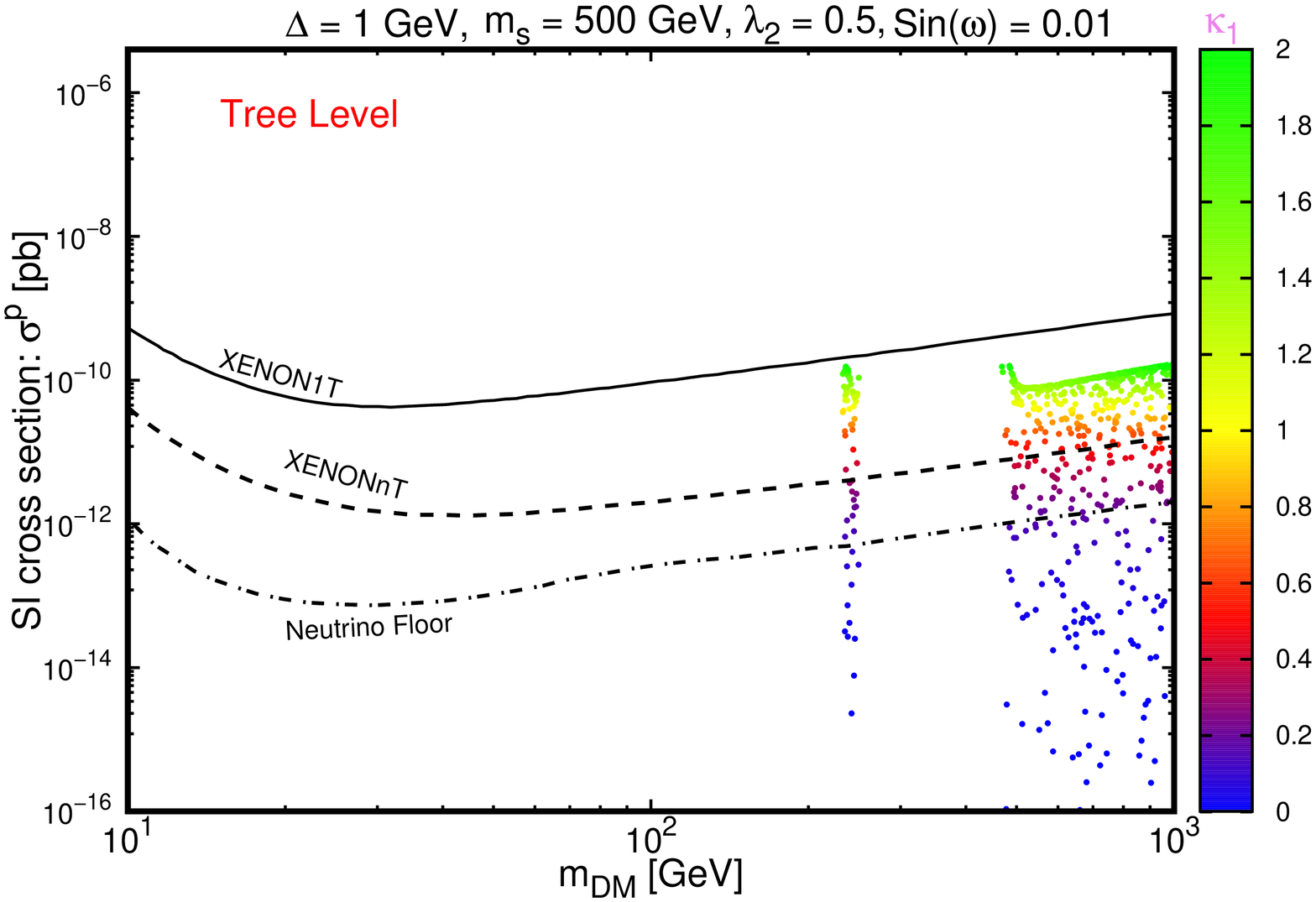}
\end{minipage}
\begin{minipage}{.55\textwidth}
\includegraphics[width=.65\textwidth,angle =-90]{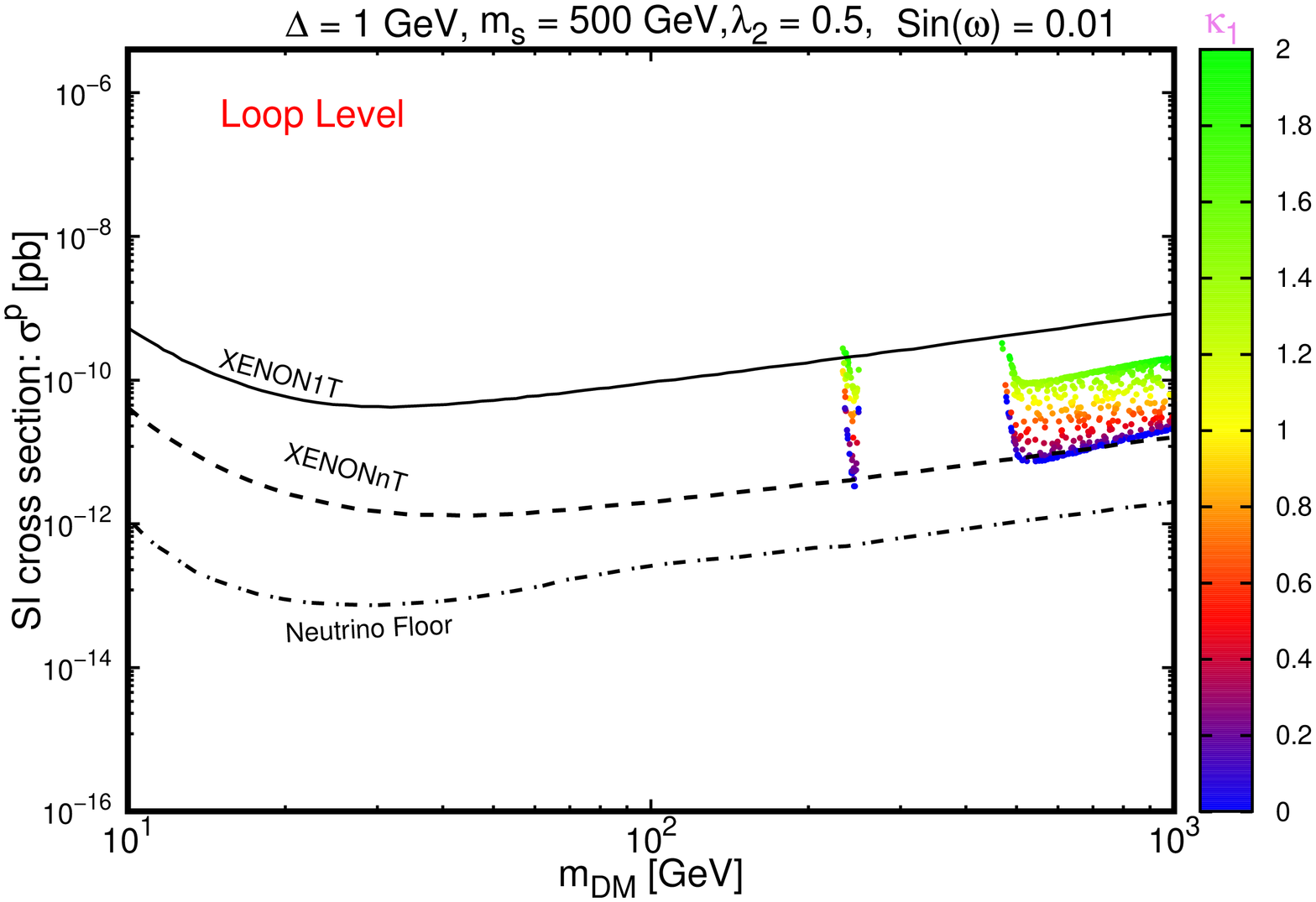}
\end{minipage}
\begin{minipage}{.55\textwidth}
\includegraphics[width=.65\textwidth,angle =-90]{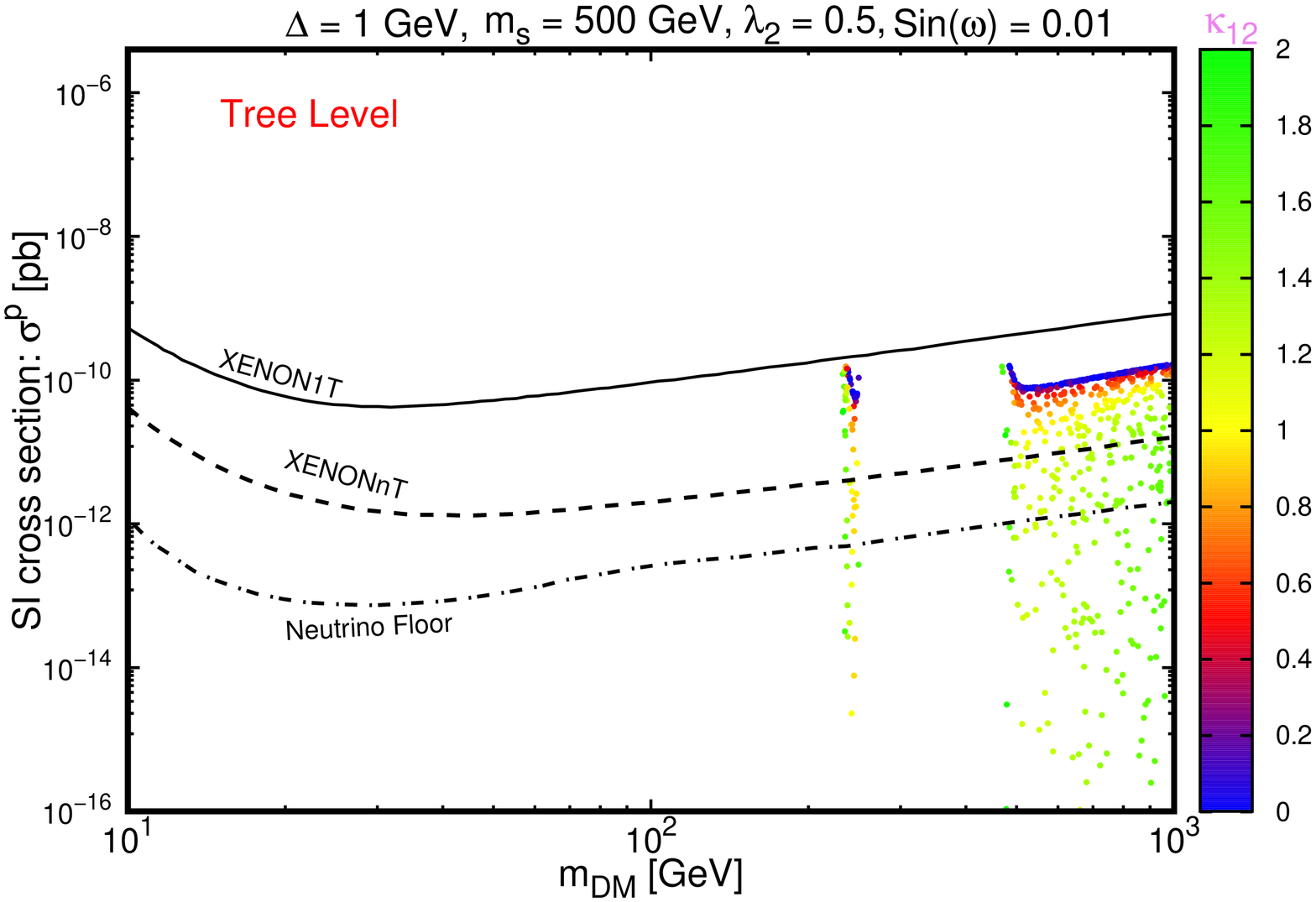}
\end{minipage}
\begin{minipage}{.55\textwidth}
\includegraphics[width=.65\textwidth,angle =-90]{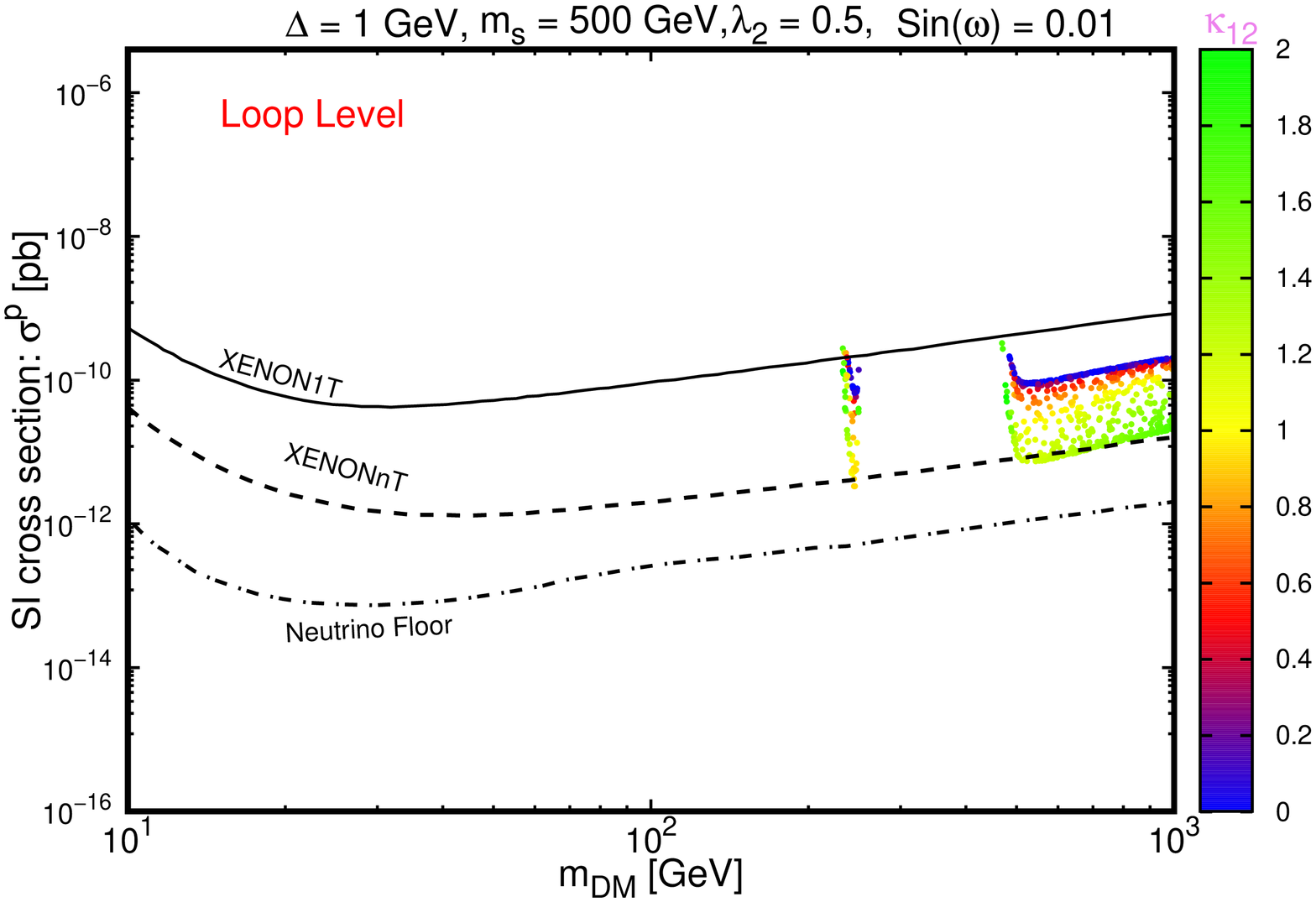}
\end{minipage}
\caption{DD cross section at one loop is shown as a function of DM mass 
for $m_s = 500$ GeV and $\Delta =1 $ GeV. Here the mixing angle is $\sin \omega = 0.01$. 
All the points respect the observed relic density. 
Upper limits from XENON1t and projected XENONnT are placed. 
As such, the neutrino floor is shown.} 
\label{CrossLoop-del1-ms500}
\end{figure}

We also consider a case where the singlet scalar mass is quite large, $m_s = 500$ GeV,
and $\sin \omega = 0.01$. In this case we take a wider range for the couplings as, 
$0< \kappa_1, \kappa_{12} < 2$.
As shown in Fig.~\ref{CrossLoop-del1-ms500}, in this case the viable parameter space is a  
resonance region around $m_\text{DM} \sim 250$ GeV and a region 
with $m_\text{DM} \gtrsim 500$ GeV. 
Now in the plots with the one loop corrections added, we see that 
viable regions below the neutrino floor go up and reside well above the neutrino floor.

\section{Conclusion}
\label{conclusion}
In this work we have considered a DM model with two fermionic WIMPs, where the 
light one is the DM candidate. The interactions between fermions and the SM particles 
are possible through a scalar-Higgs portal. 
This model is motivated given that 
a large portion of the parameter space in the simplest scenario with only 
one fermion is excluded by the current DD experiments. 
Adding the second fermion to the minimal model, opens up 
larger viable parameter space respecting bounds from the observed 
relic density and DD experiments. 

The main goal in this work has been to compute the one loop corrections to the 
DM-nucleon scattering cross section. It is found that by including these 
quantum corrections the allowed parameter space at tree level will change. 
More significantly, it happens for the regions below the limit of 
the neutrino floor on the DD cross section, where the coupling $\kappa_1$ is 
relatively smaller. This feature comes out as result of a characteristic in 
this model we explain here. The only coupling which determines the magnitude of the 
DD cross section at tree level is $\kappa_1$. The other coupling, $\kappa_{12}$, saturates
the annihilation cross section when $\kappa_1$ is much smaller. 
Now, at one loop, the DD cross section
incorporates some extra terms with pure $\kappa_{12}$ coupling. Thus, It becomes feasible to 
get a very large loop corrections in the regions with very small $\kappa_1$ and at the same 
time with large $\kappa_{12}$ .

In conclusion, when studying the viable parameter space, there are situations 
that the regions with quite small direction detection cross section lying below 
the neutrino floor might be sensitive to one loop corrections, making the quantum 
correction indispensable in these cases.

\section{Appendix A}
\label{ApenA} 
In this section we present the dark matter annihilation cross section formulas.
The co-annihilation processes are included in our analysis but their cross sections are not given here. The s-channel annihilation cross section with the SM fermions, Z-boson and W-boson in the final state are
\ba
  \sigma v_{rel} (\bar \chi_1 \chi_1 \to \bar f f) = \frac{N_c m^2_f \kappa^2_1 \sin^2 2\omega}{8\pi s v^2_H}
  \left[ (p_1.p_2)^2-2(p_1.p_2) m^2_f +2m^2_f m^2_1 -m^4_1  \right] \times
  \nonumber\\&&\hspace{-11.1cm}
 \Big|\frac{1}{s-m^2_s+im_s \Gamma_s} -\frac{1}{s-m^2_h+im_h \Gamma_h}\Big|^2 \,,
\ea

\ba
\sigma v_{rel} (\bar \chi_1 \chi_1 \to ZZ) = \frac{\kappa^2_1 \sin^2 2\omega}{16\pi s v^2_H} 
\Big[ (p_1.p_2)^3-2(p_1.p_2)^2 m^2_Z +(p_1.p_2)^2 m^2_1 + 2(p_1.p_2)m^4_Z
\nonumber\\  
  -(p_1.p_2)m^4_1 -3m^4_Z m^2_1 + 2m^2_Z m^4_1 -m^6_1  \Big] \times
\Big|\frac{1}{s-m^2_s+im_s \Gamma_s} -\frac{1}{s-m^2_h+im_h \Gamma_h}\Big|^2  \,,
\ea

\ba
\sigma v_{rel} (\bar \chi_1 \chi_1 \to W^+ W^-) = \frac{\kappa^2_1 \sin^2 2\omega}{8\pi s v^2_H} 
\Big[ (p_1.p_2)^3-2(p_1.p_2)^2 m^2_W +(p_1.p_2)^2 m^2_1 + 2(p_1.p_2)m^4_W
\nonumber\\  
  -(p_1.p_2)m^4_1 -3m^4_W m^2_1 + 2m^2_W m^4_1 -m^6_1  \Big] \times
\Big|\frac{1}{s-m^2_s+im_s \Gamma_s} -\frac{1}{s-m^2_h+im_h \Gamma_h}\Big|^2  \,.
\ea

The annihilation cross section with two singlet scalars in the final state is given by
\ba
\sigma v_{rel} (\bar \chi_1 \chi_1 \to s s) = \frac{\sqrt{1-4m^2_h/s}}{32\pi^2s}
\int d\Omega  \Big[\frac{1}{8} b^2_1 \frac{\kappa^2_1 \sin^2\omega  [p_1.p_2-m^2_1]}{(s-m^2_h)^2}
  + \frac{3}{4} b b_1 \frac{\kappa^2_1 \sin^2 \omega \cos \omega [p_1.p_2-m^2_1]}{(s-m^2_s)(s-m^2_h)}
\nonumber\\ &&\hspace{-17.2cm}
-3 b \frac{\kappa_1 \kappa^2_{12} \sin \omega \cos^3 \omega [2(p_1.p_3)m_1-(p_1.p_2)m_2 -2(p_1.p_2)m_1 +m_2m^2_1]}{(s-m^2_s)(u-m^2_2)} \nonumber\\
+ \frac{9}{8} b^2 \frac{\kappa^2_1 \cos^2\omega \sin^2\omega [p_1.p_2-m^2_1]}{(s-m^2_s)^2} 
+ b_1 \frac{\kappa_1 \kappa^2_{12} \sin \omega \cos^2 \omega [2(p_1.p_3)m_1 +(p_1.p_2)m_2 -m_2 m^2_1 -2 m^3_1]}{(s-m^2_h)(t-m^2_2)} \nonumber\\
+ b_1 \frac{\kappa^3_1 m_1 \sin \omega \cos^2 \omega [2(p_1.p_3) +(p_1.p_2) -3 m^2_1]}{(s-m^2_h)(t-m^2_1)} 
+ 3 b \frac{\kappa^3_1 m_1 \sin \omega \cos^3 \omega [2(p_1.p_3) +(p_1.p_2) -3 m^2_1]}{(s-m^2_s)(t-m^2_1)} \nonumber\\&&\hspace{-17.2cm}
- \kappa^4_1 \cos^4 \omega \frac{[2(p_1.p_3)^2-2(p_1.p_3)(p_1.p_2) -10(p_1.p_3)m^2_1 + (p_1.p_2)m^2_s + m^2_s m^2_1 +8m^4_1]}{(t-m^2_1)^2} \nonumber\\
+ \kappa^4_1 \cos^4 \omega \frac{[2(p_1.p_3)^2-2(p_1.p_3)(p_1.p_2) -2(p_1.p_3)m^2_1 + (p_1.p_2)m^2_s +4(p_1.p_2)m^2_1 + m^2_s m^2_1 -4m^4_1]}
{(t-m^2_1) (u-m^2_1)} \nonumber\\&&\hspace{-17.3cm}
-2 \frac{\kappa^2_1 \kappa^2_{12} \cos^4 \omega}{(t-m^2_1) (t-m^2_2)}  [2(p_1.p_3)^2-2(p_1.p_3)(p_1.p_2) -2(p_1.p_3)m_1 m_2 -8(p_1.p_3)m^2_1 + (p_1.p_2)m^2_s
  \nonumber\\&&\hspace{-13.5cm}
  -(p_1.p_2)m_1 m_2 +(p_1.p_2)m^2_1 + m^2_s m^2_1 +3m_2 m^3_1 + 5 m^4_1] \nonumber\\&&\hspace{-17.2cm}
+2 \frac{\kappa^2_1 \kappa^2_{12} \cos^4 \omega}{(t-m^2_1) (u-m^2_2)}  [2(p_1.p_3)^2-2(p_1.p_3)(p_1.p_2) +2(p_1.p_3)m_1 m_2 -4(p_1.p_3)m^2_1 + (p_1.p_2)m^2_s
  \nonumber\\&&\hspace{-13.5cm}
  +(p_1.p_2)m_1 m_2 +3 (p_1.p_2)m^2_1 + m^2_s m^2_1 -3 m_2 m^3_1 - m^4_1] \nonumber\\&&\hspace{-17.2cm}
- \frac{\kappa^4_{12} \cos^4 \omega}{(t-m^2_2)^2}  [2(p_1.p_3)^2-2(p_1.p_3)(p_1.p_2)-4(p_1.p_3)m_1 m_2 -6(p_1.p_3)m^2_1 + (p_1.p_2)m^2_s \nonumber\\&&\hspace{-15cm}
  -(p_1.p_2)m^2_2 + (p_1.p_2)m^2_1 + m^2_s m^2_1 + m^2_1 m^2_2 +4 m_2 m^3_1 +3 m^4_1] \nonumber\\&&\hspace{-17.2cm}
+ \frac{\kappa^4_{12} \cos^4 \omega}{(t-m^2_2)(u-m^2_2)}  [2(p_1.p_3)^2-2(p_1.p_3)(p_1.p_2)-2(p_1.p_3)m^2_1 + (p_1.p_2) m^2_s \nonumber\\&&\hspace{-15cm}
  +(p_1.p_2)m^2_2 +2 (p_1.p_2)m_1 m_2 +(p_1.p_2)m^2_1 + m^2_s m^2_1 - m^2_1 m^2_2 -2 m_2 m^3_1 - m^4_1] 
\Big]  \,,
\ea

where
\ba
b = \sin \omega \cos \omega \lambda_1 - 2 \cos^2 \omega \lambda_2 v_H - \sin^2 \omega \lambda_H v_H \,, 
\nonumber\\&&\hspace{-8.8cm}
b_1 = 3 \sin^3 \omega \lambda_1 - 2 \sin \omega \lambda_1 -6 \cos \omega \sin^2 \omega \lambda_2 v_H + 2 \cos \omega \lambda_2 v_H
  + 3 \cos \omega \sin^2 \omega \lambda_H v_H \,.
\ea

As such, the annihilation cross section with two Higgs particles in the final state is 
\ba
\sigma v_{rel} (\bar \chi_1 \chi_1 \to h h) = \frac{\sqrt{1-4m^2_h/s}}{32\pi^2s}
\int d\Omega  \Big[\frac{1}{8} b^2_2 \frac{\kappa^2_1 \cos^2\omega  [p_1.p_2-m^2_1]}{(s-m^2_s)^2}
  -\frac{3}{4}b_3 b_2 \frac{\kappa^2_1 \cos^2 \omega \sin \omega [p_1.p_2-m^2_1]}{(s-m^2_s)(s-m^2_h)}
\nonumber\\&&\hspace{-17.2cm}
+ b_2 m_1 \frac{\kappa^3_1 \cos \omega \sin^2 \omega [2p_1.p_3-3p_1.p_2 +m^2_1]}{(s-m^2_s)(u-m^2_1)} \nonumber\\&&\hspace{-17.2cm}
+ b_2 \frac{\kappa_1 \kappa^2_{12} \cos \omega \sin^2 \omega [2(p_1.p_3)m_1-(p_1.p_2)m_2 -2(p_1.p_2)m_1 +m_2m^2_1]}{(s-m^2_s)(u-m^2_2)}
\nonumber \\&&\hspace{-17.2cm}
+ \frac{9}{8} b^2_3 \frac{\kappa^2_1 \cos^2\omega \sin^2\omega [p_1.p_2-m^2_1]}{(s-m^2_h)^2} 
-3 b_3 m_1 \frac{\kappa^3_1 \cos \omega \sin^3 \omega [2p_1.p_3-3p_1.p_2 +m^2_1]}{(s-m^2_h)(u-m^2_1)} \nonumber\\&&\hspace{-17.2cm}
-3 b_3 \frac{\kappa_1 \kappa^2_{12} \cos \omega \sin^3 \omega [2(p_1.p_3)m_1-(p_1.p_2)m_2 -2(p_1.p_2)m_1 +m_2m^2_1]}{(s-m^2_h)(u-m^2_2)} \nonumber\\&&\hspace{-17.2cm}
- \kappa^4_1 \sin^4 \omega \frac{[2(p_1.p_3)^2-2(p_1.p_3)(p_1.p_2) -10(p_1.p_3)m^2_1 + (p_1.p_2)m^2_h + m^2_h m^2_1 +8m^4_1]}{(t-m^2_1)^2} \nonumber\\
+ \kappa^4_1 \sin^4 \omega \frac{[2(p_1.p_3)^2-2(p_1.p_3)(p_1.p_2) -2(p_1.p_3)m^2_1 + (p_1.p_2)m^2_h +4(p_1.p_2)m^2_1 + m^2_h m^2_1 -4m^4_1]}
{(t-m^2_1) (u-m^2_1)} \nonumber\\&&\hspace{-17.2cm}
-2 \frac{\kappa^2_1 \kappa^2_{12} \sin^4 \omega}{(t-m^2_1) (t-m^2_2)}  [2(p_1.p_3)^2-2(p_1.p_3)(p_1.p_2) -2(p_1.p_3)m_1 m_2 -8(p_1.p_3)m^2_1 + (p_1.p_2)m^2_h
  \nonumber\\&&\hspace{-13.6cm}
  -(p_1.p_2)m_1 m_2 +(p_1.p_2)m^2_1 + m^2_h m^2_1 +3m_2 m^3_1 + 5 m^4_1] \nonumber\\&&\hspace{-17.2cm}
+2 \frac{\kappa^2_1 \kappa^2_{12} \sin^4 \omega}{(t-m^2_1) (u-m^2_2)}  [2(p_1.p_3)^2-2(p_1.p_3)(p_1.p_2) +2(p_1.p_3)m_1 m_2 -4(p_1.p_3)m^2_1 + (p_1.p_2)m^2_h
  \nonumber\\&&\hspace{-13.6cm}
  +(p_1.p_2)m_1 m_2 +3 (p_1.p_2)m^2_1 + m^2_h m^2_1 -3 m_2 m^3_1 - m^4_1] \nonumber\\&&\hspace{-17.2cm}
- \frac{\kappa^4_{12} \sin^4 \omega}{(t-m^2_2)^2}  [2(p_1.p_3)^2-2(p_1.p_3)(p_1.p_2)-4(p_1.p_3)m_1 m_2 -6(p_1.p_3)m^2_1 + (p_1.p_2)m^2_h
  \nonumber\\&&\hspace{-13.6cm}
  -(p_1.p_2)m^2_2 + (p_1.p_2)m^2_1 + m^2_h m^2_1 + m^2_1 m^2_2 +4 m_2 m^3_1 +3 m^4_1] \nonumber\\&&\hspace{-17.2cm}
+ \frac{\kappa^4_{12} \sin^4 \omega}{(t-m^2_2)(u-m^2_2)}  [2(p_1.p_3)^2-2(p_1.p_3)(p_1.p_2)-2(p_1.p_3)m^2_1 + (p_1.p_2)m^2_h \nonumber\\
  +(p_1.p_2)m^2_2 +2 (p_1.p_2)m_1 m_2 +(p_1.p_2)m^2_1 + m^2_h m^2_1 - m^2_1 m^2_2 -2 m_2 m^3_1 - m^4_1] 
\Big] \,,
\ea

where
\ba
b_2 = 3  \lambda_1 \cos \omega \sin^2 \omega  - \lambda_1 \cos \omega  + 6 \lambda_2 v_H \sin^3 \omega
      -4 \lambda_2 v_H \sin \omega + 3 \lambda_H v_H \cos^2 \omega \sin \omega  \,, 
\nonumber\\&&\hspace{-14.7cm}
b_3 = \lambda_1 \cos \omega \sin \omega + 2 \lambda_2 v_H \sin^2 \omega + \lambda_H v_H \cos^2 \omega  \,.
\ea

And finally, when a singlet scalar and a Higgs particle are in the final state, the annihilation cross section reads
\ba
\sigma v_{rel} (\bar \chi_1 \chi_1 \to h s) = \frac{\sqrt{1-4m^2_h/s}}{32\pi^2s}
\int d\Omega  \Big[\frac{1}{4} b^2_1 \frac{\kappa^2_1 \cos^2\omega  [p_1.p_2-m^2_1]}{(s-m^2_s)^2}
   + \frac{1}{4} b^2_2 \frac{\kappa^2_1 \sin^2\omega  [p_1.p_2-m^2_1]}{(s-m^2_h)^2}
  \nonumber\\&&\hspace{-16cm}
  -\frac{1}{2}b_1 b_2 \frac{\kappa^2_1 \cos \omega \sin \omega [p_1.p_2-m^2_1]}{(s-m^2_s)(s-m^2_h)}
+ \frac{1}{2} b_1 m_1 \frac{\kappa^3_1 \cos^2 \omega \sin \omega [4 p_1.p_3 + 2 p_1.p_2 + m^2_h -6m^2_1 -m^2_s]}{(s-m^2_s)(t-m^2_1)}
\nonumber\\&&\hspace{-16cm}
- \frac{1}{2} b_2 m_1 \frac{\kappa^3_1 \cos \omega \sin^2 \omega [4 p_1.p_3 + 2 p_1.p_2 + m^2_h -6m^2_1 -m^2_s]}{(s-m^2_h)(t-m^2_1)}
\nonumber\\&&\hspace{-16cm}
- \frac{1}{2} b_1 m_1 \frac{\kappa^3_1 \cos^2 \omega \sin \omega [4 p_1.p_3 - 6 p_1.p_2 + m^2_h +2 m^2_1 - m^2_s]}{(s-m^2_s)(u-m^2_1)}
\nonumber\\&&\hspace{-16cm}
+ \frac{1}{2} b_2 m_1 \frac{\kappa^3_1 \cos \omega \sin^2 \omega [4 p_1.p_3 - 6 p_1.p_2 + m^2_h +2 m^2_1 - m^2_s]}{(s-m^2_h)(u-m^2_1)}
\nonumber\\&&\hspace{-16cm}
+ \frac{1}{2} b_1 \frac{\kappa_1 \kappa^2_{12} \cos^2 \omega \sin \omega [4(p_1.p_3)m_1 +2(p_1.p_2)m_2 +m^2_h m_1 -2m^2_1 m_2 - 4 m^3_1 - m_1 m^2_s]}{(s-m^2_s)(t-m^2_2)}
\nonumber\\&&\hspace{-16cm}
- \frac{1}{2} b_2 \frac{\kappa_1 \kappa^2_{12} \cos \omega \sin^2 \omega [4(p_1.p_3)m_1 +2(p_1.p_2)m_2 +m^2_h m_1 -2m^2_1 m_2 - 4 m^3_1 - m_1 m^2_s]}{(s-m^2_h)(t-m^2_2)}
\nonumber\\&&\hspace{-16cm}
+ \frac{1}{2} b_2 \frac{\kappa_1 \kappa^2_{12} \cos \omega \sin^2 \omega [4(p_1.p_3)m_1 -2(p_1.p_2)m_2  - 4 (p_1.p_2)m_1 + m^2_h m_1 +2m^2_1 m_2 - m_1 m^2_s]}{(s-m^2_h)(u-m^2_2)}
\nonumber\\&&\hspace{-16cm}
- \frac{\kappa^4_{1} \sin^2 \omega \cos^2 \omega}{(t-m^2_1)^2}  [2(p_1.p_3)^2-2(p_1.p_3)(p_1.p_2)-10(p_1.p_3)m^2_1 + (p_1.p_3)m^2_h -(p_1.p_3)m^2_s
\nonumber\\&&\hspace{-13cm}
   +(p_1.p_2)m^2_s -2 m^2_h m^2_1 +8 m^4_1 + 3 m^2_1 m^2_s]
\nonumber\\&&\hspace{-16cm}
- \frac{\kappa^4_{1} \sin^2 \omega \cos^2 \omega}{(u-m^2_1)^2}  [2(p_1.p_3)^2-2(p_1.p_3)(p_1.p_2) +6 (p_1.p_3)m^2_1 + (p_1.p_3)m^2_h  -(p_1.p_3)m^2_s
  \nonumber\\&&\hspace{-13cm}
  +(p_1.p_2)m^2_s  -8 (p_1.p_2)m^2_1 +2 m^2_h m^2_1 - m^2_1 m^2_s]
\nonumber\\&&\hspace{-16cm}
+ 2\frac{ \kappa^4_{1} \sin^2 \omega \cos^2 \omega}{(t-m^2_1)(u-m^2_1)}  [2(p_1.p_3)^2-2(p_1.p_3)(p_1.p_2) -2 (p_1.p_3)m^2_1 + (p_1.p_3)m^2_h
  \nonumber\\&&\hspace{-13cm}
  -(p_1.p_3)m^2_s +(p_1.p_2)m^2_s  +4 (p_1.p_2)m^2_1  - 4 m^4_1 +  m^2_1 m^2_s]
\nonumber\\&&\hspace{-16cm}
+ \frac{ \kappa^2_{1} \kappa^2_{12} \sin^2 \omega \cos^2 \omega}{(t-m^2_2)(u-m^2_1)}  [4(p_1.p_3)^2 -4(p_1.p_3)(p_1.p_2) -4 (p_1.p_3)m_1 m_2 + 2(p_1.p_3)m^2_h 
  \nonumber\\&&\hspace{-13cm}
  -2(p_1.p_3)m^2_s +2(p_1.p_2)m^2_s  +6 (p_1.p_2)m_1 m_2 +2 (p_1.p_2)m^2_1 - m^2_h m_1 m_2
  \nonumber\\&&\hspace{-13cm}
  + m^2_h m^2_1 - 2 m_2 m^3_1 + m_2 m_1 m^2_s - 6m^4_1 + m^2_1 m^2_s]
\nonumber\\&&\hspace{-16cm}
- \frac{ \kappa^2_{1} \kappa^2_{12} \sin^2 \omega \cos^2 \omega}{(t-m^2_1)(t-m^2_2)}  [4(p_1.p_3)^2 -4(p_1.p_3)(p_1.p_2) -4 (p_1.p_3)m_1 m_2 + 2(p_1.p_3)m^2_h 
  -16(p_1.p_3)m^2_1
\nonumber\\&&\hspace{-13cm}
 -2(p_1.p_3)m^2_s  -2 (p_1.p_2) m_1 m_2 +2 (p_1.p_2)m^2_1 +2 (p_1.p_2)m^2_s - m^2_h m_1 m_2
\nonumber\\&&\hspace{-13cm}
  -3 m^2_h m^2_1 +6 m_2 m^3_1 + m_2 m_1 m^2_s +10 m^4_1 +5 m^2_1 m^2_s]
\nonumber\\&&\hspace{-16cm}
+ \frac{ \kappa^2_{1} \kappa^2_{12} \sin^2 \omega \cos^2 \omega}{(t-m^2_1)(u-m^2_2)}  [4(p_1.p_3)^2 -4(p_1.p_3)(p_1.p_2) +4 (p_1.p_3)m_1 m_2 + 2(p_1.p_3)m^2_h 
   \nonumber\\&&\hspace{-13cm}
   -8(p_1.p_3)m^2_1 -2(p_1.p_3)m^2_s  +2 (p_1.p_2) m_1 m_2 +6(p_1.p_2)m^2_1 +2 (p_1.p_2)m^2_s
   \nonumber\\&&\hspace{-13cm}
   + m^2_h m_1 m_2  - m^2_h m^2_1 -6 m_2 m^3_1 - m_2 m_1 m^2_s -2 m^4_1 + 3 m^2_1 m^2_s]
\nonumber\\&&\hspace{-16cm}
- \frac{ \kappa^2_{1} \kappa^2_{12} \sin^2 \omega \cos^2 \omega}{(u-m^2_1)(u-m^2_2)}  [4(p_1.p_3)^2 -4(p_1.p_3)(p_1.p_2) +4 (p_1.p_3)m_1 m_2 + 2(p_1.p_3)m^2_h 
   \nonumber\\&&\hspace{-13cm}
   +8(p_1.p_3)m^2_1 -2(p_1.p_3)m^2_s -6 (p_1.p_2) m_1 m_2 -10(p_1.p_2)m^2_1 +2 (p_1.p_2)m^2_s
   \nonumber\\&&\hspace{-13cm}
   +m^2_h m_1 m_2  +3 m^2_h m^2_1 +2 m_2 m^3_1 - m_2 m_1 m^2_s -2 m^4_1 - m^2_1 m^2_s]
\nonumber\\&&\hspace{-16cm}
- \frac{ \kappa^4_{12} \sin^2 \omega \cos^2 \omega}{(t-m^2_2)^2} [2(p_1.p_3)^2 -2(p_1.p_3)(p_1.p_2) - 4 (p_1.p_3)m_1 m_2 + (p_1.p_3)m^2_h 
  \nonumber\\&&\hspace{-13cm}
  -6 (p_1.p_3) m^2_1 - (p_1.p_3) m^2_s  - (p_1.p_3)m^2_s - (p_1.p_2)m^2_2 + (p_1.p_2)m^2_1
  \nonumber\\&&\hspace{-13cm}
  + (p_1.p_2)m^2_s - m^2_h m_1 m_2  - m^2_h m^2_1 + m^2_2 m^2_1
  +4 m_2 m^3_1 + m_2 m_1 m^2_s
\nonumber\\&&\hspace{-13cm}
  +3 m^4_1 +2 m^2_1 m^2_s]
\nonumber\\&&\hspace{-16cm}
+2 \frac{ \kappa^4_{12} \sin^2 \omega \cos^2 \omega}{(t-m^2_2)(u-m^2_2)} [2(p_1.p_3)^2 -2(p_1.p_3)(p_1.p_2)  + (p_1.p_3)m^2_h 
  -2 (p_1.p_3) m^2_1
\nonumber\\&&\hspace{-13cm}
  - (p_1.p_3) m^2_s  + (p_1.p_2)m^2_s + (p_1.p_2) m^2_2 + (p_1.p_2) m^2_1 + 2(p_1.p_2) m_1 m_2
  \nonumber\\&&\hspace{-13cm}
  - m^2_2 m^2_1 -2 m_2 m^3_1 - m^4_1 + m^2_1 m^2_s]
\nonumber\\&&\hspace{-16cm}
- \frac{ \kappa^4_{12} \sin^2 \omega \cos^2 \omega}{(u-m^2_2)^2} [2(p_1.p_3)^2 -2(p_1.p_3)(p_1.p_2) + 4 (p_1.p_3)m_1 m_2
  - 4 (p_1.p_2)m_1 m_2
\nonumber\\&&\hspace{-13cm}
+ (p_1.p_3)m^2_h +2 (p_1.p_3) m^2_1 - (p_1.p_3) m^2_s + (p_1.p_2)m^2_s - (p_1.p_2)m^2_2
\nonumber\\&&\hspace{-13cm}
-3 (p_1.p_2)m^2_1 +  m^2_h m_1 m_2  + m^2_h m^2_1 + m^2_2 m^2_1
- m_2 m_1 m^2_s - m^4_1 ] \,.
\nonumber\\
\Big]
\ea

\section{Appendix B}
\label{ApenB}

Here, we provide loop functions introduced in the previous sections.
These results are obtained by using {\tt Package-X}.

\subsection{Loop Functions for Triangle Diagrams}
We have two types of triangle Feynman diagrams, each one brings in corresponding loop function. 
For type (a), we have the following functions
\begin{equation}
\begin{split}
F(m_1,m_2,m_{h_i}) &= \frac{\left(m_{h_i}^4-6 m_{h_i}^2 m_1^2 +2 m_1^4+4 m_1^3 m_2 \right) \log \left(\frac{m_{h_i}^2}{m_1^2}\right)}{4 m_1^3 (m_1-m_2)}
\\&
-\frac{\left(m_{h_i}^2-4 m_1^2\right) \sqrt{m_{h_i}^2 \left(m_{h_i}^2-4 m_1^2\right)} \log \left(\frac{\sqrt{m_{h_i}^2 \left(m_{h_i}^2-4 m_1^2\right)}
+m_{h_i}^2}{2 m_{h_i} m_1}\right)}{2 m_1^3 (m_1-m_2)}
\\&
+\frac{\left(-m_{h_i}^4+2 m_{h_i}^2 m_1^2+2 m_{h_i}^2 m_1 m_2+2 m_{h_i}^2 m_2^2- m_1^4-2 m_1^3 m_2-2 m_1 m_2^3-m_2^4\right) 
}{4 m_1^3 (m_1-m_2)} \\&
\times \log \left(\frac{m_{h_i}^2}{m_2^2}\right)
-\frac{\left(-m_{h_i}^2+m_1^2+2 m_1 m_2+ m_2^2 \right) \sqrt{\lambda \left(m_{h_i}^2, m_1^2, m_2^2\right)}}{2 m_1^3 (m_1-m_2)} \times
\\&
\log \left(\frac{\sqrt{\lambda \left(m_{h_i}^2, m_1^2, m_2^2 \right)} 
+ m_{h_i}^2- m_1^2+ m_2^2}{2 m_{h_i} m_2}\right) + \frac{5 m_1+ m_2}{2m_1} \,,
\end{split}
\end{equation}
and in case $m_1 = m_2 = m$, we have
\begin{equation}
\begin{split}
F(m,m_{h_i})=  \frac{3 \sqrt{m_{h_i}^2 \left(m_{h_i}^2-4 m^2\right)} \log \left(\frac{\sqrt{m_{h_i}^2 \left(m_{h_i}^2-4 m^2\right)}+m_{h_i}^2}{2 m m_{h_i}}\right)}{m^2} -\frac{\left(4 m^2-3 m_{h_i}^2\right) \log \left(\frac{m^2}{m_{h_i}^2}\right)}{2 m^2} + 4 \,.
\end{split}
\end{equation}
For the triangle diagrams of type (b) we introduce the loop functions
\begin{equation}
\begin{split}
G(m_{h_i},m_{h_j},m_1) &= \frac{\left(4 m_1^2-m_{h_i}^2\right) \sqrt{m_{h_i}^2 \left(m_{h_i}^2-4 m_1^2\right)} \log \left(\frac{\sqrt{m_{h_i}^2 \left(m_{h_i}^2-4 m_1^2\right)}+m_{h_i}^2}{2 m_1 m_{h_i}}\right)}{2 m_1^4 \left(m_{h_i}^2-m_{h_j}^2\right)} \\&
\frac{\left(6 m_1^2-m_{h_i}^2-m_{h_j}^2\right) \log \left(\frac{m_1^2}{m_{h_j}^2}\right)}{4 m_1^4}
-\frac{m_{h_i}^2 \left(6 m_1^2-m_{h_i}^2\right) \log \left(\frac{m_{h_i}^2}{m_{h_j}^2}\right)}{4 m_1^4 \left(m_{h_i}^2-m_{h_j}^2\right)}
\\&
+\frac{\left(4 m_1^2-m_{h_j}^2\right) \sqrt{m_{h_j}^2 \left(m_{h_j}^2-4 m_1^2\right)} \log \left(\frac{\sqrt{m_{h_j}^2 \left(m_{h_j}^2-4 m_1^2\right)}+m_{h_j}^2}{2 m_1 m_{h_j}}\right)}{2 m_1^4 \left(m_{h_j}^2-m_{h_i}^2\right)}
\\&
-\frac{1}{2 m_1^2} \,,
\end{split}
\end{equation}
and when $m_{h_i} =m_{h_j} = m$, we define $G(m,m,m_1) = G(m,m_1)$ and obtain
\begin{equation}
\begin{split}
G(m,m_1) &= \frac{\left(m_1^2-m^2\right) \sqrt{m^2 \left(m^2-4 m_1^2\right)} \log \left(\frac{\sqrt{m^2 \left(m^2-4 m_1^2\right)}+m^2}{2 m_1 m}\right)}{m_1^4 \left(4 m_1^2-m^2\right)}
\\&
\frac{1}{m_1^2} +\frac{\left(m_1^2+m^2\right) \log \left(\frac{m_1^2}{m^2}\right)}{2 m_1^4}
\end{split}
\end{equation}

\subsection{Loop Functions for Box Diagrams}
When computing the scattering amplitude of box diagrams we encounter two loop functions, $H_1$ and $H_2$. 
Here we provide explicit expressions for these functions. The function $H_1$ is
\begin{equation}
\begin{split}
H_{1}(m_{\chi_1},m_{\chi_i},m_{h_j},m_{h_k}) =& 
\\&
\hspace{-4.5cm}
\frac{\left(m_{\chi_1}^2+4 m_{\chi_1} m_{\chi_i}-5 m_{h_k}^2- m_{\chi_i}^2 \right) \sqrt{\lambda
\left(m_{\chi_1}^2,m_{h_k}^2,m_{\chi_i}^2\right)} \log \left(\frac{\sqrt{\lambda \left(m_{\chi_1}^2,m_{h_k}^2,m_{\chi_i}^2\right)}
-m_{\chi_1}^2+m_{h_k}^2+m_{\chi_i}^2}{2m_{h_k} m_{\chi_i}}\right)}{2 m_{\chi_1}^5 \left(m_{h_j}^2-m_{h_k}^2\right)}
  \\&
 \hspace{-4.5cm} 
  -\frac{ \left(m_{\chi_1}^2+4 m_{\chi_1} m_{\chi_i}-5 m_{h_j}^2 - m_{\chi_i}^2\right) \sqrt{\lambda\left(m_{\chi_1}^2,m_{h_j}^2,m_{\chi_i}^2\right)} \log \left(\frac{\sqrt{\lambda \left(m_{\chi_1}^2,m_{h_j}^2,m_{\chi_i}^2\right)}-m_{\chi_1}^2+m_{h_j}^2+m_{\chi_i}^2}{2 m_{h_j} m_{\chi_i}}\right)}{2 m_{\chi_1}^5 \left(m_{h_j}^2-m_{h_k}^2\right)}
  \\&
  \hspace{-4.5cm}
 -\frac{\left(m_{\chi_1}^4+m_{\chi_1}^3 m_{\chi_i}+m_{\chi_1}^2 m_{\chi_i}^2+m_{\chi_1} m_{h_k}^2 m_{\chi_i}+m_{\chi_1} m_{\chi_i}^3-m_{h_k}^4-2 m_{h_k}^2 m_{\chi_i}^2\right) C_0\left(0,0,m_{\chi_1}^2,m_{\chi_i},0,m_{h_k}\right)}{m^3 \left(m_{h_j}^2-m_{h_k}^2\right)}
 \\&
 \hspace{-4.5cm}
 +\frac{ \left(m_{\chi_1}^4+m_{\chi_1}^3 m_{\chi_i}+m_{\chi_1}^2 m_{\chi_i}^2+m_{\chi_1} m_{h_j}^2 m_{\chi_i}
 +m_{\chi_1} m_{\chi_i}^3-m_{h_j}^4-2 m_{h_j}^2 m_{\chi_i}^2\right) C_0 \left(0,0,m_{\chi_1}^2,m_{\chi_i},0,m_{h_j}\right)}{m_{\chi_1}^3 \left(m_{h_j}^2-m_{h_k}^2\right)}
 \\&
 \hspace{-4.5cm}
 +\frac{ \left(3 m_{\chi_1}^4-2 m_{\chi_1}^2 m_{h_k}^2+2 m_{\chi_1}^2 m_{\chi_i}^2-4 m_{\chi_1} m_{h_k}^2 m_{\chi_i}+4 m_{\chi_1} m_{\chi_i}^3+5 m_{h_k}^4-4 m_{h_k}^2 m_{\chi_i}^2-m_{\chi_i}^4\right) \log \left(\frac{m_{h_k}^2}{m_{\chi_i}^2}\right)}{4 m_{\chi_1}^5 \left(m_{h_j}^2-m_{h_k}^2\right)}
 \\&
 \hspace{-4.5cm}
 -\frac{\left(3 m_{\chi_1}^4-2 m_{\chi_1}^2 m_{h_j}^2+2 m_{\chi_1}^2 m_{\chi_i}^2-4 m_{\chi_1} m_{h_j}^2 m_{\chi_i}+4 m_{\chi_1} m_{\chi_i}^3+5 m_{h_j}^4-4 m_{h_j}^2 m_{\chi_i}^2-m_{\chi_i}^4\right) \log \left(\frac{m_{h_j}^2}{m_{\chi_i}^2}\right)}{4 m^5 \left(m_{h_j}^2-m_{h_k}^2\right)}
 \\&
 \hspace{-4.5cm}
 +\frac{5}{2 m_{\chi_1}^3} \,,
\end{split}
\end{equation}
where the K{\"a}ll\'{e}n function is $\lambda(x,y,z) = x^2+y^2+z^2-2xy-2xz-2yz$. 
The scalar function, $C_0(0,0,x,y,0,z)$ is obtained as 
\begin{equation}
\begin{split}
C_0(0,0,x,y,0,z) &= 
 -\frac{\text{DiLog}\left(\frac{2 \left(x-y^2\right)}{-\sqrt{\lambda \left(x,y^2,z^2\right)}+x-y^2-z^2},x \left(x-y^2\right)\right)}{x}
 \\&
 +\frac{\text{DiLog}\left(-\frac{2 y^2}{-\sqrt{\lambda \left(x,y^2,z^2\right)}+x-y^2-z^2},-x\right)}{x}+\frac{\text{DiLog}\left(-\frac{2 y^2}{\sqrt{\text{Kallen$\lambda $}\left(x,y^2,z^2\right)}+x-y^2-z^2},x\right)}{x}
 \\&
 -\frac{\text{DiLog}\left(\frac{2 \left(x-y^2\right)}{\sqrt{\lambda \left(x,y^2,z^2\right)}+x-y^2-z^2},-x \left(x-y^2\right)\right)}{x}+\frac{\text{Li}_2\left(\frac{y^2-x}{y^2}\right)}{x}-\frac{\pi ^2}{6 x} \,.
\end{split}
\end{equation}
And the function $H_2$ is obtained as
\begin{equation}
\begin{split}
H_{2}(m_{\chi_1},m_{\chi_i},m_{h_j},m_{h_k}) =&
\\&
\hspace{-5cm}
\frac{ \left(9 m_{\chi_1}^2+4 m_{\chi_1} m_{\chi_i}+7 m_{h_k}^2-m_{\chi_i}^2\right) \sqrt{\lambda \left(m_{\chi_1}^2,m_{h_k}^2,m_{\chi_i}^2\right)} \log \left(\frac{\sqrt{\lambda \left(m_{\chi_1}^2,m_{h_k}^2,m_{\chi_i}^2\right)}-m_{\chi_1}^2+m_{h_k}^2+m_{\chi_i}^2}{2 m_{h_k} m_{\chi_i}}\right)}{2 m_{\chi_1}^5 \left(m_{h_j}^2-m_{h_k}^2\right)}
\\&
\hspace{-5cm}
-\frac{ \left(9 m_{\chi_1}^2+4 m_{\chi_1} m_{\chi_i}+7 m_{h_j}^2-m_{\chi_i}^2\right) \sqrt{\lambda \left(m_{\chi_1}^2,m_{h_j}^2,m_{\chi_i}^2\right)} \log \left(\frac{\sqrt{\lambda \left(m_{\chi_1}^2,m_{h_j}^2,m_{\chi_i}^2\right)}-m_{\chi_1}^2+m_{h_j}^2+m_{\chi_i}^2}{2 m_{h_j} m_{\chi_i}}\right)}{2 m_{\chi_1}^5 \left(m_{h_j}^2-m_{h_k}^2\right)}
\\&
\hspace{-5cm}
-\frac{ \left(3 m_{\chi_1}^4+3 m_{\chi_1}^3 m_{\chi_i}+8 m_{\chi_1}^2 m_{h_k}^2-m^2 m_{\chi_i}^2+3 m_{\chi_1} m_{h_k}^2 m_{\chi_i}-m_{\chi_1} m_{\chi_i}^3+5 m_{h_k}^4-2 m_{h_k}^2 m_{\chi_i}^2\right) }{m_{\chi_1}^3 \left(m_{h_j}^2-m_{h_k}^2\right)}
\\&
\times C_0 \left(0,m_{\chi_1}^2,2 m_{\chi_1}^2,0,m_{h_k},m_{\chi_i}\right)
\\&
\hspace{-5cm}
+\frac{ \left(3 m_{\chi_1}^4+3 m_{\chi_1}^3 m_{\chi_i}+8 m_{\chi_1}^2 m_{h_j}^2-m^2 m_{\chi_i}^2+3 m_{\chi_1} m_{h_j}^2 m_{\chi_i}-m_{\chi_1} m_{\chi_i}^3+5 m_{h_j}^4-2 m_{h_j}^2 m_{\chi_i}^2\right) }{m_{\chi_1}^3 \left(m_{h_j}^2-m_{h_k}^2\right)}
\\&
\times C_0  \left(0,m_{\chi_1}^2,2 m^2,0,m_{h_j},m_{\chi_i}\right)
\\&
\hspace{-5cm}
-\frac{ \left(13 m_{\chi_1}^4+8 m_{\chi_1}^3 m_{\chi_i}+22 m_{\chi_1}^2 m_{h_k}^2-10 m_{\chi_1}^2 m_{\chi_i}^2+4 m_{\chi_1} m_{h_k}^2 m_{\chi_i}-4 m_{\chi_1} m_{\chi_i}^3+7 m_{h_k}^4-8 m_{h_k}^2 m_{\chi_i}^2+m_{\chi_i}^4\right)}{4 m_{\chi_1}^5 \left(m_{h_j}^2-m_{h_k}^2\right)}
\\&
\times \log \left(\frac{m_{h_k}^2}{m_{\chi_i}^2}\right)
\\&
\hspace{-5cm}
+\frac{ \left(13 m_{\chi_1}^4+8 m_{\chi_1}^3 m_{\chi_i}+22 m^2 m_{h_j}^2-10 m^2 m_{\chi_i}^2+4 m_{\chi_1} m_{h_j}^2 m_{\chi_i}-4 m_{\chi_1} m_{\chi_i}^3+7 m_{h_j}^4-8 m_{h_j}^2 m_{\chi_i}^2+m_{\chi_i}^4\right)}{4 m_{\chi_1}^5 \left(m_{h_j}^2-m_{h_k}^2\right)}
\\&
\times \log \left(\frac{m_{h_j}^2}{m_{\chi_i}^2}\right)
\\&
\hspace{-5cm}
+\frac{3}{2 m^3}+\frac{5 \left(2 m^2-m_{\chi_i}^2\right) \log \left(\frac{m_{\chi_i}^2}{m_{\chi_i}^2-2 m^2}\right)}{2 m^5}
\end{split}
\end{equation}

The scalar function, $C_0(0,x,2x,0,y,z)$ is obtained as
\begin{equation}
\begin{split}
C_0(0,x,2x,0,y,z) &=
-\frac{\text{DiLog}\left(-\frac{2 x \left(x+2 y^2-z^2\right)}{x \sqrt{\lambda \left(x,y^2,z^2\right)}-x \left(x+3 y^2-z^2\right)},x \left(x+2 y^2-z^2\right)\right)}{x}
\\&
+\frac{\text{DiLog}\left(-\frac{2 x \left(x+2 y^2-z^2\right)}{-x \sqrt{\lambda \left(x,y^2,z^2\right)}-x \left(x+3 y^2-z^2\right)},-x \left(x+2 y^2-z^2\right)\right)}{x}
\\&
-\frac{\text{DiLog}\left(-\frac{2 x \left(2 y^2-z^2\right)}{x \sqrt{\lambda \left(x,y^2,z^2\right)}-x \left(x+3 y^2-z^2\right)},x \left(2 y^2-z^2\right)\right)}{x}
\\&
-\frac{\text{DiLog}\left(-\frac{2 x \left(2 y^2-z^2\right)}{-x \sqrt{\lambda \left(x,y^2,z^2\right)}-x \left(x+3 y^2-z^2\right)},x \left(z^2-2 y^2\right)\right)}{x}
\\&
+\frac{\text{Li}_2\left(\frac{2 y^2-z^2}{2 y^2}\right)}{x}+\frac{\text{Li}_2\left(\frac{2 y^2-z^2+2 x}{2 x-z^2}\right)}{x}+\frac{\text{Li}_2\left(\frac{2 y^2-z^2}{2 y^2-z^2+2 x}\right)}{x}
\\&
-\frac{\text{Li}_2\left(\frac{2 y^2-z^2+x}{2 y^2-z^2+2 x}\right)}{x}+\frac{\log ^2\left(-\frac{2 y^2}{2 x-z^2}\right)}{2 x} \,.
\end{split}
\end{equation}
\subsection{Scalar Couplings}
The triple scalar couplings defined in section \ref{DDCS} are 
\begin{equation}
\begin{split}
c_{hhh} &=  -3\cos(\omega) \Big[ \cos(\omega) \sin(\omega) \lambda_1 + 
  2 v_H \sin^2(\omega) \lambda_2 + 2 v_H \cos^2(\omega) \lambda_H   \Big]
  \\
c_{hhs} &= 3 \cos(\omega) \sin^2(\omega) \lambda_1 
          -\cos(\omega) \lambda_1
         +6 v_H \sin^3(\omega) \lambda_2 
         -4 v_H \sin(\omega) \lambda_2 
\\&
         +6 v_H \cos^2(\omega) \sin(\omega) \lambda_H 
\\
c_{ssh} &= -3 \sin^3(\omega) \lambda_1 
          -2 \lambda_1 \sin(\omega) 
          -6 v_H \lambda_2 \cos(\omega) \sin^2(\omega)
          +2 v_H  \cos(\omega) \lambda_2 
\\&
          +6 v_H \cos(\omega) \sin^2(\omega) \lambda_H
\\
c_{sss} &= -3\sin(\omega) \Big[ \cos(\omega) \sin(\omega) \lambda_1 - 2 v_H \cos^2(\omega)
          -2 v_H \sin^2(omega) \lambda_H \Big] 
\end{split}
\end{equation}

\bibliography{ref}

\providecommand{\href}[2]{#2}\begingroup\raggedright\begin{thebibliography}{10}

\bibitem{Lee:1977ua}
B.~W. Lee and S.~Weinberg, ``{Cosmological Lower Bound on Heavy Neutrino
  Masses},'' \href{http://dx.doi.org/10.1103/PhysRevLett.39.165}{{\em Phys.
  Rev. Lett.} {\bfseries 39} (1977) 165--168}.

\bibitem{Steigman:1984ac}
G.~Steigman and M.~S. Turner, ``{Cosmological Constraints on the Properties of
  Weakly Interacting Massive Particles},''
  \href{http://dx.doi.org/10.1016/0550-3213(85)90537-1}{{\em Nucl. Phys. B}
  {\bfseries 253} (1985) 375--386}.

\bibitem{Arcadi:2017kky}
G.~Arcadi, M.~Dutra, P.~Ghosh, M.~Lindner, Y.~Mambrini, M.~Pierre, S.~Profumo,
  and F.~S. Queiroz, ``{The waning of the WIMP? A review of models, searches,
  and constraints},''
  \href{http://dx.doi.org/10.1140/epjc/s10052-018-5662-y}{{\em Eur. Phys. J. C}
  {\bfseries 78} no.~3, (2018) 203},
  \href{http://arxiv.org/abs/1703.07364}{{\ttfamily arXiv:1703.07364
  [hep-ph]}}.

\bibitem{Bergstrom:2000pn}
L.~Bergstr\"om, ``{Nonbaryonic dark matter: Observational evidence and
  detection methods},''
  \href{http://dx.doi.org/10.1088/0034-4885/63/5/2r3}{{\em Rept. Prog. Phys.}
  {\bfseries 63} (2000) 793},
  \href{http://arxiv.org/abs/hep-ph/0002126}{{\ttfamily arXiv:hep-ph/0002126}}.

\bibitem{Steigman:2012nb}
G.~Steigman, B.~Dasgupta, and J.~F. Beacom, ``{Precise Relic WIMP Abundance and
  its Impact on Searches for Dark Matter Annihilation},''
  \href{http://dx.doi.org/10.1103/PhysRevD.86.023506}{{\em Phys. Rev. D}
  {\bfseries 86} (2012) 023506},
  \href{http://arxiv.org/abs/1204.3622}{{\ttfamily arXiv:1204.3622 [hep-ph]}}.

\bibitem{Leane:2018kjk}
R.~K. Leane, T.~R. Slatyer, J.~F. Beacom, and K.~C.~Y. Ng, ``{GeV-scale thermal
  WIMPs: Not even slightly ruled out},''
  \href{http://dx.doi.org/10.1103/PhysRevD.98.023016}{{\em Phys. Rev. D}
  {\bfseries 98} no.~2, (2018) 023016},
  \href{http://arxiv.org/abs/1805.10305}{{\ttfamily arXiv:1805.10305
  [hep-ph]}}.

\bibitem{Barger:2010yn}
V.~Barger, M.~McCaskey, and G.~Shaughnessy, ``{Complex Scalar Dark Matter
  vis-\textbackslash{}`{a}-vis CoGeNT, DAMA/LIBRA and XENON100},''
  \href{http://dx.doi.org/10.1103/PhysRevD.82.035019}{{\em Phys. Rev. D}
  {\bfseries 82} (2010) 035019},
  \href{http://arxiv.org/abs/1005.3328}{{\ttfamily arXiv:1005.3328 [hep-ph]}}.

\bibitem{Gonderinger:2012rd}
M.~Gonderinger, H.~Lim, and M.~J. Ramsey-Musolf, ``{Complex Scalar Singlet Dark
  Matter: Vacuum Stability and Phenomenology},''
  \href{http://dx.doi.org/10.1103/PhysRevD.86.043511}{{\em Phys. Rev. D}
  {\bfseries 86} (2012) 043511},
  \href{http://arxiv.org/abs/1202.1316}{{\ttfamily arXiv:1202.1316 [hep-ph]}}.

\bibitem{Gross:2017dan}
C.~Gross, O.~Lebedev, and T.~Toma, ``{Cancellation Mechanism for
  Dark-Matter\textendash{}Nucleon Interaction},''
  \href{http://dx.doi.org/10.1103/PhysRevLett.119.191801}{{\em Phys. Rev.
  Lett.} {\bfseries 119} no.~19, (2017) 191801},
  \href{http://arxiv.org/abs/1708.02253}{{\ttfamily arXiv:1708.02253
  [hep-ph]}}.

\bibitem{Ghorbani:2022muk}
P.~Ghorbani, ``{Dark matter and muon g \ensuremath{-} 2 anomaly via scale
  symmetry breaking},'' \href{http://dx.doi.org/10.1007/JHEP04(2022)170}{{\em
  JHEP} {\bfseries 04} (2022) 170},
  \href{http://arxiv.org/abs/2203.03964}{{\ttfamily arXiv:2203.03964
  [hep-ph]}}.

\bibitem{Ghorbani:2014qpa}
K.~Ghorbani, ``{Fermionic dark matter with pseudo-scalar Yukawa interaction},''
  \href{http://dx.doi.org/10.1088/1475-7516/2015/01/015}{{\em JCAP} {\bfseries
  01} (2015) 015}, \href{http://arxiv.org/abs/1408.4929}{{\ttfamily
  arXiv:1408.4929 [hep-ph]}}.

\bibitem{Berlin:2015wwa}
A.~Berlin, S.~Gori, T.~Lin, and L.-T. Wang, ``{Pseudoscalar Portal Dark
  Matter},'' \href{http://dx.doi.org/10.1103/PhysRevD.92.015005}{{\em Phys.
  Rev. D} {\bfseries 92} (2015) 015005},
  \href{http://arxiv.org/abs/1502.06000}{{\ttfamily arXiv:1502.06000
  [hep-ph]}}.

\bibitem{Jia:2015uea}
L.-B. Jia, ``{Search for pseudoscalar-mediated WIMPs in t\textrightarrow{}c
  transitions with missing energy},''
  \href{http://dx.doi.org/10.1103/PhysRevD.92.074006}{{\em Phys. Rev. D}
  {\bfseries 92} no.~7, (2015) 074006},
  \href{http://arxiv.org/abs/1506.05293}{{\ttfamily arXiv:1506.05293
  [hep-ph]}}.

\bibitem{Fan:2015sza}
J.~Fan, S.~M. Koushiappas, and G.~Landsberg, ``{Pseudoscalar Portal Dark Matter
  and New Signatures of Vector-like Fermions},''
  \href{http://dx.doi.org/10.1007/JHEP01(2016)111}{{\em JHEP} {\bfseries 01}
  (2016) 111}, \href{http://arxiv.org/abs/1507.06993}{{\ttfamily
  arXiv:1507.06993 [hep-ph]}}.

\bibitem{Yang:2016wrl}
K.-C. Yang, ``{Fermionic Dark Matter through a Light Pseudoscalar Portal: Hints
  from the DAMA Results},''
  \href{http://dx.doi.org/10.1103/PhysRevD.94.035028}{{\em Phys. Rev. D}
  {\bfseries 94} no.~3, (2016) 035028},
  \href{http://arxiv.org/abs/1604.04979}{{\ttfamily arXiv:1604.04979
  [hep-ph]}}.

\bibitem{DuttaBanik:2016jzv}
A.~Dutta~Banik, M.~Pandey, D.~Majumdar, and A.~Biswas, ``{Two component
  WIMP\textendash{}FImP dark matter model with singlet fermion, scalar and
  pseudo scalar},''
  \href{http://dx.doi.org/10.1140/epjc/s10052-017-5221-y}{{\em Eur. Phys. J. C}
  {\bfseries 77} no.~10, (2017) 657},
  \href{http://arxiv.org/abs/1612.08621}{{\ttfamily arXiv:1612.08621
  [hep-ph]}}.

\bibitem{Baek:2017vzd}
S.~Baek, P.~Ko, and J.~Li, ``{Minimal renormalizable simplified dark matter
  model with a pseudoscalar mediator},''
  \href{http://dx.doi.org/10.1103/PhysRevD.95.075011}{{\em Phys. Rev. D}
  {\bfseries 95} no.~7, (2017) 075011},
  \href{http://arxiv.org/abs/1701.04131}{{\ttfamily arXiv:1701.04131
  [hep-ph]}}.

\bibitem{Ghorbani:2017qwf}
K.~Ghorbani, ``{Renormalization group equation analysis of a pseudoscalar
  portal dark matter model},''
  \href{http://dx.doi.org/10.1088/1361-6471/aa888f}{{\em J. Phys. G} {\bfseries
  44} no.~10, (2017) 105006}, \href{http://arxiv.org/abs/1702.08711}{{\ttfamily
  arXiv:1702.08711 [hep-ph]}}.

\bibitem{Ghorbani:2017jls}
P.~H. Ghorbani, ``{Electroweak Baryogenesis and Dark Matter via a Pseudoscalar
  vs. Scalar},'' \href{http://dx.doi.org/10.1007/JHEP08(2017)058}{{\em JHEP}
  {\bfseries 08} (2017) 058}, \href{http://arxiv.org/abs/1703.06506}{{\ttfamily
  arXiv:1703.06506 [hep-ph]}}.

\bibitem{YaserAyazi:2018pea}
S.~Yaser~Ayazi, A.~Mohamadnejad, and S.~P. Zakeri, ``{Search for vector-like
  quarks in a fermionic dark matter model with pseudoscalar: A resonance
  case},'' \href{http://dx.doi.org/10.1142/S0217732318501596}{{\em Mod. Phys.
  Lett. A} {\bfseries 33} no.~27, (2018) 1850159},
  \href{http://arxiv.org/abs/1804.02876}{{\ttfamily arXiv:1804.02876
  [hep-ph]}}.

\bibitem{Abe:2019wjw}
T.~Abe, M.~Fujiwara, J.~Hisano, and Y.~Shoji, ``{Maximum value of the
  spin-independent cross section in the 2HDM+a},''
  \href{http://dx.doi.org/10.1007/JHEP01(2020)114}{{\em JHEP} {\bfseries 01}
  (2020) 114}, \href{http://arxiv.org/abs/1910.09771}{{\ttfamily
  arXiv:1910.09771 [hep-ph]}}.

\bibitem{Ghorbani:2016edw}
K.~Ghorbani and L.~Khalkhali, ``{Mono-Higgs signature in a fermionic dark
  matter model},'' \href{http://dx.doi.org/10.1088/1361-6471/aa823a}{{\em J.
  Phys. G} {\bfseries 44} no.~10, (2017) 105004},
  \href{http://arxiv.org/abs/1608.04559}{{\ttfamily arXiv:1608.04559
  [hep-ph]}}.

\bibitem{DiazSaez:2021pmg}
B.~D\'\i{}az~S\'aez, P.~Escalona, S.~Norero, and A.~R. Zerwekh, ``{Fermion
  singlet dark matter in a pseudoscalar dark matter portal},''
  \href{http://dx.doi.org/10.1007/JHEP10(2021)233}{{\em JHEP} {\bfseries 10}
  (2021) 233}, \href{http://arxiv.org/abs/2105.04255}{{\ttfamily
  arXiv:2105.04255 [hep-ph]}}.

\bibitem{Abe:2019wku}
T.~Abe and R.~Sato, ``{Current status and future prospects of the
  singlet-doublet dark matter model with CP-violation},''
  \href{http://dx.doi.org/10.1103/PhysRevD.99.035012}{{\em Phys. Rev. D}
  {\bfseries 99} no.~3, (2019) 035012},
  \href{http://arxiv.org/abs/1901.02278}{{\ttfamily arXiv:1901.02278
  [hep-ph]}}.

\bibitem{Kozaczuk:2015bea}
J.~Kozaczuk and T.~A.~W. Martin, ``{Extending LHC Coverage to Light
  Pseudoscalar Mediators and Coy Dark Sectors},''
  \href{http://dx.doi.org/10.1007/JHEP04(2015)046}{{\em JHEP} {\bfseries 04}
  (2015) 046}, \href{http://arxiv.org/abs/1501.07275}{{\ttfamily
  arXiv:1501.07275 [hep-ph]}}.

\bibitem{Abe:2020obo}
T.~Abe, ``{Effect of the early kinetic decoupling in a fermionic dark matter
  model},'' \href{http://dx.doi.org/10.1103/PhysRevD.102.035018}{{\em Phys.
  Rev. D} {\bfseries 102} no.~3, (2020) 035018},
  \href{http://arxiv.org/abs/2004.10041}{{\ttfamily arXiv:2004.10041
  [hep-ph]}}.

\bibitem{Matsumoto:2018acr}
S.~Matsumoto, Y.-L.~S. Tsai, and P.-Y. Tseng, ``{Light Fermionic WIMP Dark
  Matter with Light Scalar Mediator},''
  \href{http://dx.doi.org/10.1007/JHEP07(2019)050}{{\em JHEP} {\bfseries 07}
  (2019) 050}, \href{http://arxiv.org/abs/1811.03292}{{\ttfamily
  arXiv:1811.03292 [hep-ph]}}.

\bibitem{Li:2018qip}
T.~Li, ``{Revisiting the direct detection of dark matter in simplified
  models},'' \href{http://dx.doi.org/10.1016/j.physletb.2018.05.073}{{\em Phys.
  Lett. B} {\bfseries 782} (2018) 497--502},
  \href{http://arxiv.org/abs/1804.02120}{{\ttfamily arXiv:1804.02120
  [hep-ph]}}.

\bibitem{Herrero-Garcia:2018koq}
J.~Herrero-Garcia, E.~Molinaro, and M.~A. Schmidt, ``{Dark matter direct
  detection of a fermionic singlet at one loop},''
  \href{http://dx.doi.org/10.1140/epjc/s10052-018-5935-5}{{\em Eur. Phys. J. C}
  {\bfseries 78} no.~6, (2018) 471},
  \href{http://arxiv.org/abs/1803.05660}{{\ttfamily arXiv:1803.05660
  [hep-ph]}}. [Erratum: None 82, 53 (2022)].

\bibitem{Hisano:2018bpz}
J.~Hisano, R.~Nagai, and N.~Nagata, ``{Singlet Dirac Fermion Dark Matter with
  Mediators at Loop},'' \href{http://dx.doi.org/10.1007/JHEP12(2018)059}{{\em
  JHEP} {\bfseries 12} (2018) 059},
  \href{http://arxiv.org/abs/1808.06301}{{\ttfamily arXiv:1808.06301
  [hep-ph]}}.

\bibitem{Han:2018gej}
T.~Han, H.~Liu, S.~Mukhopadhyay, and X.~Wang, ``{Dark Matter Blind Spots at
  One-Loop},'' \href{http://dx.doi.org/10.1007/JHEP03(2019)080}{{\em JHEP}
  {\bfseries 03} (2019) 080}, \href{http://arxiv.org/abs/1810.04679}{{\ttfamily
  arXiv:1810.04679 [hep-ph]}}.

\bibitem{Azevedo:2018exj}
D.~Azevedo, M.~Duch, B.~Grzadkowski, D.~Huang, M.~Iglicki, and R.~Santos,
  ``{One-loop contribution to dark-matter-nucleon scattering in the
  pseudo-scalar dark matter model},''
  \href{http://dx.doi.org/10.1007/JHEP01(2019)138}{{\em JHEP} {\bfseries 01}
  (2019) 138}, \href{http://arxiv.org/abs/1810.06105}{{\ttfamily
  arXiv:1810.06105 [hep-ph]}}.

\bibitem{Ishiwata:2018sdi}
K.~Ishiwata and T.~Toma, ``{Probing pseudo Nambu-Goldstone boson dark matter at
  loop level},'' \href{http://dx.doi.org/10.1007/JHEP12(2018)089}{{\em JHEP}
  {\bfseries 12} (2018) 089}, \href{http://arxiv.org/abs/1810.08139}{{\ttfamily
  arXiv:1810.08139 [hep-ph]}}.

\bibitem{Ghorbani:2018pjh}
K.~Ghorbani and P.~H. Ghorbani, ``{Leading Loop Effects in Pseudoscalar-Higgs
  Portal Dark Matter},'' \href{http://dx.doi.org/10.1007/JHEP05(2019)096}{{\em
  JHEP} {\bfseries 05} (2019) 096},
  \href{http://arxiv.org/abs/1812.04092}{{\ttfamily arXiv:1812.04092
  [hep-ph]}}.

\bibitem{Ertas:2019dew}
F.~Ertas and F.~Kahlhoefer, ``{Loop-induced direct detection signatures from
  CP-violating scalar mediators},''
  \href{http://dx.doi.org/10.1007/JHEP06(2019)052}{{\em JHEP} {\bfseries 06}
  (2019) 052}, \href{http://arxiv.org/abs/1902.11070}{{\ttfamily
  arXiv:1902.11070 [hep-ph]}}.

\bibitem{Li:2019fnn}
T.~Li and P.~Wu, ``{Simplified dark matter models with loop effects in direct
  detection and the constraints from indirect detection and collider search},''
  \href{http://dx.doi.org/10.1088/1674-1137/43/11/113102}{{\em Chin. Phys. C}
  {\bfseries 43} no.~11, (2019) 113102},
  \href{http://arxiv.org/abs/1904.03407}{{\ttfamily arXiv:1904.03407
  [hep-ph]}}.

\bibitem{Chao:2019lhb}
W.~Chao, ``{Direct detections of Majorana dark matter in vector portal},''
  \href{http://dx.doi.org/10.1007/JHEP11(2019)013}{{\em JHEP} {\bfseries 11}
  (2019) 013}, \href{http://arxiv.org/abs/1904.09785}{{\ttfamily
  arXiv:1904.09785 [hep-ph]}}.

\bibitem{Glaus:2019itb}
S.~Glaus, M.~M\"uhlleitner, J.~M\"uller, S.~Patel, and R.~Santos,
  ``{Electroweak Corrections to Dark Matter Direct Detection in a Vector Dark
  Matter Model},'' \href{http://dx.doi.org/10.1007/JHEP10(2019)152}{{\em JHEP}
  {\bfseries 10} (2019) 152}, \href{http://arxiv.org/abs/1908.09249}{{\ttfamily
  arXiv:1908.09249 [hep-ph]}}.

\bibitem{Borschensky:2020olr}
C.~Borschensky, G.~Coniglio, B.~J\"ager, J.~Jochum, and V.~Schipperges,
  ``{Direct detection of dark matter: Precision predictions in a simplified
  model framework},''
  \href{http://dx.doi.org/10.1140/epjc/s10052-020-08795-x}{{\em Eur. Phys. J.
  C} {\bfseries 81} no.~1, (2021) 44},
  \href{http://arxiv.org/abs/2008.04253}{{\ttfamily arXiv:2008.04253
  [hep-ph]}}.

\bibitem{Chao:2020fme}
W.~Chao, J.-g. Jiang, and M.~Su, ``{Probing Bottom-flavored Scalar Dark Matters
  at Loop Level},'' \href{http://arxiv.org/abs/2011.13619}{{\ttfamily
  arXiv:2011.13619 [hep-ph]}}.

\bibitem{Bell:2018zra}
N.~F. Bell, G.~Busoni, and I.~W. Sanderson, ``{Loop Effects in Direct
  Detection},'' \href{http://dx.doi.org/10.1088/1475-7516/2018/08/017}{{\em
  JCAP} {\bfseries 08} (2018) 017},
  \href{http://arxiv.org/abs/1803.01574}{{\ttfamily arXiv:1803.01574
  [hep-ph]}}. [Erratum: JCAP 01, E01 (2019)].

\bibitem{Abe:2018emu}
T.~Abe, M.~Fujiwara, and J.~Hisano, ``{Loop corrections to dark matter direct
  detection in a pseudoscalar mediator dark matter model},''
  \href{http://dx.doi.org/10.1007/JHEP02(2019)028}{{\em JHEP} {\bfseries 02}
  (2019) 028}, \href{http://arxiv.org/abs/1810.01039}{{\ttfamily
  arXiv:1810.01039 [hep-ph]}}.

\bibitem{Ghorbani:2018hjs}
K.~Ghorbani, ``{Split fermionic WIMPs evade direct detection},''
  \href{http://dx.doi.org/10.1007/JHEP11(2018)086}{{\em JHEP} {\bfseries 11}
  (2018) 086}, \href{http://arxiv.org/abs/1805.02098}{{\ttfamily
  arXiv:1805.02098 [hep-ph]}}.

\bibitem{CMS:2018yfx}
{\bfseries CMS} Collaboration, A.~M. Sirunyan {\em et~al.}, ``{Search for
  invisible decays of a Higgs boson produced through vector boson fusion in
  proton-proton collisions at $\sqrt{s} =$ 13 TeV},''
  \href{http://dx.doi.org/10.1016/j.physletb.2019.04.025}{{\em Phys. Lett. B}
  {\bfseries 793} (2019) 520--551},
  \href{http://arxiv.org/abs/1809.05937}{{\ttfamily arXiv:1809.05937
  [hep-ex]}}.

\bibitem{ParticleDataGroup:2020ssz}
{\bfseries Particle Data Group} Collaboration, P.~A. Zyla {\em et~al.},
  ``{Review of Particle Physics},''
  \href{http://dx.doi.org/10.1093/ptep/ptaa104}{{\em PTEP} {\bfseries 2020}
  no.~8, (2020) 083C01}.

\bibitem{XENON:2018voc}
{\bfseries XENON} Collaboration, E.~Aprile {\em et~al.}, ``{Dark Matter Search
  Results from a One Ton-Year Exposure of XENON1T},''
  \href{http://dx.doi.org/10.1103/PhysRevLett.121.111302}{{\em Phys. Rev.
  Lett.} {\bfseries 121} no.~11, (2018) 111302},
  \href{http://arxiv.org/abs/1805.12562}{{\ttfamily arXiv:1805.12562
  [astro-ph.CO]}}.

\bibitem{XENON:2020kmp}
{\bfseries XENON} Collaboration, E.~Aprile {\em et~al.}, ``{Projected WIMP
  sensitivity of the XENONnT dark matter experiment},''
  \href{http://dx.doi.org/10.1088/1475-7516/2020/11/031}{{\em JCAP} {\bfseries
  11} (2020) 031}, \href{http://arxiv.org/abs/2007.08796}{{\ttfamily
  arXiv:2007.08796 [physics.ins-det]}}.

\bibitem{Billard:2021uyg}
J.~Billard {\em et~al.}, ``{Direct Detection of Dark Matter -- APPEC Committee
  Report},'' \href{http://arxiv.org/abs/2104.07634}{{\ttfamily arXiv:2104.07634
  [hep-ex]}}.

\bibitem{Planck:2018vyg}
{\bfseries Planck} Collaboration, N.~Aghanim {\em et~al.}, ``{Planck 2018
  results. VI. Cosmological parameters},''
  \href{http://dx.doi.org/10.1051/0004-6361/201833910}{{\em Astron. Astrophys.}
  {\bfseries 641} (2020) A6}, \href{http://arxiv.org/abs/1807.06209}{{\ttfamily
  arXiv:1807.06209 [astro-ph.CO]}}. [Erratum: Astron.Astrophys. 652, C4
  (2021)].

\bibitem{Kanemura:2016lkz}
S.~Kanemura, M.~Kikuchi, and K.~Yagyu, ``{One-loop corrections to the Higgs
  self-couplings in the singlet extension},''
  \href{http://dx.doi.org/10.1016/j.nuclphysb.2017.02.004}{{\em Nucl. Phys. B}
  {\bfseries 917} (2017) 154--177},
  \href{http://arxiv.org/abs/1608.01582}{{\ttfamily arXiv:1608.01582
  [hep-ph]}}.

\bibitem{Bojarski:2015kra}
F.~Bojarski, G.~Chalons, D.~Lopez-Val, and T.~Robens, ``{Heavy to light Higgs
  boson decays at NLO in the Singlet Extension of the Standard Model},''
  \href{http://dx.doi.org/10.1007/JHEP02(2016)147}{{\em JHEP} {\bfseries 02}
  (2016) 147}, \href{http://arxiv.org/abs/1511.08120}{{\ttfamily
  arXiv:1511.08120 [hep-ph]}}.

\bibitem{Pilaftsis:1997dr}
A.~Pilaftsis, ``{Resonant CP violation induced by particle mixing in transition
  amplitudes},'' \href{http://dx.doi.org/10.1016/S0550-3213(97)00469-0}{{\em
  Nucl. Phys. B} {\bfseries 504} (1997) 61--107},
  \href{http://arxiv.org/abs/hep-ph/9702393}{{\ttfamily arXiv:hep-ph/9702393}}.

\bibitem{Patel:2016fam}
H.~H. Patel, ``{Package-X 2.0: A Mathematica package for the analytic
  calculation of one-loop integrals},''
  \href{http://dx.doi.org/10.1016/j.cpc.2017.04.015}{{\em Comput. Phys.
  Commun.} {\bfseries 218} (2017) 66--70},
  \href{http://arxiv.org/abs/1612.00009}{{\ttfamily arXiv:1612.00009
  [hep-ph]}}.

\end{thebibliography}\endgroup
\bibliographystyle{utphys}

\end{document}